\documentclass[twocolumn,showpacs,preprintnumbers,amsmath,amssymb,superscriptaddress,nofootinbib]{revtex4}
\usepackage{graphicx}% Include figure files
\usepackage{dcolumn}% Align table columns on decimal point
\usepackage{bm}% bold math
\usepackage[chatter]{rotating}
\def\bea{\begin{eqnarray}}
\def\eea{\end{eqnarray}}

\begin{document}

%\preprint{Version 2.2}

\title{Phenomenological models of two-particle correlation distributions on transverse momentum in relativistic heavy-ion collisions}

\affiliation{Department of Physics, The University of Texas at Austin, Austin, Texas 78712 USA}
\author{R. L. Ray and A. Jentsch}\affiliation{Department of Physics, The University of Texas at Austin, Austin, Texas 78712 USA}

%%%%%%%%%%%%%%%%%%%%%%%%%%%%%%%%%%%%%%%
\date{\today}

\begin{abstract}
Two-particle, pair-number correlation distributions on two-dimensional transverse momentum ($p_{t1},p_{t2}$) constructed from the particle production in relativistic heavy-ion collisions allow access to dynamical processes in these systems beyond what can be studied with angular correlations alone. Only a few measurements of this type have been reported in the literature and phenomenological models, which facilitate physical interpretation of the correlation structures, are non-existent. On-going effort at the Relativistic Heavy-Ion Collider (RHIC) will provide a significant volume of these correlation measurements in the future. In anticipation of these new data two phenomenological models are developed which describe two-dimensional 2D correlation distributions on transverse momentum. One model is based on a collision event-by-event fluctuating blast wave. The other is based on event-by-event fluctuations in fragmenting color-flux tubes and in jets. Both models are shown to be capable of accurately describing the measured single-particle $p_t$ distributions for minimum-bias Au+Au collisions at $\sqrt{s_{\rm NN}} = 200$~GeV. Both models are then applied to preliminary, charged-particle correlation measurements on 2D transverse momentum.  The capabilities of the two models for describing the overall structure of these correlations, the stability of the fitting results with respect to collision centrality, and the resulting trends of the dynamical fluctuations are evaluated. In general, both phenomenological models are capable of qualitatively describing the major correlation structures on transverse momentum and can be used to establish the required magnitudes and centrality trends of the fluctuations. Both models will be useful for interpreting the forthcoming correlation data from the RHIC.
\end{abstract}


\maketitle

\section{Introduction}
\label{SecI}

Two-particle correlations constructed from the particles produced in high-energy, heavy-ion collisions are affected by partonic and hadronic dynamics throughout the spatio-temporal evolution of the hot, dense collision system. These dynamics include soft and hard interactions as predicted by quantum chromodynamics (QCD), fragmentation and hadronization~\cite{LUND,PYTHIA,TomFrag}, partonic/hadronic collective flow~\cite{flow}, plus others~\cite{HBT,Tomreview}. For symmetric, unpolarized collision systems ({\em e.g.} p+p, Au+Au, Pb+Pb) near mid-rapidity, two-particle correlations can be completely described using the four kinematic and angular variables $p_{t1}$, $p_{t2}$ (transverse momentum), $\eta_1 - \eta_2$ (relative pseudorapidity\footnote{Pseudorapidity is defined as $\eta = -\ln[\tan(\theta/2)]$, where $\theta$ is the polar scattering angle relative to the beam direction.}), and $\phi_1 - \phi_2$ (relative azimuthal angle)~\cite{AyaCD,LHCeta1eta2,Tompp}. Correlation measurements on two-dimnsional 2D ($\eta_1 - \eta_2$,$\phi_1 - \phi_2$) angular space within 2D bins on transverse momentum space ($p_{t1},p_{t2}$)~\cite{trigassoc} should, in principle, represent all the statistically accessible information. Unfortunately, the absolute normalization of 2D angular correlations is poorly determined due to the arbitrary multiplicity fluctuations arising from finite-width multiplicity bins~\cite{STARptscale1,STARptscale2}.\footnote{A derivation of normalized 2D angular correlations of binned total $p_t$, using an angular scale-dependent mean-$p_t$ fluctuation method, is given in Refs.~\cite{STARptscale1,STARptscale2}. Application of this method to normalize the pair-number angular correlations is problematic because the finite multiplicity bin-width contributes directly to event-wise multiplicity fluctuations in the angular bins.} Measurements to date of four-dimensional 4D, two-particle correlations~\cite{LizHQ,LizThesis,PrabhatThesis,Kettlerptdependent} are therefore incomplete.

In Ref.~\cite{MCBias} it was shown that two-particle pair-number correlation distributions on $(p_{t1},p_{t2})$ can be derived from measures of non-statistical mean-$p_t$ fluctuations and that these correlations determine the average value (normalization) of the 2D angular correlations in each $(p_{t1},p_{t2})$ bin, thus allowing the experimental determination of the 4D correlations to be completed. However, experimental and theoretical efforts in correlation studies have mainly involved angular correlations, while measurements and analysis of pair-number correlations on $(p_{t1},p_{t2})$ have received much less attention. A few such measurements have been reported by the NA49 Collaboration~\cite{ReidNA49mtmt,NA49ptpt}, the CERES Collaboration~\cite{CERESptpt}, and the STAR Collaboration~\cite{ReidSTARmtmt,Ayamtmt,Tompp}. A much larger volume of preliminary $(p_{t1},p_{t2})$ correlation measurements by the STAR Collaboration exists~\cite{LizHQ,LizThesis}.

In addition to controlling the normalization of angular correlations, the measurement and analysis of $(p_{t1},p_{t2})$ correlations allow access to independent dynamical information beyond what can be gleaned from angular correlations alone. For example, in the hydrodynamic picture, event-wise fluctuations in global temperature would not be manifest in angular correlations, but would produce a distinctive ``saddle-shape'' correlation on $(p_{t1},p_{t2})$~\cite{NA49ptpt,Ayamtmt}. In fragmentation based models with jets, {\em e.g.} {\sc hijing}~\cite{HIJING}, where event-wise fluctuations occur in the angular positions and energies of the jets, analysis of angular correlations can determine the average number of jet-related pairs of particles per event. Analysis of $(p_{t1},p_{t2})$ correlations can determine the variance of the fluctuating number of jet-related pairs, an independent quantity. Pair-number correlations on $(p_{t1},p_{t2})$ in jet production models are sensitive to event-wise dynamical fluctuations in both the number and energy of the jets, thus providing access to additional dynamical information beyond that which can be studied with angular correlations.

Of equal importance is the close connection between number correlations on $(p_{t1},p_{t2})$ and model-dependent interpretation of single-particle $p_t$ spectrum data. In conventional hydrodynamic or blast-wave models~\cite{STARBWFits} $p_t$ spectrum data are analyzed with the intent of determining physical properties of the heavy-ion collision produced medium, or quark-gluon plasma (QGP). These properties include temperature, chemical potentials, and radial flow. Often, such models do not include event-wise fluctuations. They cannot produce correlations on $(p_{t1},p_{t2})$ and are therefore unphysical. The absence of fluctuations affects the shape of the $p_t$ spectrum and therefore the fits to the data, resulting in inaccurate measures of medium properties.

In high-energy minimum-bias p+p collisions a straightforward correspondence exists between angular and $(p_{t1},p_{t2})$ correlation structures for non-identified charged particles as shown in Ref.~\cite{Porter}. Two correlation peaks appear on $(p_{t1},p_{t2})$ at lower and higher $p_t$. Selecting pairs in the lower $p_t$ peak results in angular correlations consistent with longitudinal fragmentation and charge-ordering~\cite{ISR} as described by the LUND color-flux tube, or color-string model~\cite{LUND}. Selecting the higher $p_t$ pairs results in jet and dijet-like angular correlations which are well described by {\sc pythia}~\cite{PYTHIA}.

For more complex nucleus + nucleus collision systems, interpreting the correlation structures on transverse momentum coordinates is less clear, as is understanding the correspondences between correlation structures in the two, respective subspaces. For example, in 2D angular correlations~\cite{AxialCI,KoljaPaper} an azimuthal quadrupole is readily apparent, which is interpreted as pressure driven elliptic-flow in the hydrodynamic picture. Peaked correlations at relatively small opening angles are usually interpreted as jets. Back-to-back correlations in relative azimuth are interpreted as dijets or other momentum conserving processes. On the other hand, the correlation structures which have been observed so far on $(p_{t1},p_{t2})$, a saddle-shape~\cite{Ayamtmt} plus broad peak from about 1 to 2~GeV/$c$~\cite{LizHQ,LizThesis}, are not so readily interpreted. Different dynamical mechanisms, for example fluctuating jets and fragmentation versus fluctuating temperatures and radial flow, produce similar structures as will be shown in this paper.

The purpose of the present work is to develop and test two phenomenological models of relativistic heavy-ion collisions, based on distinctly different dynamical frameworks, which can be used to interpret the correlation structures and provide an efficient means for determining the nature and strength of the fluctuations, within each framework, which are required to describe the data. The required magnitudes and centrality trends of the various fluctuations within each model can be compared to that allowed by the corresponding theories, thereby testing the applicability of each theoretical framework. The phenomenologies presented here may help tease apart the underlying dynamical mechanisms and help guide theoretical developments.

The first model is based on a fluctuating blast-wave (BW)~\cite{SSH,TWH}. The second is based on fluctuating, two-component fragmentation (TCF) motivated by the success of the Kharzeev and Nardi (KN)~\cite{KN} ``soft plus hard'' two-component interaction model. The general efficacy and stability of the models are tested by fitting mathematical representations of preliminary correlation data for Au+Au collisions at $\sqrt{s_{\rm NN}}$ = 200~GeV from the STAR Collaboration~\cite{LizHQ,LizThesis}. Trends in the centrality dependences of the several fluctuating quantities in the models are presented and discussed. %Theoretical predictions for this collision system and energy, which represent the event-by-event hydrodynamic approach on the one hand and the two-component fragmentation approach on the other hand, are shown in order to establish a theoretical context for the new phenomenology and to demonstrate the status of these two theoretical frameworks with respect to $(p_{t1},p_{t2})$ correlation structures.

This paper is organized as follows. The general method for introducing dynamical fluctuations into the single- and two-particle momentum distributions is presented in Sec.~\ref{SecII}. Applications of this method for the BW and TCF models are derived in Secs.~\ref{SecIII} and \ref{SecIV}, respectively, where they are tested with respect to charged-particle $p_t$ spectra data for Au+Au collisions at $\sqrt{s_{\rm NN}}$ = 200~GeV. In Sec.~\ref{SecV} both models are further tested by fitting $(p_{t1},p_{t2})$ correlation pseudodata. The efficacy of each model, as well as the stability and centrality trends of the fluctuating quantities are also discussed in Sec.~\ref{SecV}. A summary and conclusions are given in Sec.~\ref{SecVI}.

\section{General fluctuation model}
\label{SecII}

Single-particle distributions on binned coordinates are constructed by counting all particles within a given acceptance in all collision events within a centrality class. Two-particle distributions are similarly constructed using all pairs of particles within the acceptance. If all particles in all events are emitted from equilibrated sources having the same uniform temperature $T$, using the simplest hydrodynamic picture for illustration, then the event-wise single-particle and two-particle distributions are simply the statistical samples of the same underlying {\em parent} distribution. In this case there are no correlations. To generate correlations the parent distributions must vary from event-to-event and/or within the source distribution of each event. An arbitrary i$^{\rm th}$ particle is assumed to be emitted from a region of the source having a local temperature $T_i$. If the corresponding temperatures for an arbitrary pair of particles in an event, {\em e.g.} $T_i$ and $T_j$, fluctuate independently such that the average pair-wise fluctuations about the mean temperature $\bar{T}$ vanishes, where $\langle (T_i - \bar{T})(T_j - \bar{T})\rangle_{i \neq j} = 0$, then the correlations will again vanish. Mean temperature $\bar{T}$ is the average emitting temperature for all particles in the event sample. Within this model, non-vanishing correlations can only occur when $\langle (T_i - \bar{T})(T_j - \bar{T})\rangle_{i \neq j} \neq 0$. 

For the present application each phenomenological model includes two independent sources of fluctuations in the parent distributions, either (1) temperature and transverse flow velocity, or (2) longitudinal color-flux tube energy, and jet number and energy. In this section, parameters $P$ and $Q$ are used to represent these two parameter values.

We start with a binned, single-particle density distribution on transverse momentum corresponding to an arbitrary collision event $j$, given by $\rho_{j,p_t} \equiv n_{j,p_t}/\delta_{p_t}$, where $n_{j,p_t}$ is the number of particles from the $j^{\rm th}$ event in the transverse momentum bin at $p_t$ (subscript $p_t$ is the bin index) and $\delta_{p_t}$ is the width of the bin. We then construct the event average for the total number of events $\epsilon$ in a centrality class, given by
\bea
\bar{\rho}_{p_t} & = & \frac{1}{\epsilon} \sum_{j=1}^{\epsilon} \rho_{j,p_t}
= \frac{1}{\epsilon} \sum_{j=1}^{\epsilon} \sum_{i=1}^{n_j} \kappa_{i:p_t} /\delta_{p_t},
\label{Eq1}
\eea
where $n_j$ is the number of particles in event $j$ and $\kappa_{i:p_t} = 1$ if the $i^{\rm th}$ particle is emitted into the bin at $p_t$, and is zero otherwise. Throughout this paper overlines denote event averages or other mean values.

For the phenomenological models considered here we assume that the production mechanisms are characterized by quantities $P,Q$, etc. whose values may vary within each event and from one event to the next as explained above. Each particle $(i)$ in an event $(j)$ is assumed to be produced from a region of the source characterized by discrete variables $P_{ij},Q_{ij}$. The density distribution for event $j$ generated by particles produced with discrete, source variables $P^{\prime}$ and $Q^{\prime}$ is given by
\bea
\rho_{jP^{\prime}Q^{\prime},p_t} & = & \frac{1}{\delta_{p_t}} \sum_{i=1}^{n_j} \left[ \kappa_{i:p_t} \right]_{P_{ij}=P^{\prime},Q_{ij}=Q^{\prime}}
\label{Eq2}
\eea
where $\kappa_{i:p_t} = 1$ if the source quantities for the i$^{\rm th}$ particle equal $P^{\prime}$ and $Q^{\prime}$ and the particle is emitted into the bin at $p_t$; otherwise it is zero. The $j^{\rm th}$ event distribution is therefore
\bea
\rho_{j,p_t} & = & \sum_{P^{\prime},Q^{\prime}} \rho_{jP^{\prime}Q^{\prime},p_t}
\label{Eq3}
\eea
and the event-averaged, binned distribution is given by
\bea
\bar{\rho}_{p_t} & = & \frac{1}{\epsilon} \sum_{j=1}^{\epsilon} \sum_{P^{\prime},Q^{\prime}}
\rho_{jP^{\prime}Q^{\prime},p_t} = \sum_{P^{\prime},Q^{\prime}} \frac{1}{\epsilon} \sum_{j=1}^{\epsilon}
\rho_{jP^{\prime}Q^{\prime},p_t}.
\label{Eq4}
\eea
 
The present implementation of the phenomenological models is in terms of the probability distributions for particle emission from source regions having arbitrary values $P^{\prime}$ and $Q^{\prime}$. These distributions are introduced in Eq.~(\ref{Eq4}) using a series of particle sums given by (dropping the primes)
\bea
\label{Eq5}
N_{PQ} & \equiv & \sum_{p_t} \sum_{j=1}^{\epsilon} \rho_{jPQ,p_t} \delta_{p_t}, \\
%\label{Eq5}  \\
N_P & \equiv & \sum_Q N_{PQ} %\\
\label{Eq6} \\
N & \equiv & \sum_P N_P ,
\label{Eq7}
\eea
from which Eq.~(\ref{Eq4}) becomes
\bea
\bar{\rho}_{p_t} & = & \sum_{P,Q} \frac{N}{\epsilon} \frac{N_P}{N} \frac{N_{PQ}}{N_P} 
\frac{1}{N_{PQ}} \sum_{j=1}^{\epsilon} \rho_{jPQ,p_t}.
\label{Eq8}
\eea
In Eq.~(\ref{Eq7}) $N$ is the total number of accepted particles produced in all collisions in the centrality class and $N/\epsilon \equiv \bar{N}$ is the mean multiplicity per-event. Ratio $N_P/N$ is the fraction of all particles emitted from sources with fluctuating parameter value $P$. Ratio $N_{PQ}/N_P$ is the fraction of all particles emitted from source regions with parameter value $P$ in which the other fluctuating emission quantity has value $Q$. For the models considered here we assume that the source emission parameters $P$ and $Q$ fluctuate independently of each other which allows the simplifying approximation $N_{PQ}/N_P \approx N_Q/N$. The last ratio in Eq.~(\ref{Eq8}) defines a unit-normal, binned distribution where
\bea
\hat{\rho}_{PQ,p_t} & \equiv & \frac{1}{N_{PQ}} \sum_{j=1}^{\epsilon} \rho_{jPQ,p_t}
\label{Eq9}
\eea
and $\sum_{p_t} \delta_{p_t} \hat{\rho}_{PQ,p_t} = 1.$ Throughout this paper the ``hat'' symbol denotes a unit-normalized distribution.

In the BW and TCF models the source emission parameters and the outgoing particle momentum are represented with continuous variables. The continuum limits of the above binned quantities are given by the following:
\bea
\bar{\rho}_{p_t} & \rightarrow & \bar{\rho}(p_t) \nonumber \\
N_P/N & \rightarrow & dPf(P)  \nonumber \\
N_Q/N & \rightarrow & dQg(Q)  \nonumber \\
\hat{\rho}_{PQ,p_t} & \rightarrow & \hat{\rho}(P,Q,p_t) \nonumber
\eea
where $\int dp_t \hat{\rho}(P,Q,p_t) = 1$. The single-particle density is given by
\bea
\bar{\rho}(p_t) & = & \bar{N} \int \! \int dP dQ f(P) g(Q) \hat{\rho}(P,Q,p_t).
\label{Eq10}
\eea

Similarly, the two-particle binned distribution\footnote{Throughout this paper symbol $\rho$ represents both single- and two-particle distributions. The number of particle labels distinguishes the usage.} for particles labeled 1 and 2 is given by
\bea
\bar{\rho}_{p_{t1},p_{t2}} & = & \frac{1}{\epsilon} \sum_{j=1}^{\epsilon} \frac{\bar{N}}{n_j}
\frac{n_j - 1}{n_j} \sum_{P_1,Q_1} \sum_{P_2,Q_2} \rho_{jP_1 Q_1,p_{t1}} \rho_{jP_2 Q_2,p_{t2}}
\nonumber \\
 &  & \hspace{-0.2in} = \frac{\bar{N} -1}{\bar{N}} \sum_{P_1,Q_1} \sum_{P_2,Q_2} 
\frac{1}{\epsilon} \sum_{j=1}^{\epsilon} \rho_{jP_1 Q_1,p_{t1}} \rho_{jP_2 Q_2,p_{t2}} 
\label{Eq11}
\eea
where factor $(n_j - 1)/n_j$ normalizes each event to the correct number of pairs of particles, counting both permutations, factor $\bar{N}/n_j$ eliminates statistical bias caused by multiplicity variations within the centrality bin using the $\Delta\sigma^2_{p_t:n}$ mean-$p_t$ fluctuation quantity derived in Ref.~\cite{MCBias}, and in the second line the ensemble of events is restricted to have fixed multiplicity $\bar{N}$. In Eq.~(\ref{Eq11}) particle 1 is assumed to be emitted from a region of the source where the production quantities have the values $P_1$ and $Q_1$, and similarly for particle 2. Introducing pair ratios, analogous to those in Eqs.~(\ref{Eq5})-(\ref{Eq7}), gives
\bea
N_{P_1 Q_1 P_2 Q_2} & \equiv & \sum_{p_{t1},p_{t2}} \sum_{j=1}^{\epsilon} \rho_{j P_1 Q_1 , p_{t1}}
\rho_{j P_2 Q_2 , p_{t2}} \nonumber \\
 & \times & \delta_{p_{t1}}  \delta_{p_{t2}} \\
\label{Eq12} 
N_{P_1 P_2} & \equiv & \sum_{Q_1 , Q_2} N_{P_1 Q_1 P_2 Q_2} \\
\label{Eq13}
N_{(2)} & \equiv & \sum_{P_1 , P_2} N_{P_1 P_2},
\label{Eq14}
\eea
where Eq.~(\ref{Eq11}) becomes
\bea
\bar{\rho}_{p_{t1},p_{t2}} & = & \frac{\bar{N} - 1}{\bar{N}} \sum_{P_1 , P_2}  \sum_{Q_1 , Q_2}
\frac{N_{(2)}}{\epsilon} \frac{N_{P_1 P_2}}{N_{(2)}} \frac{N_{P_1 Q_1 P_2 Q_2}}{N_{P_1 P_2}}
\nonumber \\
& \times & \frac{1}{N_{P_1 Q_1 P_2 Q_2}} \sum_{j=1}^{\epsilon} \rho_{j P_1 Q_1 , p_{t1}} \rho_{j P_2 Q_2 , p_{t2}}.
\label{Eq15}
\eea
In Eq.~(\ref{Eq14}), $N_{(2)} = \epsilon \bar{N}^2$ is the total number of pairs in the event ensemble, including self-pairs, when all events have fixed multiplicity. Assuming that fluctuation parameters $P$ and $Q$ are independent, results in $N_{P_1 Q_1 P_2 Q_2}/N_{P_1 P_2} \approx N_{Q_1 Q_2}/N_{(2)}$. The last ratio in Eq.~(\ref{Eq15}) factors into the product of unit-normalized single-particle density distributions $\hat{\rho}_{P_1 Q_1,p_{t1}} \hat{\rho}_{P_2 Q_2 ,p_{t2}}$. Clearly, if $N_{P_1 P_2} = N_{P_1} N_{P_2}$ and $N_{Q_1 Q_2} = N_{Q_1} N_{Q_2}$, then the two-particle density in Eq.~(\ref{Eq15}) factors into a product of single-particle densities, resulting in no correlations.

In the continuum limit $\bar{\rho}_{p_{t1},p_{t2}} \!\!\rightarrow \!\! \bar{\rho}(p_{t1},p_{t2})$, $N_{P_1 P_2}/N_{(2)} \!\!\rightarrow \!\!dP_1 dP_2 f(P_1,P_2)$, $N_{Q_1 Q_2}/N_{(2)} \!\!\rightarrow \!\! dQ_1 dQ_2 g(Q_1,Q_2)$, $\hat{\rho}_{P_1 Q_1,p_{t1}} \!\!\rightarrow \!\! \hat{\rho}(P_1 , Q_1 , p_{t1})$ and similarly for particle 2. The two-particle density in the continuum limit is therefore
\bea
\bar{\rho}(p_{t1},p_{t2}) & = & \bar{N}(\bar{N} - 1) \int\!\!\int\!\!\int\!\!\int dP_1 dP_2 dQ_1 dQ_2 
f(P_1,P_2) \nonumber \\
 & \times &  g(Q_1,Q_2) \hat{\rho}(P_1 , Q_1 , p_{t1}) \hat{\rho}(P_2 , Q_2 , p_{t2}).
\label{Eq16}
\eea
In the following sections explicit functional models are presented for the single-particle distributions and the emitting source parameter distributions.

\section{Blast-wave model with fluctuations}
\label{SecIII}
\subsection{Single-particle distribution}
\label{SecIIIA}

The fluctuating blast-wave model is based on the invariant phase-space source emission distribution of Schnedermann, Sollfrank and Heinz (SSH)~\cite{SSH} and as further developed by Tom\'{a}\u{s}ik, Wiedemann and Heinz~\cite{TWH}. In this model the invariant momentum distribution is calculated by integrating over the space-time coordinates of the source function $S(x,p)$, given by
\bea
E\frac{d^3N}{dp^3} & = & \frac{d^2N}{2\pi m_t dm_t dy} = \int dx^4 S(x,p) \nonumber \\
 & = & \int \tau d\tau \int d\eta_s \int rdr \int d\varphi S(x,p),
\label{Eq17}
\eea
where $x,p$ are four-vectors, $E$ is the total energy of the particle, $m_t = \sqrt{p_t^2 + m_0^2}$ is the transverse mass, and $m_0$ is assumed to be the pion rest-mass. Space-time coordinates $\tau$, $\eta_s$, $r$, and $\varphi$ are the proper time, source rapidity defined by $(1/2)\ln{[(t+z)/(t-z)]}$, transverse radius, and azimuthal angle, respectively. From Ref.~\cite{TWH} Eq.~(\ref{Eq17}) can be expressed at mid-rapidity ($y=0$) as
\bea
E\frac{d^3N}{dp^3} & = & \frac{\tau_0 m_t}{4\pi^2 \hbar^3}
\int_0^\infty rdr G(r) e^{\beta \mu_0} I_0[\beta p_t \sinh{\eta_t(r)}] \nonumber \\
  & & \hspace{-0.75in} \times \int_{-\infty}^{\infty} d\eta_s \cosh{\eta_s} H(\eta_s) \exp[-\beta m_t \cosh{\eta_t(r)} \cosh{\eta_s}] \nonumber \\
\label{Eq18}
\eea
where $\tau_0$ is the mean emission proper time, $\beta = 1/T$ is the inverse temperature, $\mu_0$ is the chemical potential, $I_0$ is a modified Bessel function, $G(r)$ and $H(\eta_s)$ are the transverse and longitudinal-rapidity source distributions, and $\eta_t(r)$ is the transverse flow rapidity. The latter is defined in terms of the transverse flow velocity $v_t(r)$, where
\bea
\eta_t(r) & = & \frac{1}{2} \ln \left( \frac{1+v_t(r)}{1-v_t(r)} \right) ,
\label{Eq19}
\eea
$v_t(r) = \tanh \eta_t(r)$, and the flow velocity profile is assumed to follow a power-law distribution given by~\cite{Fries}
\bea
v_t(r) & = & a_0 \varrho^{n_{\rm flow}}, \hspace{0.1in} \varrho \equiv r/R_0,
\label{Eq20}
\eea
where $R_0$ is the transverse radius parameter of the source.

In deriving Eq.~(\ref{Eq18}) Bjorken boost invariant expansion~{\cite{Bjorken,TWH} was assumed, which is conventional in BW models, where longitudinal flow rapidity equals $\eta_s$. The source distribution was assumed to be uniform on azimuth, {\em e.g.} no $\cos (2\phi)$ dependence, because the final-state particle-pair yield in the present application is integrated over relative azimuth intervals of either $\pi$ or 2$\pi$ where such correlations average to zero. We also assumed the following in order to simplify the model, to focus on the dominant sources of fluctuations in the $p_t$ distribution, and to simplify the numerical integrations: (1) the Maxwell-Boltzmann limit for the emission function, (2) a constant chemical potential $\mu(r) \approx \mu_0$, (3) a constant source distribution $G(r)$ from $r = 0$ to maximum radius $R_0$, and (4) the shape of source distribution $H(\eta_s)$ is taken from measured $dN_{\rm ch}/d\eta$ distributions. For the latter, $H(\eta_s)$ is taken to be symmetric about $\eta_s = 0$ for symmetric collision systems and is represented by a modified Woods-Saxon distribution given by
\bea
H(\eta_s) & = & H(|\eta_s|) = {\cal N}_s \frac{1+w|\eta_s|^2}{1+\exp[(|\eta_s|-\eta_{sr})/\eta_{st}]},
\label{Eq21}
\eea
where ${\cal N}_s$ is a normalization constant and parameters $w$, $\eta_{sr}$ (source range) and $\eta_{st}$ (source end-point thickness) were fitted to the $dN_{\rm ch}/d\eta$ distributions for minimum-bias Au+Au collisions at $\sqrt{s_{\rm NN}}$ = 200~GeV reported by the PHOBOS Collaboration~\cite{PHOBOS}. Parameter values $w$ = 0.02, $\eta_{sr}$ = 3.45 and $\eta_{st}$ = 0.73 approximately describe the shapes of these data at each measured centrality.

For applications to correlations on transverse momentum it is beneficial to display results on transverse rapidity, given by $y_t = \ln [(p_t + m_t)/m_0]$ at mid-longitudinal rapidity, where $p_t = m_0 \sinh(y_t)$. Plotting the correlations on transverse rapidity, rather than $p_t$, enhances the visual access to correlation structures at both lower and higher $p_t$. In addition, transverse rapidity is an additively boost-invariant coordinate which facilitates studies of transverse fragmentation, {\em i.e.} jets. The single-particle distribution on $y_t$ at $y=0$ (longitudinal mid-rapidity) is given by
\bea
\frac{d^2N}{dy_td\eta} & = & 2\pi p_t \frac{dp_t}{dy_t} \frac{dy}{d\eta} \left(
\frac{d^2N}{2\pi m_t dm_t dy} \right) \nonumber \\
 & = & 2\pi p_t^2 \left( \frac{d^2N}{2\pi m_t dm_t dy} \right)
\label{Eq22}
\eea
where $m_t dm_t = p_t dp_t$, and $\eta = \lim_{m_0 \rightarrow 0} y$ is pseudorapidity. Jacobians $dp_t/dy_t$ and $dy/d\eta$ equal $m_t$ and $p_t/m_t$, respectively, at mid-rapidity. The quantity in parentheses in Eq.~(\ref{Eq22}) is either taken from experiment or calculated in the blast-wave model.

A collection of collision events within a centrality bin can be expected to have fluctuating properties due to fluctuating initial-conditions~\cite{EPOS,glasma} and the stochastic nature of the system evolution from the initial impact to final kinetic decoupling. Within the context of the BW model we would therefore expect the source geometry, freeze-out temperature, and transverse flow to fluctuate from event-to-event. Furthermore, due to non-uniform initial conditions, the temperature and flow fields within each collision environment might also vary relative to the smooth, analytic distribution assumed in Eq.~(\ref{Eq18}). Fluctuations in $\tau_0$, $G$, $H$, $\mu_0$, $\beta$ and $\eta_{t}$ are therefore possible.

To account for these fluctuations we calculate the ensemble average of event-wise fluctuating BW distributions for non-identified, charged-particles within mid-rapidity acceptance $\Delta\eta$ [{\em e.g.} $\Delta\eta = 2$ for the STAR Time Projection Chamber (TPC) tracking detector~\cite{TPC}]. The measured and BW model charged-particle density distributions are related as follows, 
\bea
\bar{\rho}_{\rm ch}(y_t) & = & \Delta\eta  \frac{d^2N_{\rm ch,exp}} {dy_t d\eta}
\nonumber \\
 & = &
\Delta\eta  \frac{1}{\epsilon} \sum_{j=1}^{\epsilon} \frac{d^2N_{{\rm BW},j}}{dy_t d\eta}
+ \delta\bar{\rho}(y_t) \nonumber \\
 & \equiv & \frac{1}{\epsilon} \sum_{j=1}^{\epsilon} \rho_{{\rm BW},j} (y_t) + \delta\bar{\rho}(y_t) \nonumber \\
 & \equiv & \bar{\rho}_{\rm BW}(y_t) + \delta\bar{\rho}(y_t),
\label{Eq23}
\eea
where the measured charged-particle distribution is introduced in the first line. In Eq.~(\ref{Eq23}) the summation includes $\epsilon$ collision events within a centrality event-class and $\delta\bar{\rho}(y_t)$ is the residual between the BW model and the spectrum data. Quantities $\bar{\rho}_{\rm ch}(y_t)$ and $\bar{\rho}_{\rm BW}(y_t)$ give the event-average number of charged-particles per $y_t$ bin and are normalized to the measured number of charged particles produced within the acceptance, $y_t \in [y_{t_{\rm min}},y_{t_{\rm max}}]$, $\Delta\eta$ and $2\pi$ in azimuth.

Event averaging over $\tau_0$ and $\mu_0$ do not affect the shape of the distribution $\bar{\rho}_{\rm BW}(y_t)$, and calculations show that fluctuations in $G(r)$, or in radius $R_0$, and in $H(\eta_s)$ produce minor effects relative to those generated by fluctuations in $\beta$ and $\eta_t(r)$. We therefore fix $\tau_0$, $\mu_0$, $G(r)$ and $H(\eta_s)$ and only allow $\beta$ and $\eta_t(r)$ to fluctuate from event-to-event as well as within the source distribution of each collision. Flow fluctuations are introduced by allowing the transverse flow rapidity to fluctuate about its nominal value where in the following calculations $\eta_t(r)$ in Eq.~(\ref{Eq19}) is replaced with $\eta_{t0} \eta_t(r)$, where $\eta_{t0}$ is a random variable sampled from a peaked distribution whose variance is an adjustable parameter.

The BW distribution in Eq.~(\ref{Eq23}), with fluctuating temperature and transverse flow, is given by
\bea
\bar{\rho}_{\rm BW}(y_t) & = & \bar{N} \int d\beta f(\beta,\bar{\beta},q_{\beta}) \nonumber \\
 & \times & \int d\eta_{t0} g(\eta_{t0}, \bar{\eta}_{t0}, \sigma_{\eta_{t}})
            \hat{\rho}(\beta, \eta_{t0}, y_t)
\label{Eq24}
\eea
using the steps in Sec.~\ref{SecII}, where fluctuations in inverse temperature and transverse flow rapidity sample the probability densities $f(\beta,\bar{\beta},q_{\beta})$ and $g(\eta_{t0}, \bar{\eta}_{t0}, \sigma_{\eta_{t}})$, respectively. Both are assumed to be peaked distributions whose mean and variances are determined by parameters $\bar{\beta},q_{\beta},\bar{\eta}_{t0}$ and $\sigma_{\eta_{t}}$.

In applying the blast-wave model with fluctuating $\beta$ and $\eta_{t0}$ it was assumed that the regions of the source where $\beta$ and $ \eta_{t0}$ are greater than, or smaller than the respective means, are uniformly and randomly distributed. With this assumption the summations in Eq.~(\ref{Eq8}), for arbitrary values of $\beta$ and $\eta_{t0}$, uniformly sample the entire source volume such that the resulting invariant momentum distribution is given by Eq.~(\ref{Eq18}) when calculated with those specific $\beta$ and $\eta_{t0}$ values. Calculations of the emitted particle $p_t$ spectrum from sources with either correlated $\beta$ and $\eta_{t0}$ fluctuations, or with position correlated $\beta,\eta_{t0}$ fluctuations require microscopic models or Monte Carlo simulations, {\em e.g.} {\sc epos}~\cite{EPOS} and {\sc nexspherio}~\cite{NEXSPHERIO}, both of which are well beyond the scope and intent of the present phenomenological study.

%In applying the blast-wave model we require that the fluctuations of $\beta$ and $\eta_{t0}$, when occurring within an event system, are sufficiently small in relative amplitude and slowly varying with respect to source positions $r$ and $\eta_s$ that the emitted particle distributions on $y_t$ from hot or cold spots, and from rapidly or slowly expanding regions are adequately described by Eq.~(\ref{Eq2}), which was derived assuming uniform tempertaure and uniform $r$-dependent transverse flow. Calculations of the emitted $y_t$ spectrum from sources having variable tempertaure and transverse flow require microscopic models, {\em e.g.} {\sc epos}~\cite{EPOS} and {\sc nexspherio}~\cite{NEXSPHERIO}, and is well beyond the scope and intent of the present phenomenological model study.

In Ref.~\cite{Ayamtmt} it was shown that the transverse momentum spectrum data from relativistic heavy-ion collisions can be accurately described for $p_t < 5$~GeV/$c$ when the inverse temperature $\beta$ of a Maxwell-Boltzmann (MB) distribution, $\exp [-\beta (m_t - m_0)]$, is convoluted with a gamma distribution. The unit-normal gamma distribution is given by
\bea
f_{\gamma}(\beta,\bar{\beta},q_{\beta}) & = & \frac{q_{\beta}}{\bar{\beta}\Gamma(q_{\beta})}
\left( \frac{\beta q_{\beta}}{\bar{\beta}} \right) ^{q_{\beta} - 1} e^{-\beta q_{\beta}/\bar{\beta}}
\label{Eq25}
\eea
where $\bar{\beta}$ is the mean and $1/q_{\beta}$ is the relative variance $\sigma^2_{\beta}/\bar{\beta}^2$. The above convolution integral gives~\cite{Ayamtmt}
\bea
\int_0^\infty d\beta f_{\gamma}(\beta,\bar{\beta},q_{\beta}) e^{-\beta(m_t - m_0)} \nonumber \\
&  &  \hspace{-1.0in} =
\left[ 1 + \bar{\beta}(m_t - m_0)/q_{\beta} \right] ^{-q_{\beta}},
\label{Eq26}
\eea
a Levy distribution~\cite{Levy}.

The transverse flow rapidity scale parameter was assumed to follow a similar peaked distribution except with a suppressed long-range tail which helps the numerical integrations converge. The distribution was chosen to be a modified Gaussian given by
\bea
g(\eta_{t0}, \bar{\eta}_{t0}, \sigma_{\eta_{t}}) & = & {\cal N}_g \eta_{t0}
\exp \left[ - \frac{1}{2} \left( \frac{\eta_{t0} - \bar{\eta}_{t0}}{\sigma_{\eta_{t}}}
\right) ^2 \right]
\label{Eq27}
\eea
where $\bar{\eta}_{t0} = 1$ (fixed) and ${\cal N}_g$ normalizes the distribution to unity over the domain $\eta_{t0} \in [0,\infty]$.

\begin{widetext}
The final form of the fluctuating blast-wave single-particle distribution is given by
\bea
\bar{\rho}_{\rm BW}(y_t) & = & \bar{N} \int_0^{\infty} d\beta f_{\gamma}(\beta,\bar{\beta},q_{\beta})
     \int_0^{\infty} d\eta_{t0} g(\eta_{t0}, \bar{\eta}_{t0}, \sigma_{\eta_{t}})
  \hat{\rho}_{\rm BW}(\beta,\eta_{t0}, y_t) 
\label{Eq28}
\eea
where 
\bea
\hat{\rho}_{\rm BW}(\beta,\eta_{t0}, y_t) & = & 
{\cal N} m_t p^2_t \frac{\tau_0 R_0^2 G_0 \Delta\eta }{2\pi\hbar^3}
\int_0^1 \!\! \varrho d \varrho e^{\beta \mu_0} I_0[\beta p_t \sinh\eta_t(\varrho)]
%\nonumber \\ & \times &
 \int_{-\infty}^{\infty} \hspace{-0.15in} d\eta_s \cosh\eta_s H(\eta_s) e^{-\beta m_t \cosh\eta_t(\varrho) \cosh\eta_s}.
\label{Eq29}
\eea
Constant ${\cal N}$ ensures that $\hat{\rho}_{\rm BW}(\beta,\eta_{t0}, y_t)$ is normalized to unity in the domain $y_t \in [y_{t_{\rm min}},y_{t_{\rm max}}]$. To compare with experiment, Eq.~(\ref{Eq29}) was calculated at the $y_t$ bin centers.
In Eq.~(\ref{Eq29}) the $\eta_s$ integration was done numerically for discrete values of $\beta m_t \cosh{\eta_t(\varrho)}$ and saved for later interpolation during the three-dimensional 3D numerical integration over variables $\beta$, $\eta_{t0}$ and $\varrho$. Integration limits and step sizes were studied to ensure sufficiently accurate convergence in the calculated $y_t$ spectrum relative to the statistical errors in the data. The fit parameters in the single-particle BW model are $\bar{\beta}$ and $q_{\beta}$ in Eq.~(\ref{Eq25}), $a_0$ and $n_{\rm flow}$ in Eq.~(\ref{Eq20}), and $\sigma_{\eta_{t}}$ in Eq.~(\ref{Eq27}) where $\bar{N}$ is taken from data. These fit parameters control the mean temperature and transverse flow profile plus the temperature and flow fluctuations.
\end{widetext}

The blast-wave model was applied to the charged particle $p_t$ spectra data for Au + Au minimum-bias collisions at $\sqrt{s_{\rm NN}} = 200$~GeV measured by the STAR Collaboration~\cite{STARspectra} for collision centralities 0-5\%, 5-10\%, 10-20\%, 20-30\%, 30-40\%, 40-60\% and 60-80\%. These data were fitted within the $y_t$ range from 1.34 to 4.36, corresponding to $p_t$ from 0.25 to 5.5~GeV/$c$. Three sets of fits were done in which (1) the full BW model was used where the five parameters above were freely varied, (2) a non-flowing ($a_0 = 0$), thermal fluctuation model was used, and (3) a non-fluctuating, pure BW model was used where $q_{\beta} = \sigma_{\eta_{t}} = 0$ while $\bar{\beta}$, $a_0$ and $n_{\rm flow}$ were freely varied. Best fits were based on minimum chi-square.

Quantitative descriptions of the data were obtained for all centralities using the full blast-wave. Examples are shown in Fig.~\ref{Fig1} for the 60-80\%, 20-30\% and 0-5\% centralities where fits produced by the full BW, the non-flowing thermal fluctuation BW, and the non-fluctuating BW are shown by the solid, dashed and dotted curves, respectively. The BW model fit parameter values for all centralities and for each of the three model scenarios are listed in Table~\ref{TableI}. The residuals, $\delta \bar{\rho}(y_t)$ in Eq.~(\ref{Eq23}), for the full BW model fits are of order 5\% or less throughout the $y_t$ and centrality ranges studied here.

The full BW model accurately describes the data over the entire $y_t$ range considered in this analysis. The non-flowing, thermal fluctuation BW model overestimates the mode (peak position) but accurately describes the data at larger $y_t$. The non-fluctuating BW model overestimates the peak position by an even larger amount and underestimates the data at low $y_t$ less than 1.5 and at the largest $y_t$ bin considered here.

Typical, non-fluctuating blast-wave model fits to $p_t$ spectrum data produce results where the temperature decreases and the average flow velocity increases with centrality~\cite{STARBWFits}. In the present BW model application, the average flow velocity increases slightly with centrality, but the fitted temperature also increases. It should be noted that in the present application the fitting is performed over a larger $p_t$ range than is usually addressed with blast-wave models~\cite{STARBWFits} and the additional effects of fluctuations are included. 

The results in Table~\ref{TableI} illustrate the risk associated with relying on non-fluctuating models to infer physical properties of the medium. The temperatures and transverse flow velocities inferred with the non-fluctuating BW model fits are approximately twice and one-half, respectively, the values inferred with the full, fluctuating BW model. At a minimum, event-wise fluctuating BW models, or event-by-event hydrodynamic models, {\em e.g.} {\sc epos}~\cite{EPOS} and {\sc nexspherio}~\cite{NEXSPHERIO}, should be used in such analyses. Ideally, both the spectrum and correlation data should be fit simultaneously.

%%%%%%%%%%%%%%%%%%%%%%%%%%%%%%%%%%
\begin{table*}[htb]
\caption{Blast-wave fit model parameters for the 200 GeV Au+Au minimum-bias $p_t$ spectrum data from STAR~\cite{STARspectra} for the full BW model, the non-flowing BW, and the non-fluctuating BW as explained in the text. Data were fitted in the $y_t$ range from 1.34 to 4.36 using 30 data points at each centrality. Temperature ($T$ in GeV) equals $1/\beta$. Average transverse flow velocity $\bar{v}_t$ equals $2a_0/(n_{\rm flow} + 2)$ in units where $c$ = 1.}
\label{TableI}
\begin{tabular}{c|ccccccc|ccc|ccccc}
\hline \hline
Centrality & \multicolumn{7}{c}{Full BW} & \multicolumn{3}{c}{No Flow BW} & \multicolumn{5}{c}{No Fluct. BW} \\
  (\%)  & $T$~(GeV) & $q_{\beta}$ & $a_0$ & $n_{\rm flow}$ & $\sigma_{\eta_{t}}$ & $\bar{v}_t$ & $\frac{\chi^2}{\rm DoF}$ & $T$~(GeV) & $q_{\beta}$ & $\frac{\chi^2}{\rm DoF}$ & $T$~(GeV) & $a_0$ & $n_{\rm flow}$ & $\bar{v}_t$ & $\frac{\chi^2}{\rm DoF}$ \\
\hline
0-5  & 0.110 & 20.2 & 0.68 & 0.49 & 0.051 & 0.55 & 2.10 & 0.172 & 16.1 & 2.38 & 0.184 & 0.76 & 5.8 & 0.20 & 6.54 \\
5-10 & 0.112 & 19.8 & 0.66 & 0.47 & 0.065 & 0.53 & 1.92 & 0.169 & 15.6 & 2.23 & 0.180 & 0.76 & 5.0 & 0.22 & 5.45 \\
10-20& 0.110 & 18.7 & 0.68 & 0.57 & 0.033 & 0.53 & 1.34 & 0.166 & 14.8 & 2.04 & 0.180 & 0.77 & 5.2 & 0.21 & 4.00 \\
20-30& 0.105 & 17.5 & 0.70 & 0.60 & 0.018 & 0.54 & 1.01 & 0.162 & 14.1 & 1.66 & 0.186 & 0.78 & 7.0 & 0.17 & 4.12 \\
30-40& 0.103 & 16.7 & 0.70 & 0.64 & 0.043 & 0.53 & 1.26 & 0.157 & 13.5 & 1.85 & 0.172 & 0.80 & 5.5 & 0.21 & 4.72 \\
40-60& 0.100 & 15.3 & 0.47 & 0.04 & 0.38  & 0.46 & 0.64 & 0.144 & 12.2 & 1.15 & 0.172 & 0.82 & 7.5 & 0.17 & 3.00 \\
60-80& 0.082 & 12.9 & 0.75 & 0.89 & 0.02  & 0.52 & 0.56 & 0.129 & 11.1 & 1.08 & 0.162 & 0.84 & 8.2 & 0.16 & 3.13 \\
\hline \hline
\end{tabular}
\end{table*}
%%%%%%%%%%%%%%%%%%%%%%%%%%%%%%%%%%%%

%%%%%%%%%%%%%%%%%%%%%%%%%%%%%%
\begin{figure*}[t]
\includegraphics[keepaspectratio,width=2.5in]{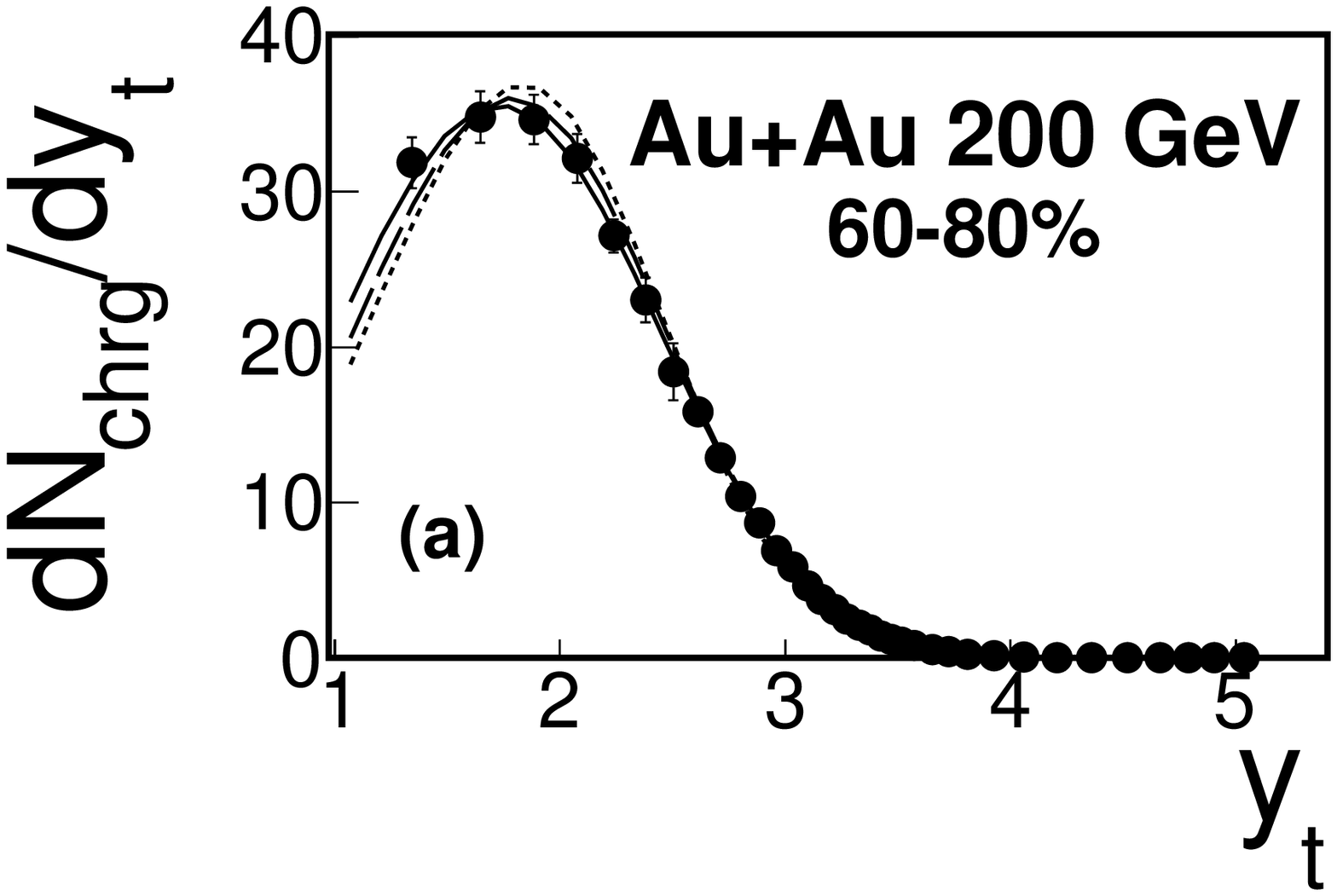}
%\put(-130,40){\bf (a)}
\includegraphics[keepaspectratio,width=2.2in]{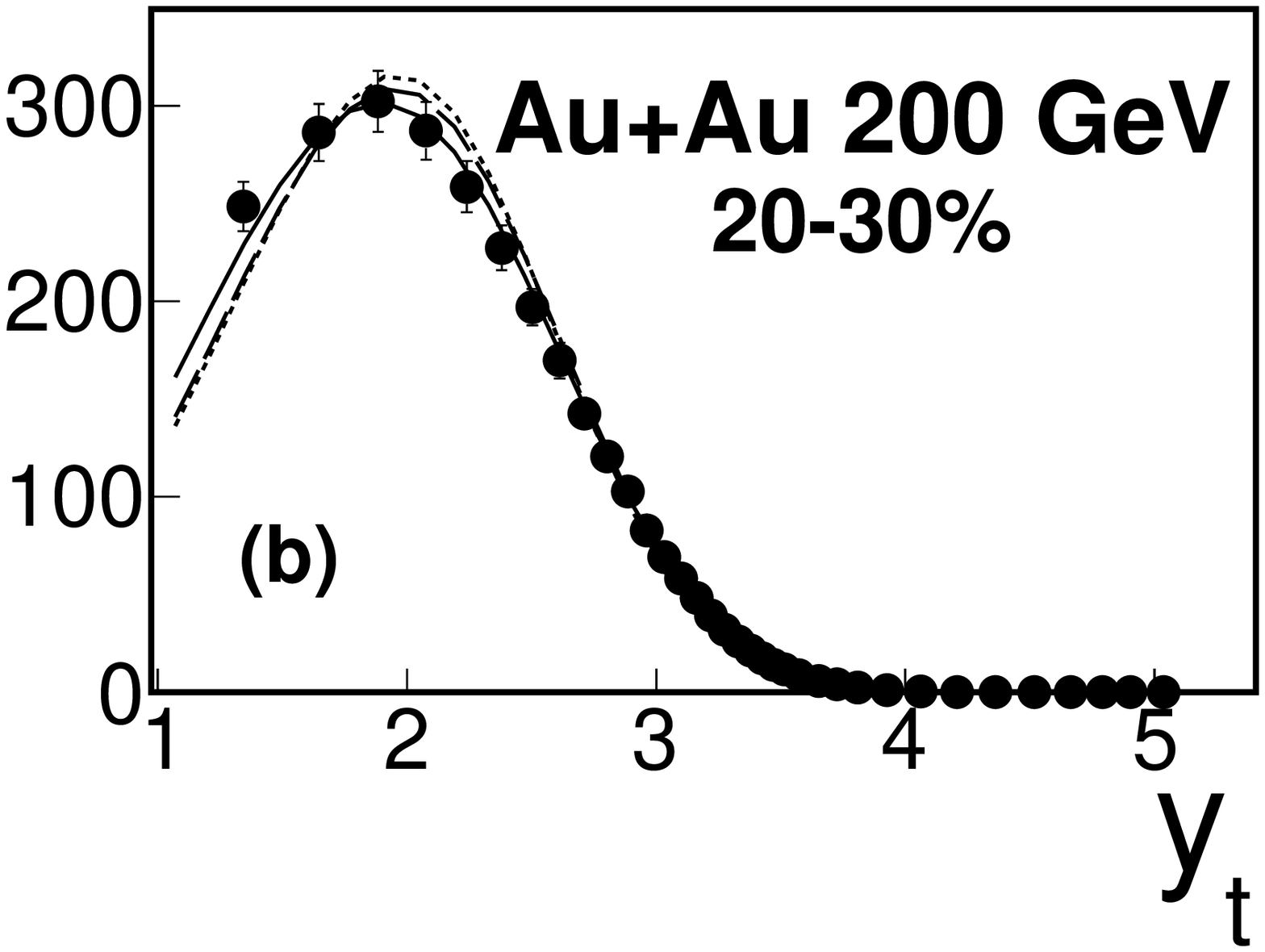}
%\put(-130,40){\bf (b)}
\includegraphics[keepaspectratio,width=2.2in]{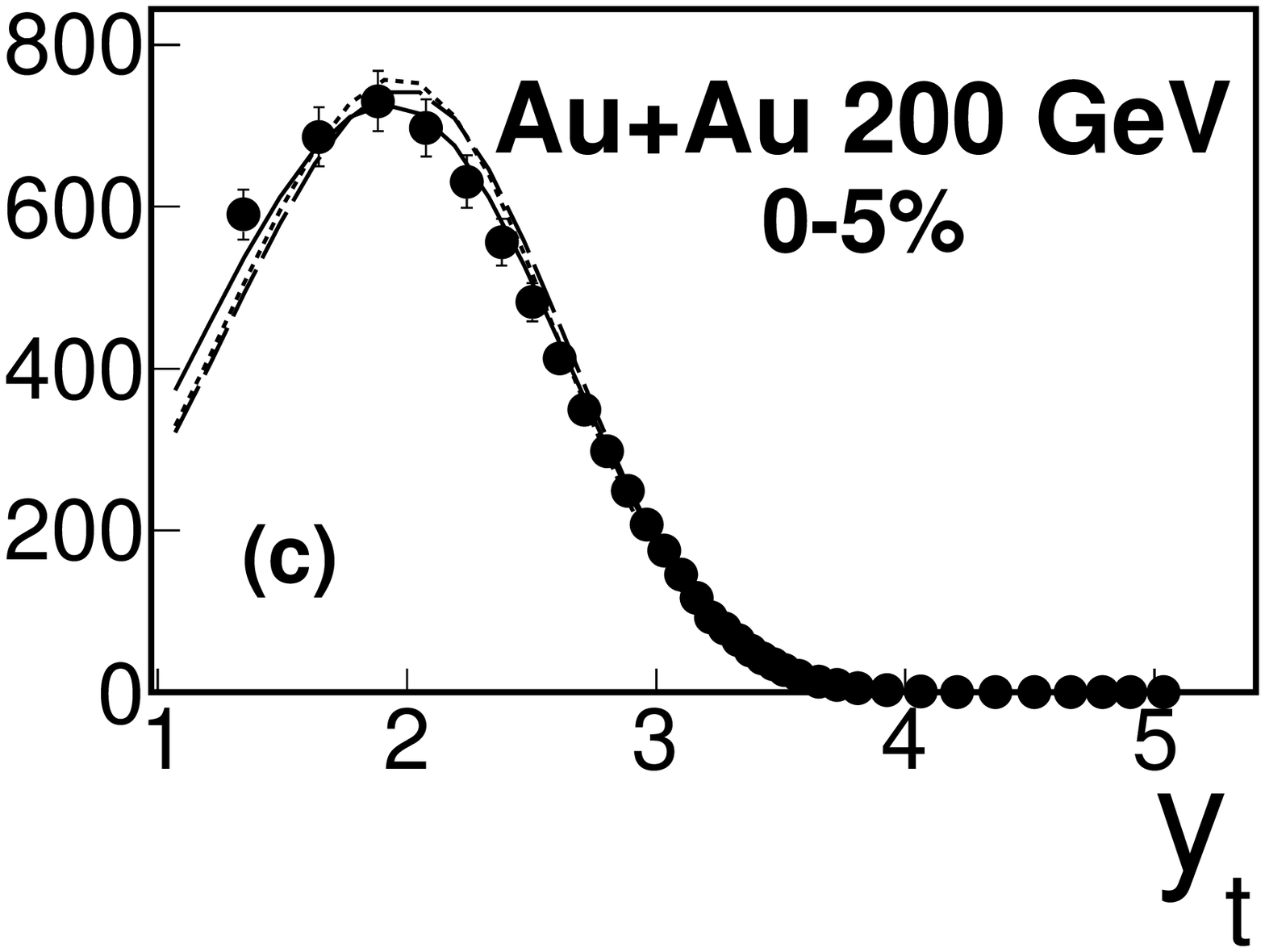} \\
%\put(-130,40){\bf (c)} \\
\includegraphics[keepaspectratio,width=2.5in]{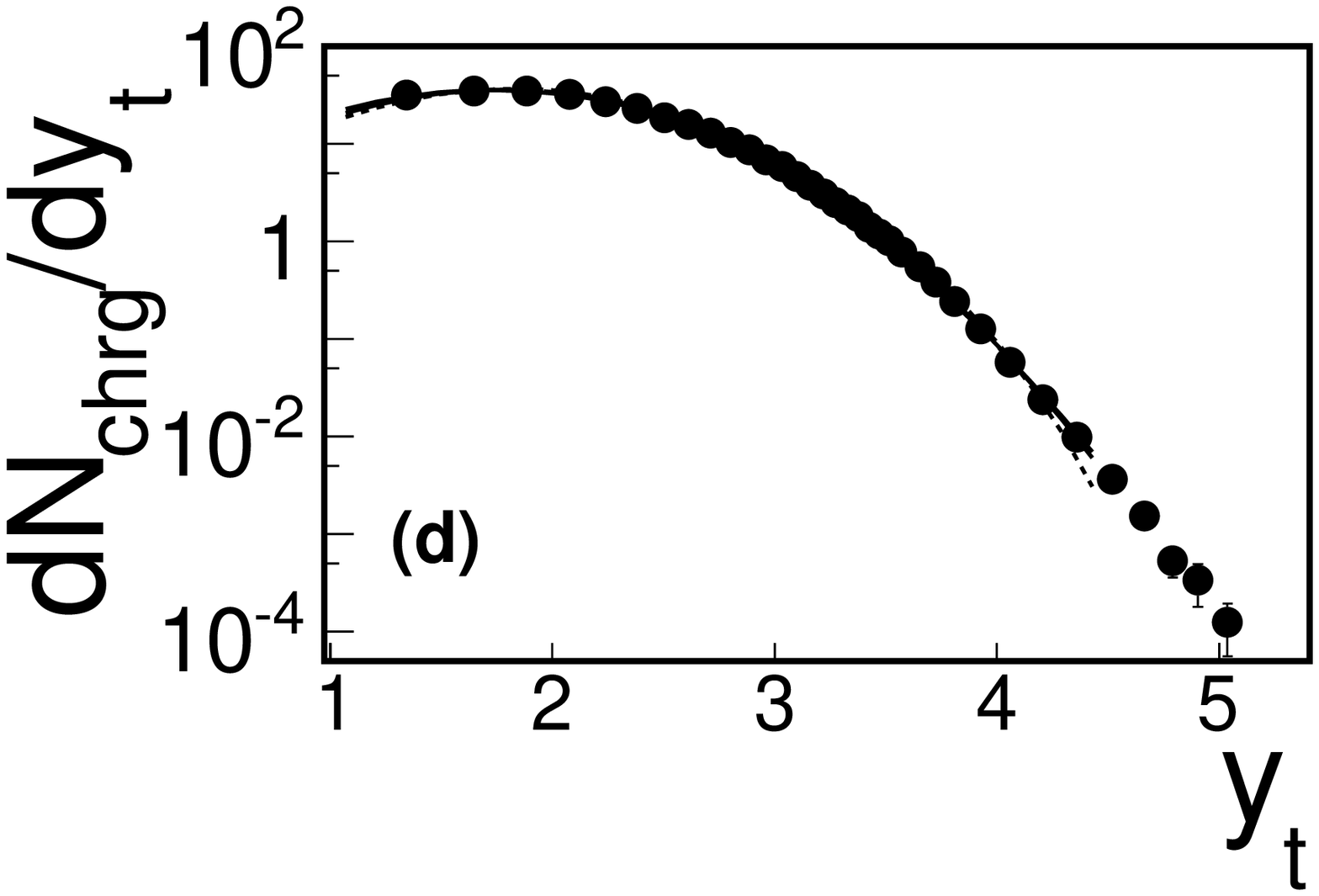}
%\put(-130,40){\bf (d)}
\includegraphics[keepaspectratio,width=2.2in]{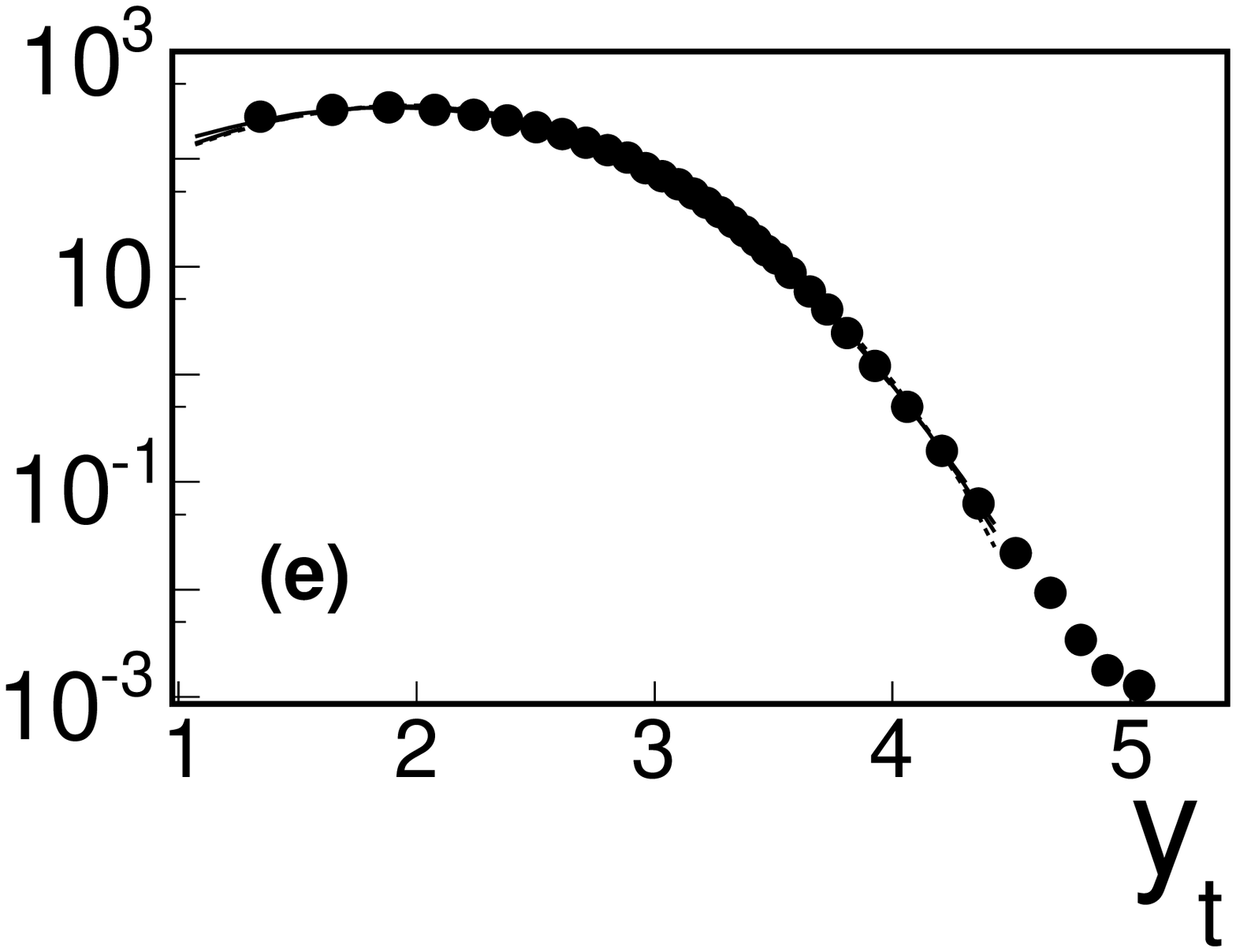}
%\put(-130,40){\bf (e)}
\includegraphics[keepaspectratio,width=2.2in]{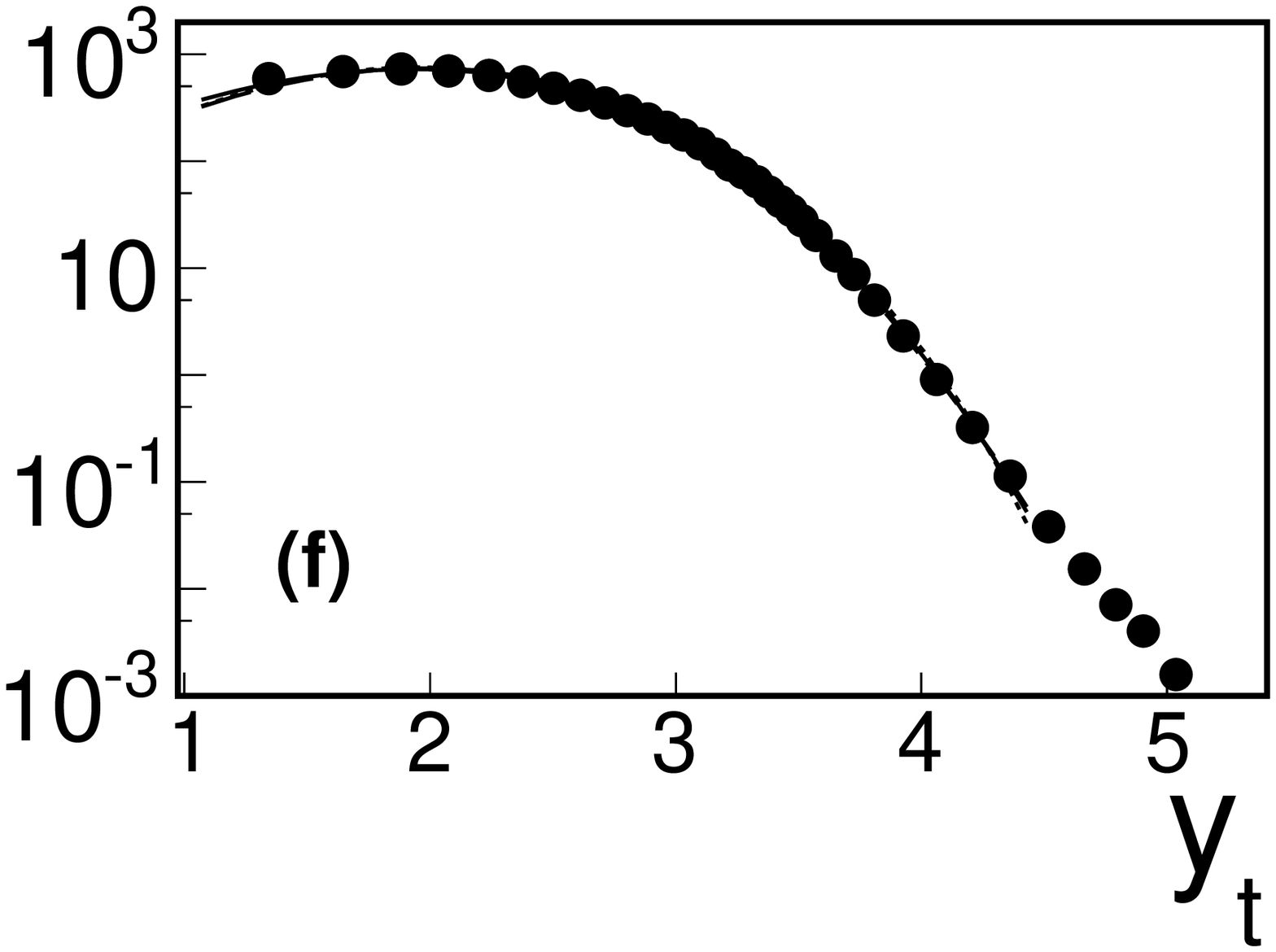}
%\put(-130,40){\bf (f)}
\caption{\label{Fig1}
Fluctuating blast-wave model fits to the 200 GeV Au+Au minimum-bias $p_t$ spectrum data from STAR~\cite{STARspectra} showing only the 60-80\%, 20-30\% and 0-5\% centrality results. Yields are shown as quantity $dN_{\rm ch}/dy_t$ versus transverse rapidity $y_t$ as defined in the text, assuming pseudorapidity acceptance $\Delta\eta = 2$. Linear (upper row) and semi-log (lower row) plots are shown for the same data and curves to allow visual access to both lower and higher $y_t$ fit results. Fit results assuming the full, non-flowing, and non-fluctuating BW models are shown by the solid, dashed and dotted curves, respectively, as explained in the text.}
\end{figure*}
%%%%%%%%%%%%%%%%%%%%%%%%%%%%%%%%%%

\subsection{Two-particle distribution}
\label{SecIIIB}

Two-particle distributions were calculated by summing over all pairs of particles from the same collision (same-event pairs denoted ``se'') for all events within a given centrality range. In the BW model, arbitrary pairs are emitted from  two, arbitrary regions of the source which are characterized by inverse temperature and transverse flow rapidity parameters $(\beta_1,\eta_{t0_1})$ and $(\beta_2,\eta_{t0_2})$, respectively. Correlations arise when the distributions of $(\beta_1,\eta_{t0_1})$ versus $(\beta_2,\eta_{t0_2})$ are correlated~\cite{Ayamtmt} (see Sec.~\ref{SecII}). Correlated fluctuations between $\beta$ and $\eta_{t0}$ are not considered here; only $(\beta_1,\beta_2)$ and $(\eta_{t0_1},\eta_{t0_2})$ correlated fluctuations are included in the present model, both for computational simplicity and in lieu of credible models of 4D $(\beta_1,\eta_{t0_1},\beta_2,\eta_{t0_2})$ correlated fluctuations.

The two-particle, same-event BW density distribution, using Eqs.~(\ref{Eq11}) and (\ref{Eq23}), is given by
\bea
\bar{\rho}_{\rm BW,se}(y_{t1},y_{t2}) & = & \frac{\bar{N} - 1}{\bar{N}} \frac{1}{\epsilon} \sum_{j=1}^{\epsilon}
\left[ \rho_{{\rm BW},j}(y_{t1}) + \delta\bar{\rho}(y_{t1}) \right] \nonumber \\
 & \times &  \left[ \rho_{{\rm BW},j}(y_{t2}) + \delta\bar{\rho}(y_{t2}) \right] \nonumber \\
 & & \hspace{-1.2in} = \frac{\bar{N} - 1}{\bar{N}} \frac{1}{\epsilon} \sum_{j=1}^{\epsilon}
\left[  \rho_{{\rm BW},j}(y_{t1}) \rho_{{\rm BW},j}(y_{t2}) + \rho_{{\rm BW},j}(y_{t1})  \delta\bar{\rho}(y_{t2})
\right.
\nonumber \\
 & & \left. \hspace{-0.7in} + \rho_{{\rm BW},j}(y_{t2})  \delta\bar{\rho}(y_{t1}) +
\delta\bar{\rho}(y_{t1}) \delta\bar{\rho}(y_{t2}) \right].
\label{Eq30}
\eea
The event averages in the second and third terms are calculated as in the preceding subsection. The last term is simply $[(\bar{N}-1)/\bar{N}] \delta\bar{\rho}(y_{t1}) \delta\bar{\rho}(y_{t2})$. The first term can be expanded as in Sec.~\ref{SecII} and is given by 
\bea
 \bar{\rho}^{\prime}_{\rm BW,se}(y_{t1},y_{t2}) &  \equiv & \bar{N}(\bar{N}-1) \int \!\! \int d\beta_1 d\beta_2
f(\beta_1,\beta_2) \nonumber \\
& & \hspace{-0.75in} \times \int \!\! \int d\eta_{t0_1} d\eta_{t0_2} g(\eta_{t0_1},\eta_{t0_2}) \nonumber \\
& & \hspace{-0.75in} \times  \hat{\rho}_{\rm BW}(\beta_1,\eta_{t0_1},y_{t1})
\hat{\rho}_{\rm BW}(\beta_2,\eta_{t0_2},y_{t2}).
\label{Eq31}
\eea

In the absence of temperature correlations $f(\beta_1,\beta_2)$ is simply a product of gamma distributions for particles 1 and 2. This uncorrelated product of gamma distributions can be expressed in terms of the sum and difference variables~\cite{Ayamtmt} $\beta_{\Sigma} = \beta_1 + \beta_2$ and $\beta_{\Delta} = \beta_1 - \beta_2$, and is given by
\bea
f_{\gamma}(\beta_1,\bar{\beta},q_{\beta}) f_{\gamma}(\beta_2,\bar{\beta},q_{\beta}) & = &
f_{\gamma}(\beta_{\Sigma},2\bar{\beta},2q_{\beta}) \tilde{f}(\beta_{\Sigma},\beta_{\Delta},q_{\beta}),
\nonumber \\
\label{Eq32}
\eea
where
\bea
\tilde{f}(\beta_{\Sigma},\beta_{\Delta},q_{\beta}) & = &
\frac{\Gamma(2q_{\beta})}{\Gamma(q_{\beta})^2}
\frac{1}{2^{2(q_{\beta} - 1)}} \frac{1}{\beta_{\Sigma}}
\left( 1 - \frac{\beta_{\Delta}^2}{\beta_{\Sigma}^2} \right) ^{q_{\beta}-1} \nonumber \\
\label{Eq33}
\eea
and $\Gamma$ is the gamma function. Parameters $\bar{\beta}$ and relative variance $\sigma^2_{\beta}/\bar{\beta}^2 = 1/q_{\beta}$ were determined by fitting the single-particle distributions (Table~\ref{TableI}).

When the source temperatures at arbitrary emission sites are correlated, the 2D distribution of $\beta_1$ and $\beta_2$ values for all particle pairs in the event collection has positive covariance as shown in the diagram in Fig.~\ref{Fig2b}. Correlated temperature emission can be introduced in Eq.~(\ref{Eq33}) by allowing the relative variances along the $\beta_{\Sigma}$ and $\beta_{\Delta}$ directions to independently vary. We therefore define
\bea
f(\beta_1,\beta_2) & \equiv & 
f_{\gamma}(\beta_{\Sigma},2\bar{\beta},2q_{\beta_\Sigma}) \tilde{f}(\beta_{\Sigma},\beta_{\Delta},q_{\beta_\Delta}).
\label{Eq34}
\eea
The correlation data will be fitted by adjusting the relative variances along the $\beta_{\Sigma}$ and $\beta_{\Delta}$ directions, $\sigma^2_{\beta_{\Sigma,\Delta}}/\bar{\beta}^2$, as shown in Fig.~\ref{Fig2b}. The shifts in the relative variances are defined by parameters $\Delta(1/q)_{\Sigma}$ and $\Delta(1/q)_{\Delta}$, respectively, where
\bea
\Delta(1/q)_{\Sigma} & = & 1/q_{\beta_\Sigma} - 1/q_{\beta} \nonumber \\
\Delta(1/q)_{\Delta} & = & 1/q_{\beta_\Delta} - 1/q_{\beta}.
\label{Eq35}
\eea
If $\Delta(1/q)_{\Sigma} > \Delta(1/q)_{\Delta}$, then the $\beta$ emissions are correlated and if $\Delta(1/q)_{\Sigma} < \Delta(1/q)_{\Delta}$, then they are anti-correlated. Results are more conveniently reported as the combinations
\bea
\Delta(1/q)_{\rm Vol} & \equiv & [\Delta(1/q)_{\Sigma} + \Delta(1/q)_{\Delta} ]/2 \nonumber \\
 & = & \frac{(\sigma^2_{\beta_\Sigma} - \sigma^2_{\beta})}{2\bar{\beta}^2}
   +   \frac{(\sigma^2_{\beta_\Delta} - \sigma^2_{\beta})}{2\bar{\beta}^2} \nonumber \\
\Delta(1/q)_{\rm cov} & \equiv & [\Delta(1/q)_{\Sigma} - \Delta(1/q)_{\Delta} ]/2, \nonumber \\
 & = & \frac{\sigma^2_{\beta_\Sigma} - \sigma^2_{\beta_\Delta}}{2\bar{\beta}^2}
\label{Eq36}
\eea
where $\Delta(1/q)_{\rm Vol}$ measures the overall (volume) change in width of the 2D $(\beta_1,\beta_2)$ distribution and $\Delta(1/q)_{\rm cov}$ indicates the covariance.

%%%%%%%%%%%%%%%%%%%%%%%%%%%%%
\begin{figure}[t]
\includegraphics[keepaspectratio,width=3.5in]{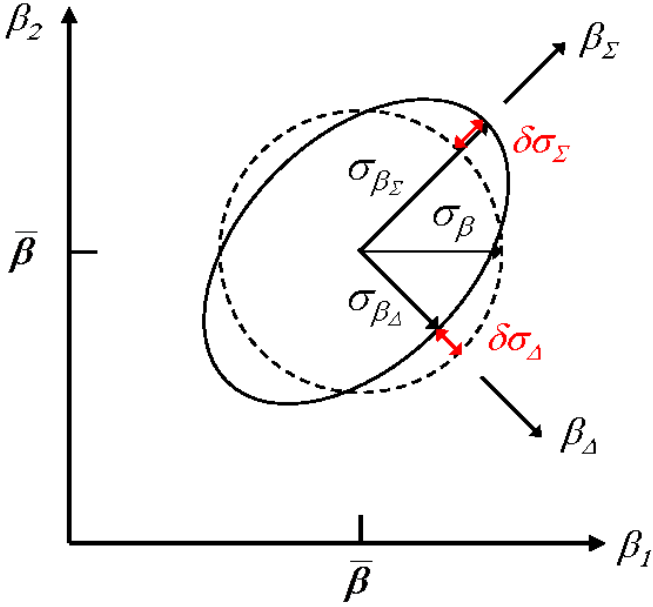}
\caption{\label{Fig2b}
Diagrammatic representation of a 2D scatter plot of the inverse temperatures ($\beta_1$ and $\beta_2$) for all pairs of particles (1,2) emitted from all events in a typical centrality bin. The dashed circle (solid ellipse) represents the 1-$\sigma$ contour of an uncorrelated (correlated) distribution. The mean and 1$\sigma$ width of the uncorrelated distribution are denoted by $\bar{\beta}$ and $\sigma_{\beta}$. Widths and changes in widths along the $\beta_{\Sigma}$ and $\beta_{\Delta}$ directions for the correlated distribution are denoted by $\sigma_{\beta_\Sigma},\delta\sigma_{\Sigma}$ and $\sigma_{\beta_\Delta},\delta\sigma_{\Delta}$, respectively.}
\end{figure}
%%%%%%%%%%%%%%%%%%%%%%%%%%%%%%%%%%

Similarly, the transverse flow rapidity scale parameter 2D distribution can be written as a product of the single-particle distributions $g(\eta_{t0},\bar{\eta}_{t0},\sigma_{\eta_t})$. The product can be expressed in terms of sum and difference variables  $\eta_{t0_{\Sigma,\Delta}} = \eta_{t0_1} \pm \eta_{t0_2}$, and the variances along $\eta_{t0_{\Sigma}}$ and $\eta_{t0_{\Delta}}$ can be varied in order to fit the correlation data. The resulting correlated, transverse-flow rapidity scale parameter distribution is given by
\begin{widetext}
\begin{equation}
g(\eta_{t0_1},\eta_{t0_2})  \equiv  
g_2(\eta_{t0_1},\eta_{t0_2},\bar{\eta}_{t0},\sigma_{\eta_{t\Sigma}},\sigma_{\eta_{t\Delta}})
= {\cal N}_{g_2} \eta_{t0_1} \eta_{t0_2} \exp \left\{ -\frac{1}{2} \left[
\left( \frac{\eta_{t0_\Sigma} - 2 \bar{\eta}_{t0}}{\sqrt{2} \sigma_{\eta_{t\Sigma}}} \right) ^2
+ \left( \frac{\eta_{t0_\Delta}}{\sqrt{2} \sigma_{\eta_{t\Delta}}  } \right) ^2
\right] \right\},
\label{Eq37}
\end{equation}
where ${\cal N}_{g_2}$ normalizes the 2D distribution to unity, the widths are defined as
$\sigma_{\eta_{t\Sigma}}  \equiv  \sigma_{\eta_t} + \Delta_{\eta_t}/2$ and $\sigma_{\eta_{t\Delta}}  \equiv  \sigma_{\eta_t} - \Delta_{\eta_t}/2$, $\bar{\eta}_{t0}$ is fixed to 1, and parameter $\sigma_{\eta_t}$ was determined by fitting the single-particle distributions (Table~\ref{TableI}). 

Using Eqs.~(\ref{Eq29}), (\ref{Eq34}) and (\ref{Eq37}) in Eq.~(\ref{Eq31}) gives the 4D integration result
\bea
\bar{\rho}^{\prime}_{\rm BW,se}(y_{t1},y_{t2}) % & = & \nonumber \\
%\hspace{-1.0in} & = & 
 & = & \bar{N}(\bar{N}-1) \int \!\! \int d\beta_1 d\beta_2
f_{\gamma}(\beta_{\Sigma},2\bar{\beta},2q_{\beta_\Sigma})
\tilde{f}(\beta_{\Sigma},\beta_{\Delta},q_{\beta_\Delta}) % \nonumber \\
%\hspace{-1.0in} &  & \times
\int \!\! \int d\eta_{t0_1} d\eta_{t0_2}
g_2(\eta_{t0_1},\eta_{t0_2},\bar{\eta}_{t0},\sigma_{\eta_{t\Sigma}},\sigma_{\eta_{t\Delta}})
\nonumber \\
&  \times & \hat{\rho}_{\rm BW}(\beta_1, \eta_{t0_1},y_{t1})
            \hat{\rho}_{\rm BW}(\beta_2, \eta_{t0_2},y_{t2}).
\label{Eq38}
\eea
The same numerical integration ranges and step sizes used for the single-particle BW calculation were used in the numerical integration in Eq.~(\ref{Eq38}).
\end{widetext}

\subsection{Two-particle correlation}
\label{SecIIIC}

By definition, the two-particle correlations contained in the two-particle, BW distribution in Eq.~(\ref{Eq38}) equal the difference between it and the product of marginals, where
\bea
\bar{\rho}_{\rm BW,marg}(y_{t1}) & \equiv & \frac{1}{\bar{N}-1} \int dy_{t2}
\bar{\rho}^{\prime}_{\rm BW,se}(y_{t1},y_{t2})
\label{Eq39}
\eea
with normalization $\int dy_{t1} \bar{\rho}_{\rm BW,marg}(y_{t1}) = \bar{N}$. To ensure consistency with the single-particle measurements we also require the marginal of the entire two-particle distribution in Eq.~(\ref{Eq30}) to equal the measured charge distribution $\bar{\rho}_{\rm ch}(y_t)$ in Eq.~(\ref{Eq23}). However, in order to fit the correlation data the variances in the inverse temperature and transverse flow rapidity, $\Delta(1/q)_{\rm Vol,cov}$ and $\Delta_{\eta_t}$, were freely varied resulting in marginals which may not precisely equal $\bar{\rho}_{\rm BW}(y_t)$ in Eq.~(\ref{Eq23}). This condition requires an adjusted residual $\delta\bar{\rho}^{\prime}(y_t)$ defined by
\bea
\delta\bar{\rho}^{\prime}(y_t) & \equiv & \bar{\rho}_{\rm ch}(y_t) - \bar{\rho}_{\rm BW,marg}(y_{t}).
\label{Eq40}
\eea
The adjusted residual is normalized such that $\int dy_t \delta\bar{\rho}^{\prime}(y_t) = 0$ because both $\bar{\rho}_{\rm ch}$ and $\bar{\rho}_{\rm BW,marg}$ are normalized to $\bar{N}$. Acceptable BW correlation model fits should not only describe the correlation data but should maintain a small residual such that $\delta\bar{\rho}^{\prime}(y_t) \ll  \bar{\rho}_{\rm ch}(y_t)$.

\begin{widetext}
The complete two-particle distribution, whose marginal equals the measured single-particle charge distribution, must be adjusted from the original form in Eq.~(\ref{Eq30}). The adjusted distribution is given by
\bea
\bar{\rho}_{\rm BW,se}(y_{t1},y_{t2}) & = & \bar{\rho}^{\prime}_{\rm BW,se}(y_{t1},y_{t2})
 \nonumber \\
\hspace{-0.5in}  & & + \frac{\bar{N}-1}{\bar{N}} \left[ \bar{\rho}_{\rm BW,marg}(y_{t1})
\delta\bar{\rho}^{\prime}(y_{t2}) + \bar{\rho}_{\rm BW,marg}(y_{t2}) \delta\bar{\rho}^{\prime}(y_{t1}) 
+ \delta\bar{\rho}^{\prime}(y_{t1}) \delta\bar{\rho}^{\prime}(y_{t2}) \right],
\label{Eq41}
\eea
where the pair normalization factor $(\bar{N}-1)/\bar{N}$ from Eq.~(\ref{Eq30}) was applied to the last three terms.
The uncorrelated reference pair distribution is defined as the product of marginals of $\bar{\rho}_{\rm BW,se}(y_{t1},y_{t2})$ in Eq.~(\ref{Eq41}), which is given by
\bea
\bar{\rho}_{\rm BW,ref}(y_{t1},y_{t2}) & = & \frac{\bar{N}-1}{\bar{N}} \left[
\bar{\rho}_{\rm BW,marg}(y_{t1}) \bar{\rho}_{\rm BW,marg}(y_{t2}) 
 + \bar{\rho}_{\rm BW,marg}(y_{t1}) \delta\bar{\rho}^{\prime}(y_{t2})
+ \bar{\rho}_{\rm BW,marg}(y_{t2}) \delta\bar{\rho}^{\prime}(y_{t1}) \right. \nonumber \\
  & & \left. +  \delta\bar{\rho}^{\prime}(y_{t1}) \delta\bar{\rho}^{\prime}(y_{t2}) \right]
  =  \frac{\bar{N}-1}{\bar{N}} \bar{\rho}_{\rm ch}(y_{t1}) \bar{\rho}_{\rm ch}(y_{t2})
\label{Eq42}
\eea
where the pair normalization factor $(\bar{N}-1)/\bar{N}$ must also be applied to the reference as shown in Ref.~\cite{MCBias}.
The per-pair normalized correlation is finally given by
\bea
\frac{\Delta\bar{\rho}_{\rm BW}}{\bar{\rho}_{\rm BW,ref}}(y_{t1},y_{t2}) & \equiv &
\frac{\bar{\rho}_{\rm BW,se}(y_{t1},y_{t2}) - \bar{\rho}_{\rm BW,ref}(y_{t1},y_{t2})}{\bar{\rho}_{\rm BW,ref}(y_{t1},y_{t2})} 
 = \frac{\bar{\rho}^{\prime}_{\rm BW,se}(y_{t1},y_{t2}) - \frac{\bar{N}-1}{\bar{N}}
\bar{\rho}_{\rm BW,marg}(y_{t1}) \bar{\rho}_{\rm BW,marg}(y_{t2})}
{\bar{\rho}_{\rm BW,ref}(y_{t1},y_{t2})}.
 \nonumber \\
\label{Eq43}
\eea
\end{widetext}

\subsection{Correlation prefactor}
\label{SecIIID}

The final BW correlation quantity to be compared with data includes a prefactor corresponding to that applied to the data~\cite{LizThesis}. In general, the purpose of a correlation prefactor is to replace the pair ratio in Eq.~(\ref{Eq43}), which is required in data analysis to correct for efficiency and acceptance, with a quantity better suited to the study of specific scaling trends, {\em e.g.} binary scaling, per-trigger scaling, etc. A prefactor may also be required by the specific charge-pair combinations used, and the relative pseudorapidity and/or azimuthal angle selections.

In the present model the specific purposes of the correlation prefactor are: (1) To convert the number of correlated pairs per final-state {\em pair} quantity in Eq.~(\ref{Eq43}) to a number of correlated pairs per final-state {\em particle} ratio as in Pearson's correlation coefficient~\cite{LizThesis,Pearson}. (2) To scale this ``pairs per singles'' ratio to account for the fact that only one-half of the available charged-particle pairs are included when selecting only the away-side pairs whose relative azimuth angle $|\phi_1 - \phi_2|$ exceeds $ \pi/2$. (Away-side pairs were selected for the analytical model fitting in Ref.~\cite{LizHQ}, and in the present analysis, in order to suppress contributions from HBT correlations~\cite{Ayamtmt}.) (3) To provide an overall normalization which facilitates tests of binary scaling in the correlation structures. The last requirement can be achieved by using the soft-QCD process particle yield, as estimated in the Kharzeev-Nardi~\cite{KN} two-component model. In the KN model, soft-QCD yields are proportional to $N_{\rm part}$,  where $N_{\rm part}$ is the number of participant nucleons in the heavy-ion collision. If the number of correlated pairs in the numerator is proportional to the number of binary nucleon + nucleon (N+N) interactions, $N_{\rm bin}$, then the resulting correlation quantity will be proportional to $N_{\rm bin}/N_{\rm part}$. Ratio $N_{\rm bin}/N_{\rm part}$ is proportional to centrality measure $\nu \equiv N_{\rm bin}/(N_{\rm part}/2)$~\cite{AxialCI}. Correlation structures which scale with $N_{\rm bin}$ will linearly increase with centrality measure $\nu$, and can therefore be readily identified.

For the present study we use a charge-independent (CI, all charge-pair combinations), away-side azimuth (AS, $|\phi_1 - \phi_2| > \pi/2$), soft-process particle production prefactor, ${\cal P}^{\rm AS-CI}_{\rm Fac,soft}(y_{t1},y_{t2})$. The final correlation quantity is given by
\bea
\frac{\Delta\bar{\rho}_{\rm BW}} {\sqrt{\bar{\rho}_{\rm soft}}}(y_{t1},y_{t2}) & \equiv &
{\cal P}^{\rm AS-CI}_{\rm Fac,soft}(y_{t1},y_{t2})
\frac{\Delta\bar{\rho}_{\rm BW}}{\bar{\rho}_{\rm BW,ref}}(y_{t1},y_{t2})
 \nonumber \\
\label{Eq44}
\eea
where the prefactor is defined and calculated in Appendix A and the last quantity is given in Eq.~(\ref{Eq43}).

\section{Two-component fragmentation model with fluctuations}
\label{SecIV}
\subsection{Single-particle distribution}
\label{SecIVA}

The two-component fragmentation model presented here is based on the two-component multiplicity production model of Kharzeev and Nardi~\cite{KN}, discussed briefly in the preceding section. In this model particle production is assumed to be dominated by two processes which scale with either $N_{\rm part}$ or $N_{\rm bin}$. The relevance of this model in the description of the peaked correlation structures on ($p_{t1},p_{t2}$) from p+p collisions was discussed in Refs.~\cite{Tompp,Porter}. In the KN model the particle yield $N$ within some $(\eta,\phi)$ acceptance is given by
\bea
N & = & n_{pp} (1-x_{\rm KN}) N_{\rm part}/2 + n_{pp} x_{\rm KN} N_{\rm bin}
\label{Eq45}
\eea
where $n_{pp} = 4.95$ is the charged-particle yield in $\sqrt{s}$ = 200~GeV non-singly diffractive, minimum-bias p+p collisions at mid-rapidity within acceptance $\Delta\eta = 2$, full $2\pi$ azimuth and $p_t >$ 0.15~GeV/$c$~\cite{Trainorspectra}. Parameter $x_{\rm KN}$ is approximately 0.1~\cite{Trainorspectra} for charged-particle production in $\sqrt{s_{\rm NN}}$ = 200~GeV minimum-bias Au+Au collisions within the preceding acceptance.

In the present application we assume the $N_{\rm part}$-scaling production derives from soft-QCD, longitudinal fragmentation of color-flux tubes~\cite{glasma}. Similarly, the $N_{\rm bin}$-scaling production corresponds to semi-hard (few GeV) and hard (few tens of GeV) QCD, transversely fragmenting partons, or jets. For the present application the $N_{\rm bin}$-scaling production is dominated by the lower energy, semi-hard part of the spectrum~\cite{TomFrag}. Fluctuations are included in the following: (1) the $p_t$-distribution shape, {\em e.g.} overall slope parameter $\beta_{cs}$, for the charged-particle production from each longitudinally fragmenting color-string~\cite{LUND,glasma}; (2) the energy of each semi-hard scattered parton and resulting jet; (3) the relative number of ``soft'' and ``semi-hard'' produced particles per event.

For a collection of collision events within a centrality bin, the mean charged-particle yield in this model is given by
\bea
\bar{\rho}_{\rm ch}(y_t) & = & \bar{\rho}_{\rm s}(y_t) + \bar{\rho}_{\rm h}(y_t) + \delta\bar{\rho}(y_t)
\label{Eq46}
\eea
for ``soft,'' ``hard,'' and residual components, respectively. The soft-component production occurs via fragmentation of longitudinal color-strings~\cite{LUND} which are assumed to produce MB $p_t$ distributions with fluctuating slope parameter $\beta_{cs}$. As in the BW model, we assume the probability distribution of parameter $\beta_{cs}$ is given by a gamma distribution, such that
\bea
\bar{\rho}_s(y_t) & = & \bar{N}_s \int d\beta_{cs} f_{\gamma}(\beta_{cs},\bar{\beta}_{cs},q_{\beta_{cs}})
\hat{\rho}_{s}(\beta_{cs},y_t)
\label{Eq47}
\eea
using the steps in Sec.~\ref{SecII}, where unit-normalized density $\hat{\rho}_{s}(\beta_{cs},y_t) \propto \exp [-\beta_{cs} (m_t - m_0)]$. The resulting Levy distribution in Eq.~(\ref{Eq47}) can be equated to the soft-production particle spectrum estimated in Appendix A, given by
\bea
\bar{\rho}_s(y_t) & = & \Delta\eta \frac{d^2 N_{\rm ch,soft}}{dy_t d\eta}.
\label{Eq48}
\eea
The mean multiplicity $\bar{N}_s$ in Eq.~(\ref{Eq47}) is determined by the parameters in Table~\ref{TableV} in Appendix A.

The semi-hard component yield is produced by fragmenting partons (jets) whose total energy fluctuates from jet-to-jet. The jet energy is represented in terms of the maximum possible transverse rapidity, $y_{\rm max}$, of its final-state fragment particles. The probability distribution of $y_{\rm max}$ is given by QCD power-law distribution $\hat{g}(y_{\rm max})$, defined in Ref.~\cite{TomFrag} and given below. The fragment distribution on $y_t$ for given jet parameter $y_{\rm max}$ is $\hat{\rho}_h(y_{\rm max},y_t)$, which is also defined below. Using the steps in Sec.~\ref{SecII}, the event-average semi-hard single-particle yield distribution is given by
\bea
\bar{\rho}_h(y_t) & = &
 \bar{N}_h \int_0^{\infty} dy_{\rm max} \hat{g}(y_{\rm max}) \hat{\rho}_{h}(y_{\rm max},y_t) \nonumber \\
 & \equiv & \bar{\rho}_{[g]}(y_t),
\label{Eq49}
\eea
where in the last line we define the convolution integral with symbol $\bar{\rho}_{[g]}(y_t)$ for later use.

Quantity $\hat{g}(y_{\rm max})$ is the probability distribution for producing particles from a jet with maximum fragment rapidity $y_{\rm max}$ in a N+N collision. In Ref.~\cite{TomFrag} this quantity is given by a QCD power-law distribution with low momentum cut-off, multiplied by a quadratic yield increase factor $(y_{\rm max} - y_{\rm min})^2$ where $y_{\rm min}$ is an empirical fitting parameter given in Ref.~\cite{TomFrag}. The quadratically increasing yield results from the approximate shape invariant evolution of the distribution of jet fragments observed at LEP in inclusive $e^+ + e^- \rightarrow {\rm jet}(Q^2)+X$ production over a wide range of jet energies~\cite{TomFrag,LEPJets}. Probability distribution $\hat{g}(y_{\rm max})$ is therefore proportional to~\cite{TomFrag}
\bea
\hat{g}(y_{\rm max}) & \propto & \nonumber \\
 &  &  \hspace{-0.75in} \frac{1}{2} \sigma_{\rm dijet} (n_{\rm QCD} - 2)
\left\{ \tanh \left( \frac{y_{\rm max} - y_{\rm cut}}{\xi_{\rm cut}} \right) + 1 \right\}
\nonumber \\
 &  &  \hspace{-0.75in} \times  e^{-(n_{\rm QCD} - 2)(y_{\rm max} - y_{\rm cut})} (y_{\rm max} - y_{\rm min})^2,
\label{Eq50}
\eea
where $\sigma_{\rm dijet}$ = 2.5~mb at $\sqrt{s}$ = 200~GeV, and from Ref.~\cite{TomFrag} $n_{\rm QCD}$ = 7.5, $y_{\rm min}$ = 0.35, and low momentum cut-off parameters are $y_{\rm cut}$ = 3.75 and $\xi_{\rm cut}$ = 0.1.

Particle distribution $\hat{\rho}_{h}(y_{\rm max},y_t)$ is proportional to the distribution deduced in Ref.~\cite{TomFrag} for $e^+ + e^- \rightarrow$~jet + $X$ multiplied by a low momentum jet-fragment suppression factor determined by analyzing the jet fragment distributions from $p + \bar{p} \rightarrow {\rm jet}+X$ collisions~\cite{FNALjets}. Quantity $\hat{\rho}_{h}(y_{\rm max},y_t)$ from Ref.~\cite{TomFrag} is proportional to
\bea
\hat{\rho}_{h}(y_{\rm max},y_t) & \propto & \tanh \left( \frac{y_t - y_0}{\xi_y} \right)
\frac{u^{\lambda -1} (1-u)^{\omega -1}}{B(\lambda,\omega)}
\label{Eq51}
\eea
for $y_{\rm max} \geq y_t \geq y_0$,  where the last factor is a normalized beta distribution with
\bea
u & \equiv & \frac{y_t - y_{\rm min}}{y_{\rm max} - y_{\rm min}}, ~~ u \in [0,1].
\label{Eq52}
\eea
Quantity $B(\lambda,\omega) = \Gamma(\lambda)\Gamma(\omega)/\Gamma(\lambda + \omega)$ where $\Gamma$ is the gamma function.

\begin{widetext}
Collecting terms, the above semi-hard process single-particle distribution becomes
\bea
 \bar{\rho}_{h}(y_t) & = & \bar{N}_h {\cal N}_h \int_0^{\infty} dy_{\rm max}
\frac{1}{2} \sigma_{\rm dijet} (n_{\rm QCD} - 2) 
\left\{ \tanh \left( \frac{y_{\rm max} - y_{\rm cut}}{\xi_{\rm cut}} \right) + 1 \right\}
\nonumber \\
 & \times & e^{-(n_{\rm QCD} - 2)(y_{\rm max} - y_{\rm cut})} (y_{\rm max} - y_{\rm min})^2
  \tanh \left( \frac{y_t - y_0}{\xi_y} \right)|_{y_t \geq y_0}
\frac{u^{\lambda -1} (1-u)^{\omega -1}}{B(\lambda,\omega)}|_{y_{\rm min} \leq y_t \leq y_{\rm max}}
\label{Eq53}
\eea
where ${\cal N}_h$ normalizes the integral of $\bar{\rho}_{h}(y_t)$ over all $y_t$ bins to $\bar{N}_h$. Quantities in Eq.~(\ref{Eq53}) are calculated at the mid-points of the $y_t$ bins when comparing with data. 
\end{widetext}

The two-component fragmentation model was applied to the charged-particle $p_t$ spectrum data discussed in Sec.~\ref{SecIII}. The semi-hard process particle production model in Eq.~(\ref{Eq53}) was fitted to the difference distribution $\bar{\rho}_{\rm ch}(y_t) - \bar{\rho}_{s}(y_t)$ in the $y_t$ range from 1.34 to 4.36 as before by varying parameters $n_{\rm QCD}$, the jet production cut-off $y_{\rm cut}$, the soft-fragment cut-off $y_0 = \xi_y$, and fragment distribution parameters $\lambda$ and $\omega$ in Eq.~(\ref{Eq51}). Better fits were achieved by variation of the shape of the fragment distribution, via parameters $\lambda$ and $\omega$, than were obtained by varying the cut-off parameter $y_0$. The latter parameter was subsequently fixed to zero.

Best fits were attained via $\chi^2$-minimization. Quantitative descriptions of the semi-hard component spectrum at the maximum peak and in the higher momentum tails were achieved for each centrality. Example fits to $\bar{\rho}_{\rm ch}(y_t)$ are shown in Fig.~\ref{Fig2} for the 60-80\%, 20-30\% and 0-5\% centralities. The TCF model parameters are listed for all centralities in Table~\ref{TableII}. Parameter $n_{\rm QCD}$ increases slightly and smoothly with centrality, increasing above the value (7.5) estimated in Ref.~\cite{TomFrag}. Jet production cut-off parameter $y_{\rm cut}$ is approximately constant and larger than the value (3.75) in Ref.~\cite{TomFrag}. The modifications of the fragment distribution (beta distribution in Eq.~(\ref{Eq51}))  relative to the nominal shape from Ref.~\cite{TomFrag} are also shown in the lower row of panels. The trends imply a softening of the fragment distribution (suppression at higher $p_t$) coupled with a suppression at lower-momentum relative to that observed in $e^+ + e^- \rightarrow$~jet + $X$, which was discussed in Ref.~\cite{TomFrag} and which quickly develops with increasing collision centrality. The residuals vary from $\leq 2$\% to $\leq 4$\% of the charged-particle distribution from peripheral to most-central collisions, respectively, for $y_t < 3$. The residuals increase in relative magnitude at larger $y_t > 3$, varying from $\leq 3$\% to $\leq 7$\% of the charged-particle distribution from peripheral to most-central collisions, respectively.

Conventional, theoretical applications of the TCF framework include event-wise fluctuations which cause the $p_t$ spectra to vary event-by-event resulting in non-zero correlations on transverse momentum. The Monte Carlo code {\sc hijing}~\cite{HIJING}, which combines the LUND model~\cite{LUND} and {\sc pythia}~\cite{PYTHIA}, includes fluctuating particle production from fragmenting color-strings and minijets. {\sc ampt}~\cite{AMPT} incorporates event-wise fluctuating initial conditions from {\sc hijing}, then includes stochastic parton propagation and interactions followed by hadronization. 

%%%%%%%%%%%%%%%%%%%%%%%%%%%%%%%%%%
\begin{table}[htb]
\caption{TCF model fit parameters for the 200 GeV Au+Au minimum-bias $p_t$ spectra data from STAR~\cite{STARspectra} within $\Delta\eta = 2$ units acceptance. Data were fit in the $y_t$ range from 1.34 to 4.36 using 30 data points at each centrality. Fit quality was insensitive to soft-fragment cut-off parameter $y_0$ which was subsequently set to zero.}
\label{TableII}
\begin{tabular}{cccccccc}
\hline \hline
Centrality(\%) & $\bar{N}_s$ & $\bar{N}_h$ & $n_{\rm QCD}$ & $y_{\rm cut}$ & $\lambda$  &   $\omega$   &   $\chi^2$/DoF \\
\hline
0-5   & 649.2  & 479.5 &  9.1  &  4.3  &  5.6  &  6.9  &  1.632 \\
5-10  & 555.1  & 354.9 &  9.0  &  4.3  &  5.9  &  7.0  &  1.823 \\
10-20 & 433.1  & 256.7 &  8.55 &  4.35 & 5.95  &  7.3  &  1.042 \\
20-30 & 308.4  & 160.0 &  8.35 &  4.4  &  6.2  &  7.6  &  0.957 \\
30-40 & 215.2  &  98.8 &  8.1  &  4.45 &  6.25 &  7.85 &  1.266 \\
40-60 & 110.8  &  53.4 &  7.6  &  4.3  &  4.7  &  6.4  &  0.665 \\
60-80 &  36.1  &  16.6 &  7.15 &  4.25 &  3.5  &  5.7  &  0.410 \\
\hline \hline
\end{tabular}
\end{table}
%%%%%%%%%%%%%%%%%%%%%%%%%%%%%%%%%%%%

%%%%%%%%%%%%%%%%%%%%%%%%%%%%%
\begin{figure*}[t]
\includegraphics[keepaspectratio,width=2.5in]{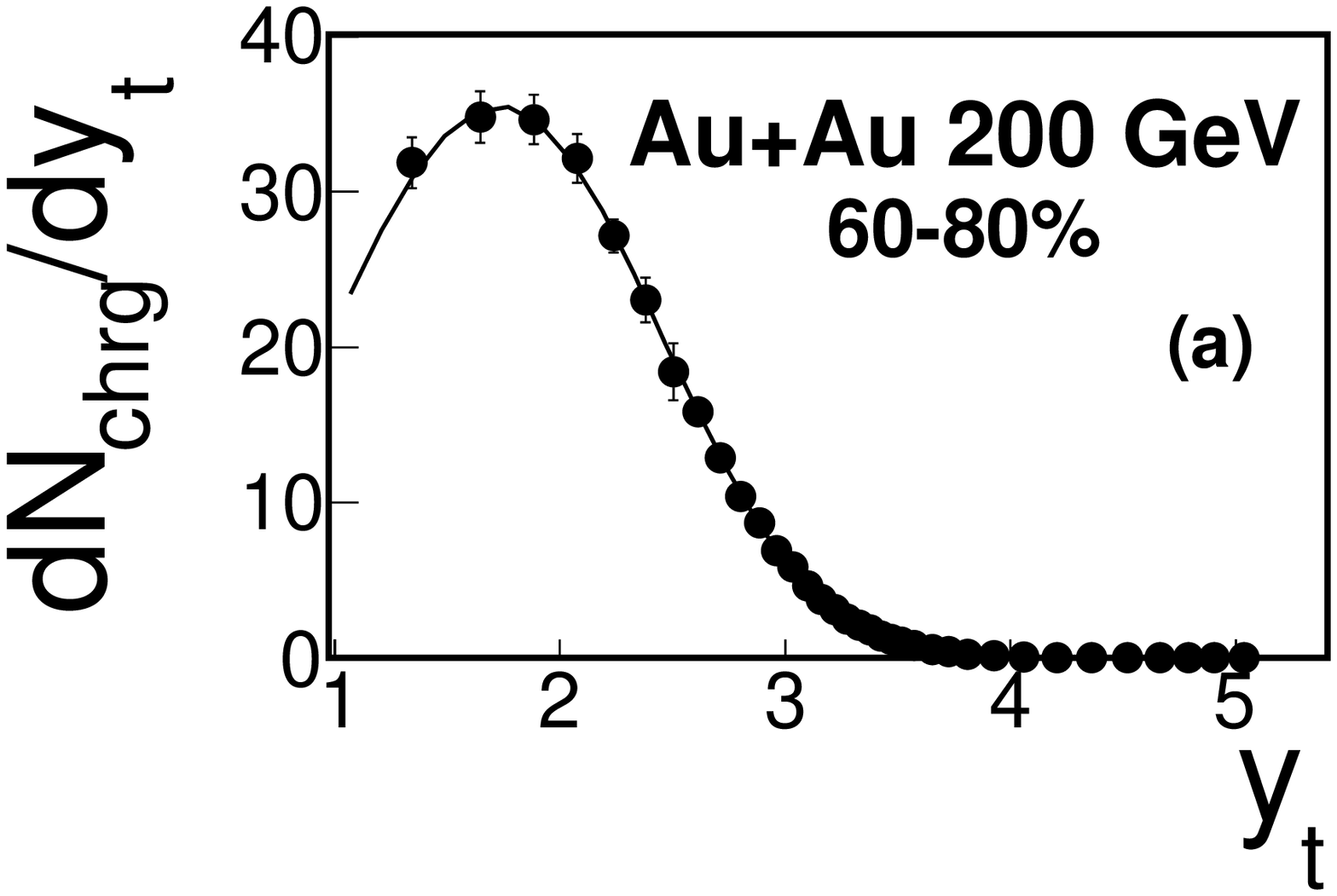}
%\put(-40,70){\bf (a)}
\includegraphics[keepaspectratio,width=2.2in]{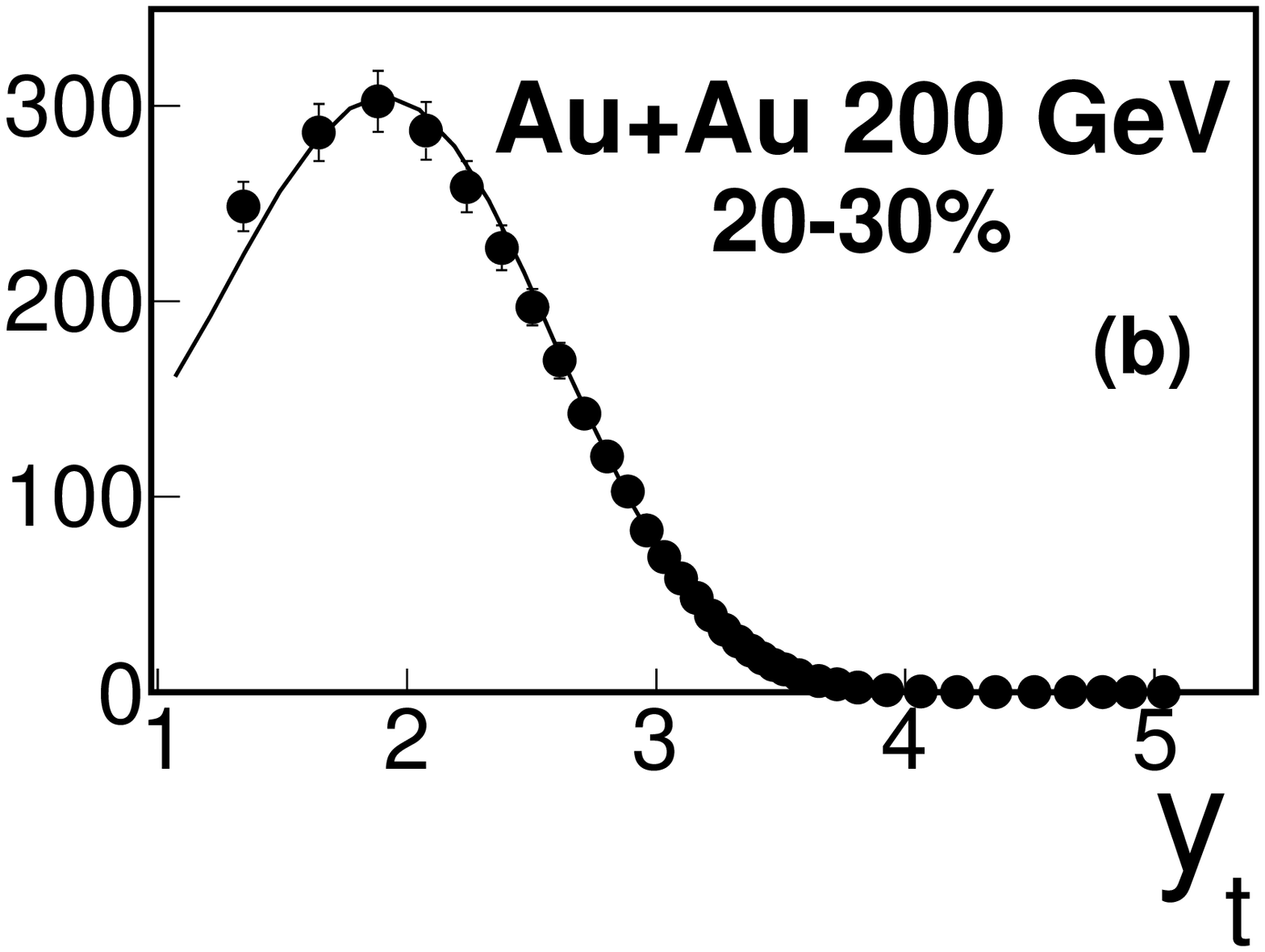}
%\put(-40,70){\bf (b)}
\includegraphics[keepaspectratio,width=2.2in]{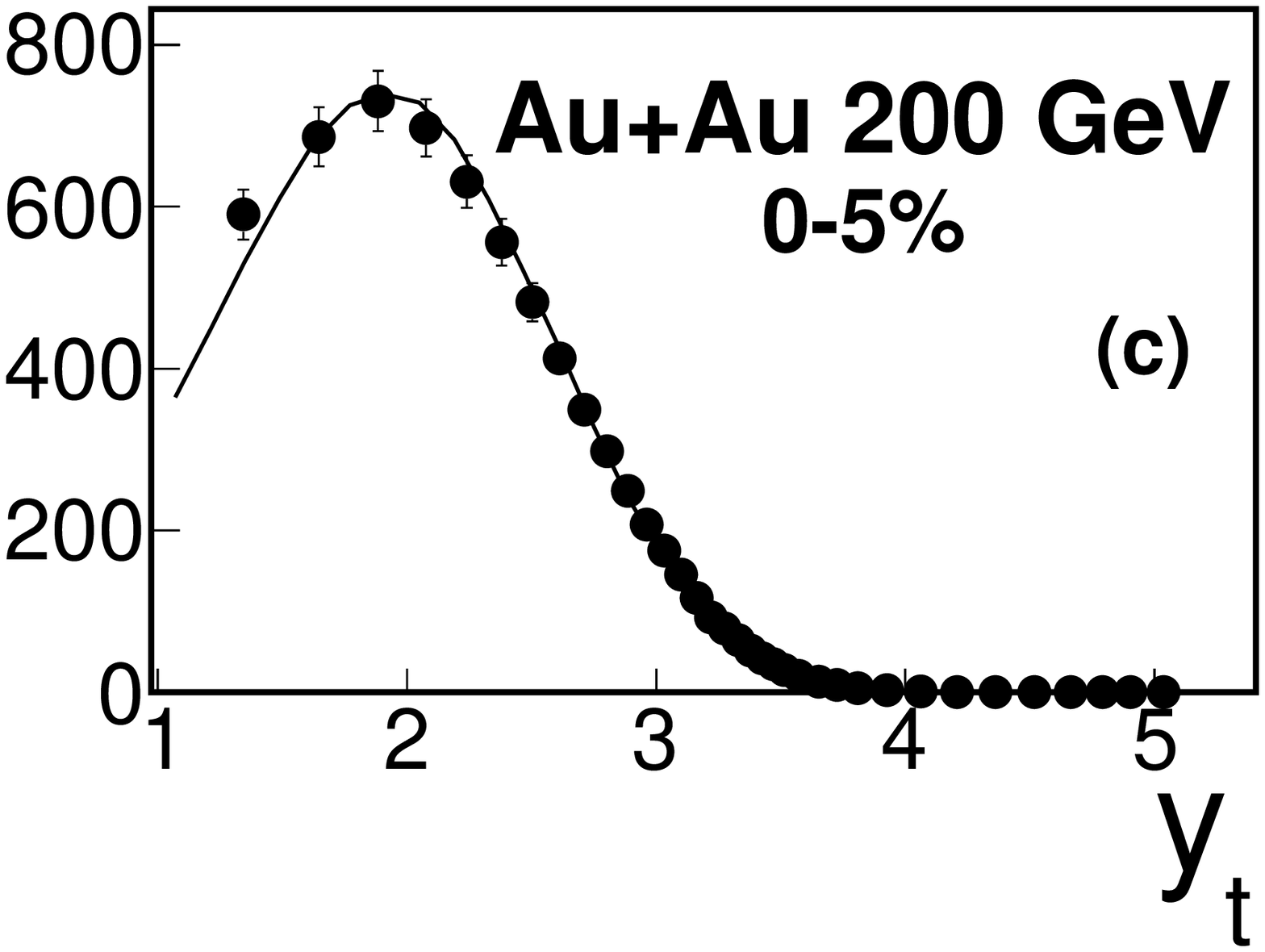} \\
%\put(-40,70){\bf (c)} \\
\includegraphics[keepaspectratio,width=2.5in]{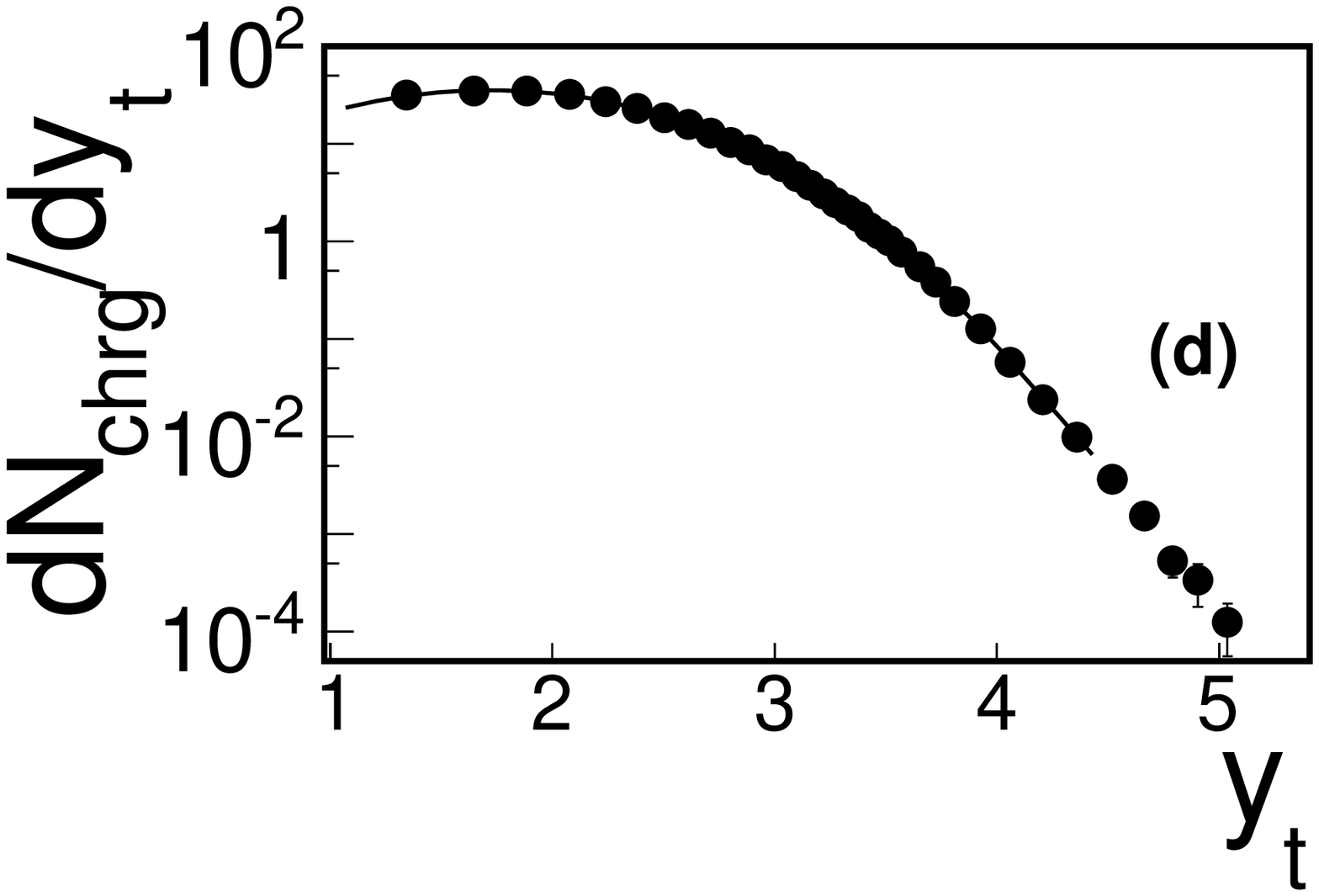}
%\put(-40,70){\bf (d)}
\includegraphics[keepaspectratio,width=2.2in]{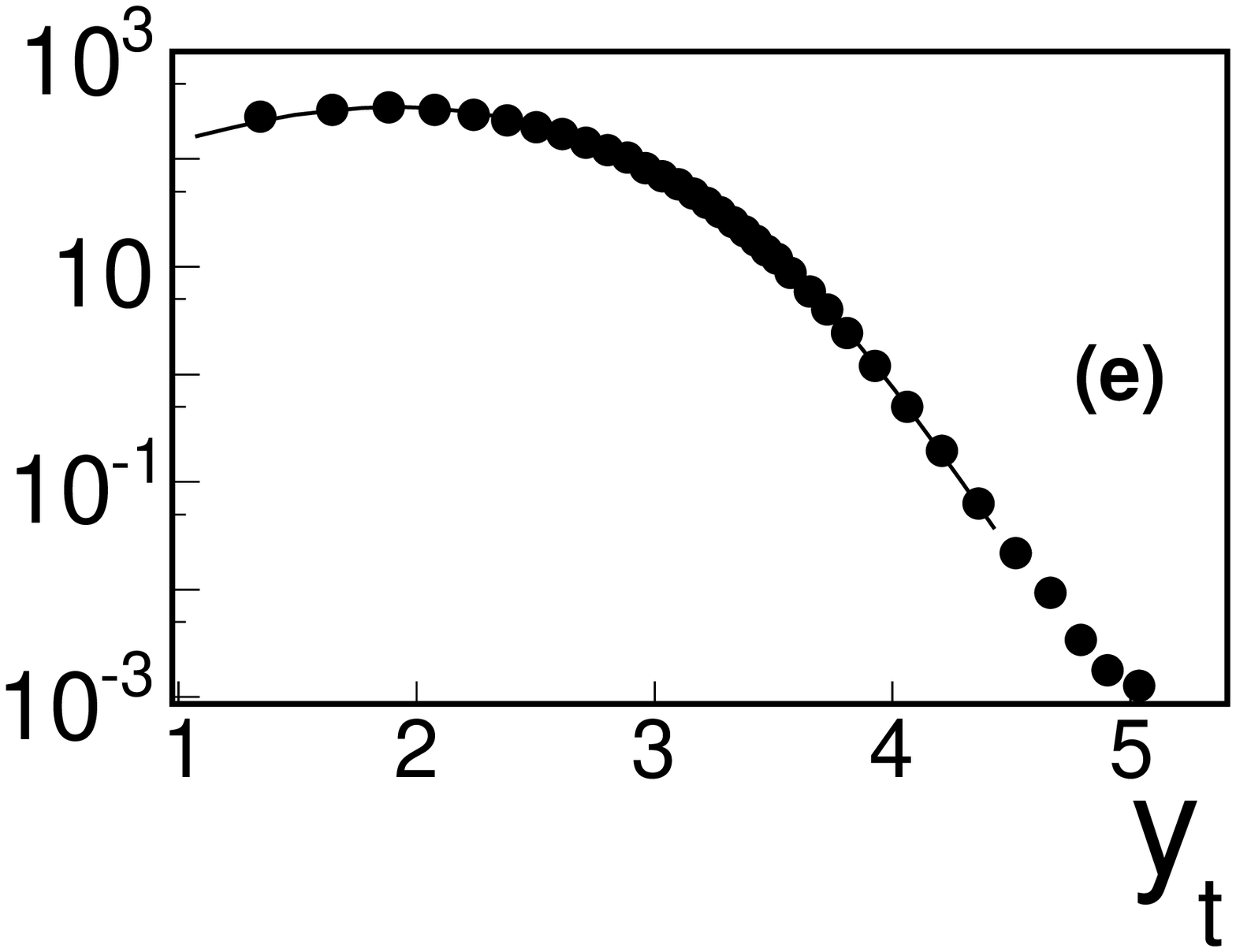}
%\put(-40,70){\bf (e)}
\includegraphics[keepaspectratio,width=2.2in]{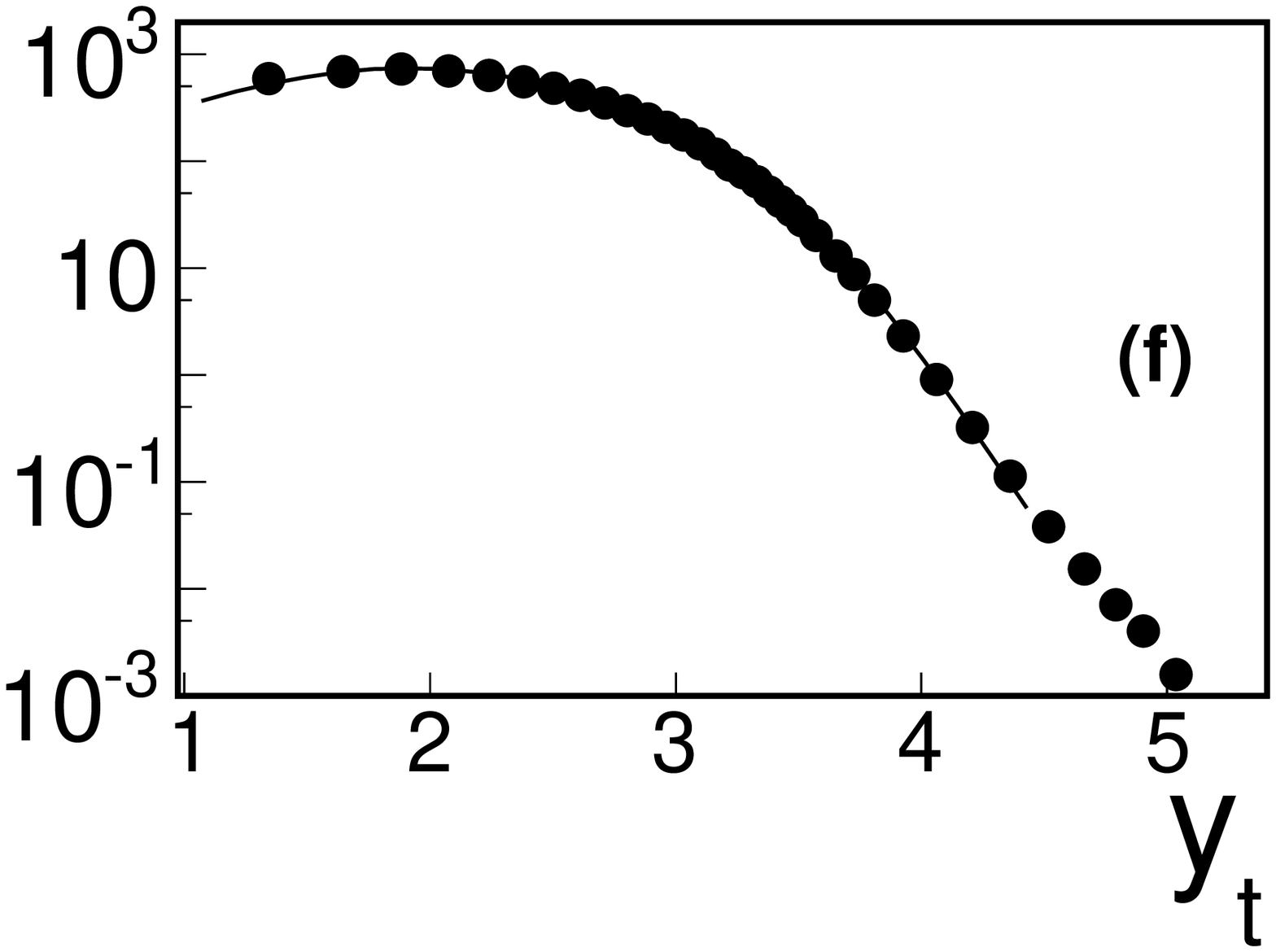} \\
%\put(-40,70){\bf (f)} \\
\includegraphics[keepaspectratio,width=2.5in]{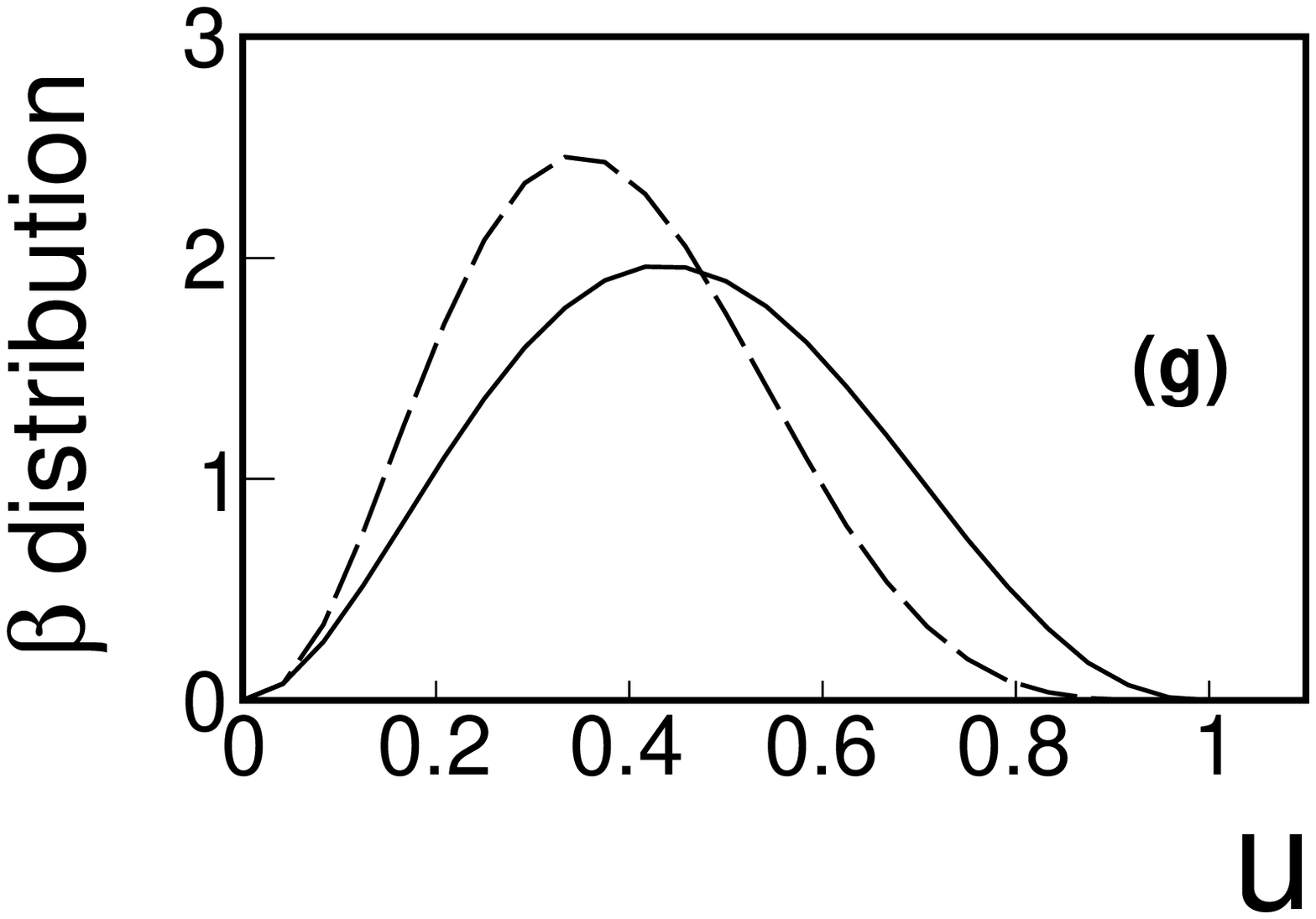}
%\put(-40,70){\bf (g)}
\includegraphics[keepaspectratio,width=2.2in]{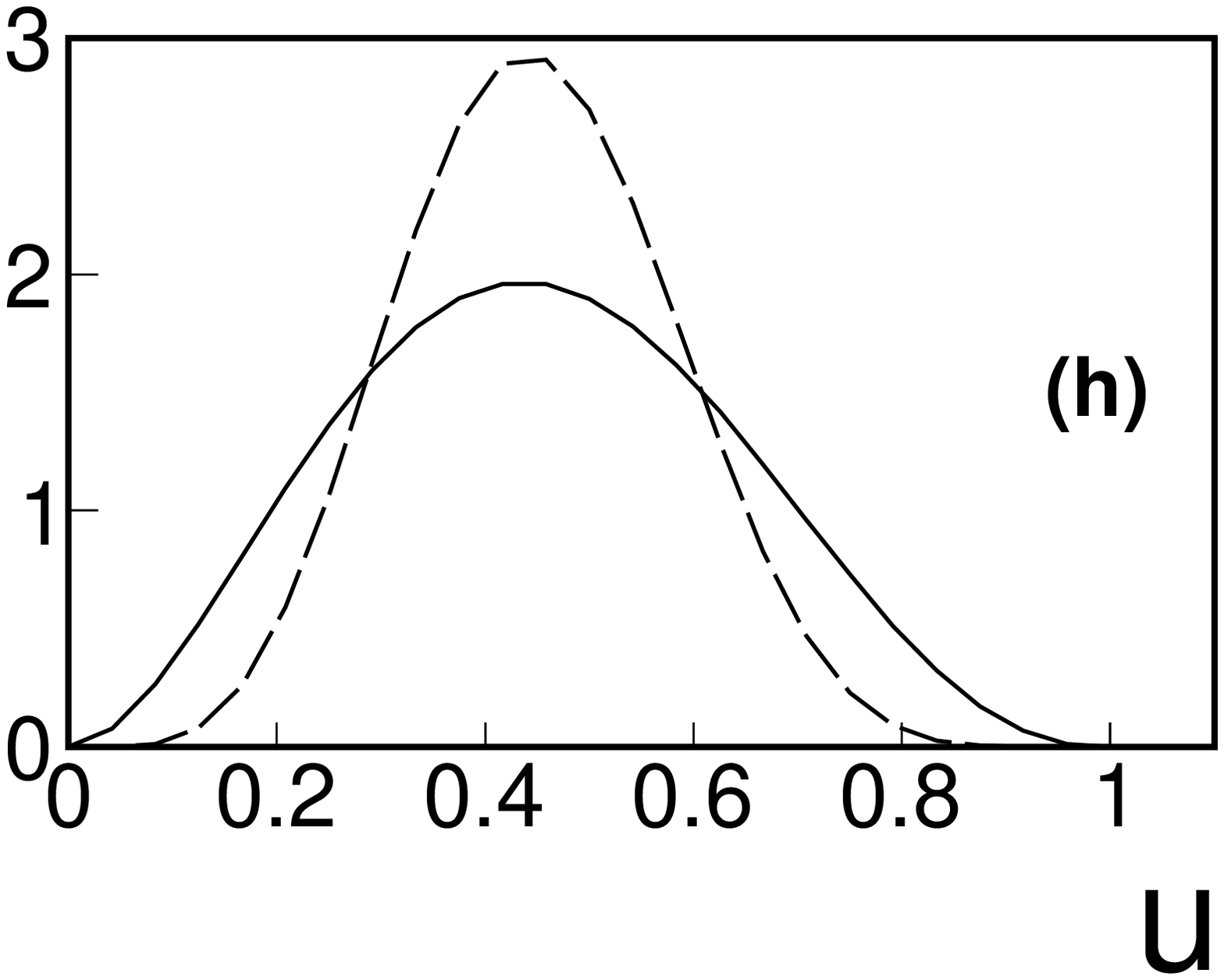}
%\put(-40,70){\bf (h)}
\includegraphics[keepaspectratio,width=2.2in]{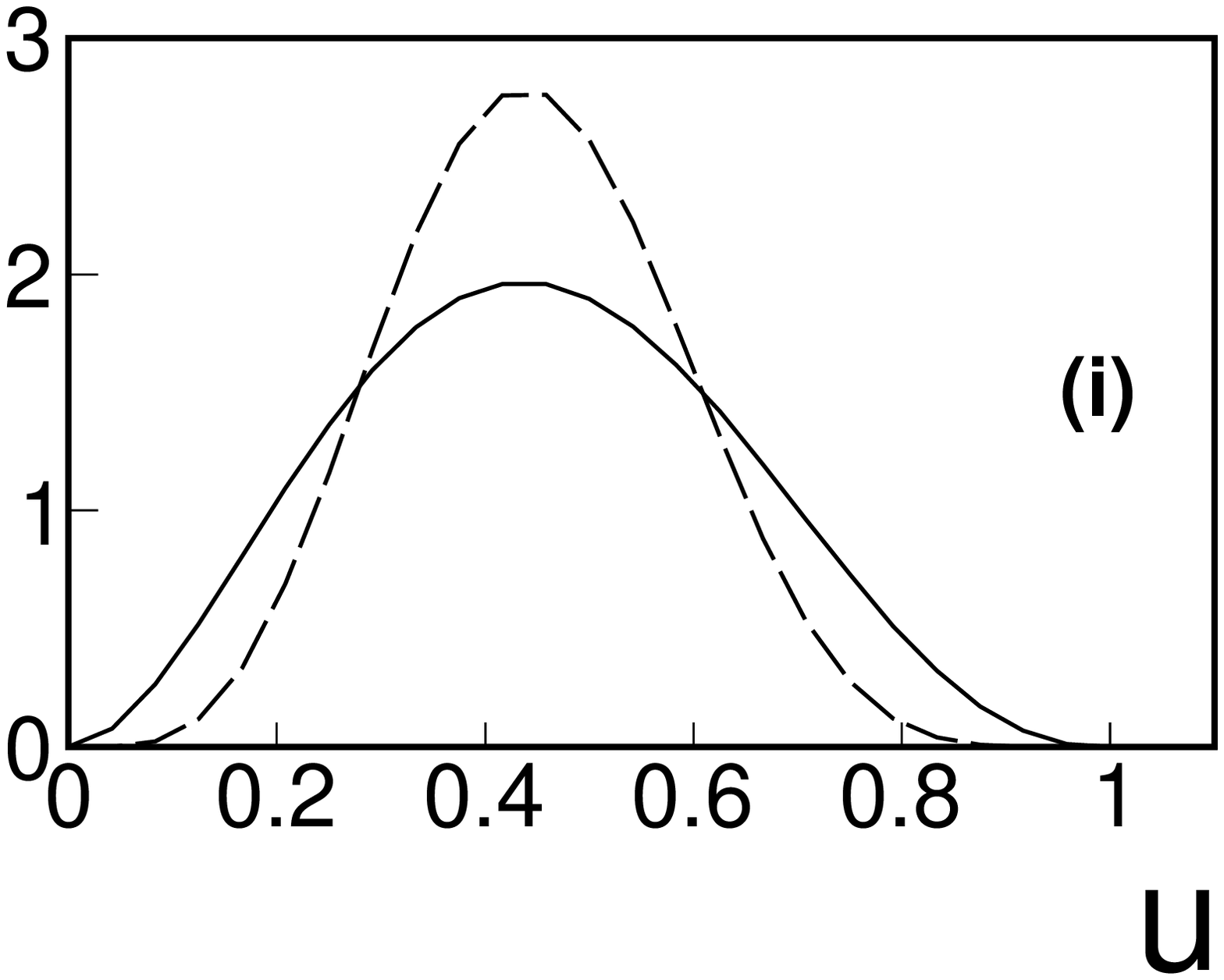}
%\put(-40,70){\bf (i)}
\caption{\label{Fig2}
Fluctuating TCF model fits to the 200 GeV Au+Au minimum-bias $p_t$ spectrum data from STAR~\cite{STARspectra} (see text) showing only the 60-80\%, 20-30\% and 0-5\% centrality results. Combined soft plus semi-hard process yields are shown as quantity $dN_{\rm ch}/dy_t$ versus transverse rapidity $y_t$, assuming pseudorapidity acceptance $\Delta\eta = 2$. Linear (upper row) and semi-log (middle row) plots are shown for the same data and model fits to allow visual access to the fit quality in both the lower and higher $y_t$ ranges. Modifications to the fragment distribution for each centrality are shown in the lower row of panels where the nominal (solid lines)~\cite{TomFrag} and fitted (dashed lines) normalized beta distributions are plotted versus transverse rapidity scaling variable $u$ in Eq.~(\ref{Eq52}).}
\end{figure*}
%%%%%%%%%%%%%%%%%%%%%%%%%%%%%%%%%%

\subsection{Two-particle distribution}
\label{SecIVB}

In the two-component fragmentation model the two-particle distribution is generated by emissions from multiple color-strings and from multiple jets in each heavy-ion collision. These processes are characterized by $p_t$ slope parameters $\beta_{cs_1}$ and $\beta_{cs_2}$ and by jet parameters $y_{{\rm max}_1}$ and $y_{{\rm max}_2}$ for arbitrary particles 1 and 2, respectively. Correlations arise when the event-average probability distributions on $(\beta_{cs_1},\beta_{cs_2})$ and $(y_{{\rm max}_1},y_{{\rm max}_2})$ are correlated. For example, a pair of particles emitted from the same color-string, or from the same jet, are correlated in the sense that they share the same $\beta_{cs}$ or $y_{\rm max}$, respectively. In the present application correlated fluctuations between color-string $\beta_{cs}$ and jet $y_{\rm max}$ are not included as these are defined to be independent processes in this model. The two-particle same-event pair-distribution in this model is given by
\bea
\bar{\rho}_{\rm TCF,se}(y_{t1},y_{t2}) & = &  \nonumber \\
 & & \hspace{-1.0in} \frac{\bar{N}-1}{\bar{N}} \frac{1}{\epsilon} \sum_{j=1}^{\epsilon}
\left[ \rho_{s,j}(y_{t1})+\rho_{h,j}(y_{t1})+\delta\bar{\rho}(y_{t1})\right] \nonumber \\
 &  & \hspace{-1.0in} \times  \left[ \rho_{s,j}(y_{t2})+\rho_{h,j}(y_{t2})+\delta\bar{\rho}(y_{t2})\right] \\
\label{Eq54}
 & & \hspace{-1.0in} = \frac{\bar{N}-1}{\bar{N}} \frac{1}{\epsilon} \sum_{j=1}^{\epsilon}
\left\{ \rho_{s,j}(y_{t1}) \rho_{s,j}(y_{t2}) + \rho_{h,j}(y_{t1}) \rho_{h,j}(y_{t2}) \right. \nonumber \\
 & & \hspace{-1.0in} + \left( \rho_{s,j}(y_{t1}) \rho_{h,j}(y_{t2}) + \rho_{s,j}(y_{t2}) \rho_{h,j}(y_{t1}) 
\right) \nonumber \\
 & & \hspace{-1.0in} \left. +  \delta\bar{\rho}(y_{t1}) \left( \rho_{s,j}(y_{t2})+\rho_{h,j}(y_{t2}) \right)
\right. \nonumber \\
 & & \hspace{-1.0in} \left.   + \delta\bar{\rho}(y_{t2}) \left( \rho_{s,j}(y_{t1})+\rho_{h,j}(y_{t1}) \right)
    +  \delta\bar{\rho}(y_{t1}) \delta\bar{\rho}(y_{t2})  \right\} \nonumber \\
 & \equiv & \bar{\rho}_{ss} + \bar{\rho}_{hh} + \bar{\rho}_{sh} + \bar{\rho}_{hs} + \bar{\rho}_{\delta}
\label{Eq55}
\eea

The color-string term, $\bar{\rho}_{ss}$, can be expanded as in Sec.~\ref{SecII}, where
\bea
\bar{\rho}_{ss}(y_{t1},y_{t2}) & \equiv & \frac{\bar{N}-1}{\bar{N}} \frac{N_{(2)s}}{\epsilon}
\int \!\! \int d\beta_{cs_1} d\beta_{cs_2} f(\beta_{cs_1},\beta_{cs_2}) \nonumber \\
 & \times & \hat{\rho}_s (\beta_{cs_1},y_{t1}) \hat{\rho}_s (\beta_{cs_2},y_{t2}).
\label{Eq56}
\eea
The average number of soft-particle pairs is
\bea
\frac{N_{(2)s}}{\epsilon} & = & \bar{N}_s^2 + \sigma_s^2
\label{Eq57}
\eea
where $\sigma_s^2$ is the variance of the event-wise fluctuation in the number of particles emitted by color-string fragmentation. Correlated distribution $f(\beta_{cs_1},\beta_{cs_2})$ is the same as in Eq.~(\ref{Eq34}) and $\hat{\rho}_s (\beta_{cs},y_{t})$ is the unit-normalized MB distribution introduced in Eq.~(\ref{Eq47}). The integral in Eq.~(\ref{Eq56}) is given in Ref.~\cite{Ayamtmt} in terms of single-particle MB distributions on transverse mass $m_t$, where single-particle distributions on kinematic variables $p_t$, $m_t$ and $y_t$ are related by
\begin{equation}
\begin{split}
& \frac{d^2N}{dy_t d\eta}  =  2\pi p_t \frac{dp_t}{dy_t} \frac{d^2N}{2\pi p_t dp_t d\eta}
 = 2\pi p_t m_t \frac{d^2N}{2\pi m_t dm_t d\eta}
\label{Eq58}
\end{split}
\end{equation}
where $dp_t/dy_t \rightarrow m_t$ at mid-rapidity. The resulting two-particle distribution is given by
\bea
\bar{\rho}_{ss} & = & \frac{\bar{N}-1}{\bar{N}} \left( \bar{N}_s^2 + \sigma_s^2 \right) {\cal N}_{ss} {\cal J}
\left( 1 + \frac{\bar{\beta}_{cs} m_{t\Sigma}}{2q_{\beta_{cs\Sigma}}} \right) ^{-2q_{\beta_{cs\Sigma}}}
\nonumber \\
 & \times & \left[ 1 - \left( \frac{\bar{\beta}_{cs} m_{t\Delta}}{2q_{\beta_{cs\Delta}} + \bar{\beta}_{cs} m_{t\Sigma}}
\right) ^2 \right] ^{-q_{\beta_{cs\Delta}}}
\nonumber \\
 & \equiv & \frac{\bar{N}-1}{\bar{N}} \left( \bar{N}_s^2 + \sigma_s^2 \right)
\hat{\rho}_{\rm 2D-Levy}(y_{t1},y_{t2}).
\label{Eq59}
\eea
where ${\cal N}_{ss}$ is a normalization factor and ${\cal J} = 4\pi^2 p_{t1} m_{t1} p_{t2} m_{t2}$ is the Jacobian which transforms the 2D distribution on transverse mass to transverse rapidity. The unit-normalized 2D Levy distribution is defined in the last line of Eq.~(\ref{Eq59}) which is calculated at the mid-points of the $y_t$ bins when comparing to data. Also in the preceding equation kinematic variables $m_{t\Sigma} = m_{t1} + m_{t2} -2m_0$ and $m_{t\Delta} = m_{t1} - m_{t2}$ were introduced. Relative variance difference quantities $\Delta(1/q)_{cs\Sigma,\Delta}$ and $\Delta(1/q)_{cs,{\rm Vol,cov}}$ are used in the fitting in analogy with similar quantities defined in Eqs.~(\ref{Eq35}) and (\ref{Eq36}).

The hard-scattering term $\bar{\rho}_{hh}$ in Eq.~(\ref{Eq55}) is similarly expanded as
\bea
\bar{\rho}_{hh}(y_{t1},y_{t2}) & = & \frac{\bar{N}-1}{\bar{N}} \frac{N_{(2)h}}{\epsilon}
\int \!\! \int dy_{{\rm max}_1} dy_{{\rm max}_2} \nonumber \\
 & & \hspace{-0.75in} \times \hat{g}(y_{{\rm max}_1},y_{{\rm max}_2})
\hat{\rho}_h(y_{{\rm max}_1},y_{t1}) \hat{\rho}_h(y_{{\rm max}_2},y_{t2})
\label{Eq60}
\eea
where the mean number of hard-scattering particle pairs is $\bar{N}_h^2 + \sigma_h^2$, where $\sigma_h^2 = \sigma_s^2$ when event multiplicities are constrained to fixed total $\bar{N}$. The unit-normalized, single-particle distribution $\hat{\rho}_h(y_{\rm max},y_{t})$ was defined in Eq.~(\ref{Eq51}).

For the correlated distribution $\hat{g}(y_{{\rm max}_1},y_{{\rm max}_2})$, a simplified functional form was assumed in order to reduce computational demands. The simplified function, $\hat{g}_2$ combines an uncorrelated (factorized) component and a fully correlated (diagonal) component defined by
\bea
\hat{g}_2(y_{{\rm max}_1},y_{{\rm max}_2}) & = & (1-\zeta)\hat{h}(y_{{\rm max}_1})\hat{h}(y_{{\rm max}_2})
\nonumber \\
 & + &  \zeta \hat{b}(y_{{\rm max}_1}) \delta(y_{{\rm max}_1} - y_{{\rm max}_2}),
\label{Eq61}
\eea
where $0 \leq \zeta \leq 1$ is a fitting parameter, $\delta$ is the Dirac delta-function,
\bea
\hat{b}(y_{{\rm max}_1}) & = & \lim_{y_{{\rm max}_2} \rightarrow y_{{\rm max}_1}} 
\hat{g}^{\prime}(y_{{\rm max}_1}) \hat{g}^{\prime}(y_{{\rm max}_2}),
\label{Eq62}
\eea
and $\hat{g}^{\prime}(y_{\rm max})$ has the same form as $\hat{g}(y_{\rm max})$ in Eq.~(\ref{Eq50}), but can have different parameter values. In taking the limit in the above equation the product of the two hyperbolic tangent cut-off functions in both instances of $\hat{g}^{\prime}$ is approximated by a single cut-off function with variable parameter $y^{\star}_{\rm cut}$. In addition, exponential argument $2(n_{\rm QCD} - 2)(y_{\rm max} - y_{\rm cut})$ is re-written as $(n^{\star}_{\rm QCD} - 2)(y_{\rm max} - y^{\star}_{\rm cut})$ where $n^{\star}_{\rm QCD}$ is freely varied in the fitting. The correlated portion of $\hat{g}_2(y_{{\rm max}_1},y_{{\rm max}_2})$ becomes
\bea
\hat{b}(y_{{\rm max}_1}) \delta(y_{{\rm max}_1} - y_{{\rm max}_2}) & = & \nonumber \\
 & & \hspace{-1.0in} {\cal N}_p \frac{1}{2} \left\{ \tanh \left( \frac{y_{{\rm max}_1} - y^{\star}_{\rm cut}}{\xi_{\rm cut}} \right)
+1 \right\} \nonumber \\
 & & \hspace{-1.0in} \times  e^{-(n^{\star}_{\rm QCD} - 2)(y_{{\rm max}_1} - y^{\star}_{\rm cut})}
(y_{{\rm max}_1} - y_{\rm min})^4 \nonumber \\
 & & \hspace{-1.0in} \times \delta(y_{{\rm max}_1} - y_{{\rm max}_2})
\label{Eq63}
\eea
with normalization constant ${\cal N}_p$. Correlations are generated in this model when $0 < \zeta \leq 1$ and may be modified by allowing $n^{\star}_{\rm QCD} \neq 2(n_{\rm QCD} - 1)$ and/or $y^{\star}_{\rm cut} \neq y_{\rm cut}$ where $n_{\rm QCD}$ and $y_{\rm cut}$ are determined by fitting the single particle $p_t$ spectra described in Sec.~\ref{SecIVA}. 

It is essential that the single-particle projection (marginal) of $\bar{\rho}_{hh}$ equal the single-particle, semi-hard component $\bar{\rho}_h(y_t)$ (Eq.~(\ref{Eq49})) in order to maintain equality between the single-particle projection of the full, two-particle distribution in Eq.~(\ref{Eq54}) and the measured charge distribution. This can be accomplished by requiring that
\bea
\int dy_{{\rm max}_2} \hat{g}_2(y_{{\rm max}_1},y_{{\rm max}_2}) & = & \hat{g}(y_{{\rm max}_1}),
\label{Eq64}
\eea
which in turn requires that function $\hat{h}(y_{\rm max})$ in Eq.~(\ref{Eq61}) be determined by
\bea
\hat{h}(y_{\rm max}) & = & \left[ \hat{g}(y_{\rm max}) - \zeta \hat{b}(y_{\rm max}) \right]/(1-\zeta),
\zeta < 1
\label{Eq65}
\eea
as parameters $n^{\star}_{\rm QCD}$ and $y^{\star}_{\rm cut}$ in $\hat{b}(y_{\rm max})$ vary.

Substituting the above quantities into Eq.~(\ref{Eq60}), where $\hat{g}(y_{{\rm max}_1},y_{{\rm max}_2}) \longrightarrow \hat{g}_2(y_{{\rm max}_1},y_{{\rm max}_2})$, results in the purely hard-scattering contribution to the same-event pair-distribution given by
\begin{widetext}
\bea
\bar{\rho}_{hh}(y_{t1},y_{t2}) & = & \frac{\bar{N}-1}{\bar{N}} (\bar{N}_h^2 + \sigma_s^2)
  \left[ (1-\zeta)
\int dy_{{\rm max}_1} \hat{h}(y_{{\rm max}_1}) \hat{\rho}_h(y_{{\rm max}_1},y_{t1}) 
\int dy_{{\rm max}_2} \hat{h}(y_{{\rm max}_2}) \hat{\rho}_h(y_{{\rm max}_2},y_{t2}) \right.
\nonumber \\
   &  & \left. + \zeta \int dy_{\rm max} \hat{b}(y_{\rm max})  \hat{\rho}_h(y_{\rm max},y_{t1}) \hat{\rho}_h(y_{\rm max},y_{t2})  \right]
\nonumber \\
 & \equiv & \frac{\bar{N}-1}{\bar{N}} (\bar{N}_h^2 + \sigma_s^2)
\left[ (1-\zeta) \hat{\rho}_{\rm [h]}(y_{t1}) \hat{\rho}_{\rm [h]}(y_{t2})
 + \zeta \hat{\rho}_{\rm 2D[b]}(y_{t1},y_{t2}) \right], 
\label{Eq66}
\eea
\end{widetext}
where $ \hat{\rho}_{\rm [h]}$ and $\hat{\rho}_{\rm 2D[b]}$ in the last line are defined by the integrals in the first two lines of the equation. As usual, the above quantities are calculated at the $y_t$ bin mid-points.

The color-string, hard-scattering cross terms do not contribute to the correlations when $\beta_{cs}$ and $y_{\rm max}$ fluctuations are independent. These terms are readily given by
\bea
\bar{\rho}_{sh} & = & \frac{\bar{N}-1}{\bar{N}} [1 - \sigma_s^2/(\bar{N}_s \bar{N}_h)]
\bar{\rho}_s(y_{t1}) \bar{\rho}_{\rm [g]}(y_{t2})
\label{Eq67}
\eea
using Eqs.~(\ref{Eq47}) and (\ref{Eq49}) where the event-averaged number of ``string-jet'' pairs equals $(\bar{N}_s \bar{N}_h - \sigma_s^2)$ if the event multiplicity is fixed. Cross term $\bar{\rho}_{hs}$ is calculated by interchanging labels 1 and 2 in Eq.~(\ref{Eq67}).

The remaining terms include products of the residual $\delta\bar{\rho}(y_t)$ with either $\bar{\rho}_s$, $\bar{\rho}_{\rm [g]}$ or itself, and are collected into one term given by
\bea
\bar{\rho}_{\delta} & \equiv & \frac{\bar{N}-1}{\bar{N}} \left\{ \delta\bar{\rho}(y_{t1}) 
\left[ \bar{\rho}_s(y_{t2}) + \bar{\rho}_{\rm [g]}(y_{t2}) \right] \right.
\nonumber \\
 & & \left. +  \delta\bar{\rho}(y_{t2}) \left[ \bar{\rho}_s(y_{t1})
+ \bar{\rho}_{\rm [g]}(y_{t1}) \right] \right.
\nonumber \\
 & & \left. + \delta\bar{\rho}(y_{t1}) \delta\bar{\rho}(y_{t2}) 
\right\} .
\label{Eq68}
\eea
Combining terms $\bar{\rho}_{ss}$, $\bar{\rho}_{hh}$, $\bar{\rho}_{sh}$, $\bar{\rho}_{hs}$ and $\bar{\rho}_{\delta}$ gives $\bar{\rho}_{\rm TCF,se}$ in Eq.~(\ref{Eq54}).

\subsection{Two-particle correlation and prefactor}
\label{SecIVC}

The single-particle projection (marginal) of the two-particle distribution in Eq.~(\ref{Eq55}) is given by
\bea
\bar{\rho}_{\rm TCF,marg}(y_{t1}) & = & \frac{1}{\bar{N}-1} \int dy_{t2} \bar{\rho}_{\rm TCF,se}(y_{t1},y_{t2})
\nonumber \\
 &  & \hspace{-1.3in} = \bar{N}_s \int dy_{t2} \hat{\rho}_{\rm 2D-Levy}(y_{t1},y_{t2}) 
  +  \bar{\rho}_{\rm [g]}(y_{t1}) + \delta\bar{\rho}(y_{t1}).
\label{Eq69}
\eea
If $|\Delta (1/q)_{cs,{\rm Vol}}| \ll 1$, then the integral of $\hat{\rho}_{\rm 2D-Levy}$ over $y_{t2}$ is accurately given by $\bar{\rho}_s(y_{t1})/\bar{N}_s$. The per-pair normalized correlation quantity is given by
\bea
\frac{\Delta\bar{\rho}_{\rm TCF}}{\bar{\rho}_{\rm ref}}(y_{t1},y_{t2}) & = & \nonumber \\
 & & \hspace{-1.2in} \frac{\bar{\rho}_{\rm TCF,se}(y_{t1},y_{t2}) - \frac{\bar{N}-1}{\bar{N}}\bar{\rho}_{\rm TCF,marg}(y_{t1})
\bar{\rho}_{\rm TCF,marg}(y_{t2})}{\bar{\rho}_{\rm ref}(y_{t1},y_{t2})},
\nonumber \\
\label{Eq70}
\eea
analogous to Eq.~(\ref{Eq43}) for the blast-wave, where $\bar{\rho}_{\rm ref}$ is defined as the product of marginals [see Eq.~(\ref{Eq42})]. It is given by
\bea
\bar{\rho}_{\rm ref}(y_{t1},y_{t2}) & = & \frac{\bar{N}-1}{\bar{N}}
\bar{\rho}_{\rm TCF,marg}(y_{t1}) \bar{\rho}_{\rm TCF,marg}(y_{t2}).
\nonumber \\
\label{Eq71}
\eea
Using the soft-process prefactor in Appendix A for the charge-independent, away-side azimuth pair correlations gives the final correlation for the two-component fragmentation model:
\bea
\frac{\Delta\bar{\rho}_{\rm TCF}}{\sqrt{\bar{\rho}_{\rm soft}}}(y_{t1},y_{t2}) & = & {\cal P}^{\rm AS-CI}_{\rm Fac,soft}(y_{t1},y_{t2})
\frac{\Delta\bar{\rho}_{\rm TCF}}{\bar{\rho}_{\rm ref}}(y_{t1},y_{t2}).
\nonumber \\
\label{Eq72}
\eea

Finally, it is instructive to expand $\Delta\bar{\rho}_{\rm TCF}(y_{t1},y_{t2})$ in terms of the separate sources of correlations built into the model. Inserting Eqs.~(\ref{Eq59}) and (\ref{Eq66})-(\ref{Eq69}) into $\Delta\bar{\rho}_{\rm TCF}$ in Eq.~(\ref{Eq70}) gives
\begin{widetext}
\bea
\Delta\bar{\rho}_{\rm TCF}(y_{t1},y_{t2}) & = & \frac{\bar{N}-1}{\bar{N}}
\left[ (\bar{N}_s^2 + \sigma_s^2) \left( \hat{\rho}_{\rm 2D-Levy}(y_{t1},y_{t2})
- \bar{\rho}_s(y_{t1}) \bar{\rho}_s(y_{t2})/\bar{N}^2_s \right) \right. \nonumber \\
 & + & (\bar{N}_h^2 + \sigma_s^2) \left( (1-\zeta) \hat{\rho}_{\rm [h]}(y_{t1})
\hat{\rho}_{\rm [h]}(y_{t2}) + \zeta \hat{\rho}_{\rm 2D[b]}(y_{t1},y_{t2})
- \hat{\rho}_{\rm [g]}(y_{t1}) \hat{\rho}_{\rm [g]}(y_{t2}) \right) \nonumber \\
 & & \left. + \sigma_s^2 \left(\bar{\rho}_s(y_{t1})/\bar{N}_s -  \hat{\rho}_{\rm [g]}(y_{t1}) \right)
\left(\bar{\rho}_s(y_{t2})/\bar{N}_s -  \hat{\rho}_{\rm [g]}(y_{t2}) \right) \right]
\label{Eq73}
\eea
\end{widetext}
where $\hat{\rho}_{[g]} = \bar{\rho}_{[g]}/\bar{N}_h$. Soft string-fragmentation induced correlations are represented in the first term and are controlled in the model via relative covariance parameter $\Delta(1/q)_{cs,{\rm cov}} \equiv (1/2)[\Delta(1/q)_{cs\Sigma} - \Delta(1/q)_{cs\Delta}] = (1/q_{\beta_{cs\Sigma}} - 1/q_{\beta_{cs\Delta}})/2$, where the correlations scale with $(\bar{N}_s^2 + \sigma_s^2)$. The semi-hard scattering, correlated fragmentation contributions are represented in the second term and are controlled by parameter $\zeta \in [0,1]$ and modulated by the two-particle, correlated semi-hard scattering probability distribution parameters $y_{\rm cut}^{\star}$ and $n_{\rm QCD}^{\star}$ in function $\hat{b}(y_{\rm max})$ [see Eq.~(\ref{Eq63})]. The semi-hard scattering correlations scale with $(\bar{N}_h^2 + \sigma_s^2)$. The semi-hard scattering versus soft string-fragmentation multiplicity fluctuation variance, $\sigma_s^2$, independently generates correlations when the soft and semi-hard particle distribution shapes differ as given by the third term in Eq.~(\ref{Eq73}).

\section{Phenomenological model correlation results}
\label{SecV}

The BW and TCF models were fitted to analytical representations of preliminary $(y_{t1},y_{t2})$ charged-particle correlation data from STAR~\cite{LizHQ,LizThesis} described in Appendix B. We refer to these representations as ``pseudodata.'' The fitting results are shown and discussed with respect to the efficacy of each model and the stability and systematic centrality dependencies of the model parameters. The centrality trends of the BW and TCF fitting parameters and some implications of those trends are discussed in the following subsections. 

\subsection{Blast-wave model description of correlations}
\label{SecVA}

The AS-CI correlation pseudodata were fitted with the fluctuating blast-wave model in  Eq.~(\ref{Eq44}) using fit parameters $\Delta(1/q)_{\rm Vol}$ and $\Delta(1/q)_{\rm cov}$ in Eq.~(\ref{Eq36}) plus the transverse flow correlation parameter $\Delta_{\eta_t}$ introduced just after Eq.~(\ref{Eq37}). Other parameters of the model including $\bar{\beta} = 1/T$, $q_\beta$, $a_0$, $n_{\rm flow}$ and $\sigma_{\eta_t}$ were determined by fitting the single-particle $p_t$ spectrum data (see Table~\ref{TableI}) and were kept fixed. Fit parameters and statistical fitting errors are listed in Table~\ref{TableIII}. Pseudodata, BW model fits, and residuals (pseudodata - model) are shown for three example centrality bins (60-80\%, 20-30\% and 0-5\%) in Fig.~\ref{Fig3}. The results show smooth, monotonic centrality dependence from most-peripheral to most-central. The general features of the correlation structures, {\em e.g.} saddle-shape and peak near  $(y_{t1},y_{t2}) = (3,3)$, are qualitatively reproduced by the model, however the (3,3) peak amplitude is underestimated by about 20-30\%. Residuals are somewhat smaller than the data overall, differing mainly at lower $y_t$ and near the (3,3) peak.

The best determined fit parameter (smallest uncertainty) is the inverse temperature co-variation $\Delta(1/q)_{\rm cov}$ which is always positive, corresponding to positive correlations in the temperature fluctuations, and which displays a monotonic decrease with centrality. Of the three fit parameters, $\Delta(1/q)_{\rm cov}$ has the smallest relative errors and displays the smoothest centrality trend. The overall $(\beta_1,\beta_2)$ distribution expansion/contraction parameter $\Delta(1/q)_{\rm Vol}$ tends to decrease ({\em i.e.} reduced fluctuations) with more-central collisions. It has larger, relative errors and greater variability than $\Delta(1/q)_{\rm cov}$. The transverse flow rapidity correlation fit parameters $\Delta\eta_t$ are non-negative, indicating positive flow correlations, but have relatively large uncertainties and erratic centrality dependence meaning that correlated transverse flow fluctuations are poorly determined with these fits.

An essential requirement of the BW correlation model is that the single-particle $p_t$ distribution be preserved throughout the fitting process. In the BW model small, non-zero values of $\Delta(1/q)_{\rm Vol}$, which are beneficial in fitting the correlations, cause the marginal of the two-particle distribution [Eq.~(\ref{Eq39})] to differ from the single-particle BW model fit to the $p_t$ spectrum data. For the present fits however the projections were consistent with the single-particle BW fits to within a few percent for all centralities except the most-peripheral 60-80\% for $y_t \geq 3$.

It is interesting to examine the degree of correlation in the inverse temperature and transverse flow rapidity sampled by arbitrary pairs of particles. Ratio $\sigma_\beta^2/\bar{\beta}^2$ is the relative variance of the inverse temperature distribution [Eq.~(\ref{Eq25})] for the single-particle distribution in Eq.~(\ref{Eq24}). Defining $\delta\sigma_{\Sigma,\Delta}$ as the change in widths of the two-particle $(\beta_1,\beta_2)$ distribution [see Fig.~\ref{Fig2b}] along the $\beta_{\Sigma,\Delta} = \beta_1 \pm \beta_2$ directions, respectively, where $\delta\sigma_{\Sigma,\Delta} \equiv \sigma_{\beta_{\Sigma,\Delta}} - \sigma_\beta$, we estimate the average, relative expansion or contraction of the $(\beta_1,\beta_2)$ distribution as
\bea
\frac{\delta\sigma_\Sigma + \delta\sigma_\Delta}{2\bar{\beta}} & \approx & (\sqrt{q_\beta}/2)\Delta(1/q)_{\rm Vol},
\label{Eq74}
\eea
assuming $\delta\sigma_{\Sigma,\Delta}/\bar{\beta} \ll 1$. Similarly, the average, relative co-variation in the $(\beta_1,\beta_2)$ distribution is estimated by
\bea
\frac{\delta\sigma_\Sigma - \delta\sigma_\Delta}{2\bar{\beta}} & \approx & (\sqrt{q_\beta}/2)\Delta(1/q)_{\rm cov}.
\label{Eq75}
\eea
The average, relative co-variation in the two-particle, transverse flow rapidity is equal to $\Delta_{\eta_t}/(2\sigma_{\eta_t})$. These three quantities are listed in Table~\ref{TableIII}. The results show that, within this fluctuating BW model and for these AS-CI pseudodata, thermal fluctuation widths vary from about +2\% increased overall fluctuation relative to that for single-particle production in peripheral collisions to about $-$0.4\% (reduced fluctuations) in most-central collisions. The relative covariance decreases monotonically with centrality from about 0.4\% to 0.1\% from peripheral to most-central collisions. Transverse flow covariances are non-negative but display large variability with respect to collision centrality, showing no clear trend in these fitting results. These small, relative changes in widths of the inverse temperature and flow fluctuations imply that intra-event $\beta,\eta_{t0}$ fluctuation magnitudes exceed the mean differences in the inter-event fluctuations as discussed in Ref.~\cite{Ayamtmt}. In other words, event-to-event fluctuations in the mean temperature and transverse flow are small relative to fluctuations within each collision system.

The BW model fits to the $p_t$ spectrum data provide an estimate of the variance in the distribution of inverse $p_t$-slope parameters, {\em e.g.} temperature in the BW approach. It is informative to compare these empirical fluctuation magnitudes to that expected for fully equilibrated (uniform temperature), relativistic hadron-gas models at kinematic decoupling, or ``freeze-out,'' when event-wise fluctuations in participant nucleon number alone are included. In relativistic hadron-gas models the energy density $\varepsilon$ is proportional to the freeze-out temperature to the fourth power~\cite{Mueller}, $\varepsilon \propto T^4$. In hydrodynamic models the total energy available for hydrodynamic expansion among the interacting partons is proportional to the number of participant nucleons. At mid-rapidity the energy density is therefore proportional to $N_{\rm part}^{1/3}$~\cite{Bjorken}. In hydrodynamic models the energy density at freeze-out includes both thermal and collective modes, but it is still reasonable to assume that $T^4 \propto N_{\rm part}^{1/3}$. Event-wise fluctuations in $N_{\rm part}$ among collisions having the same centrality, for example as defined by the impact parameter or multiplicity, produce temperature fluctuations which, in turn, produce two-particle correlations on transverse momentum.

Numerical estimates can be carried out using the BW fit values for temperature from Table~\ref{TableI} and $N_{\rm part}$ from Table~\ref{TableV}. A proportionality constant, $\alpha = \partial T^4/\partial N_{\rm part}^{1/3} \approx$ 0.000015~GeV$^4$ is estimated from the results if the most-peripheral bin is excluded. The resulting relation
\bea
\frac{\partial N_{\rm part}}{\partial T} & = & 12 T^3 N_{\rm part}^{2/3} / \alpha \approx \frac{\delta N_{\rm part}}{\delta T},
\label{Eq76}
\eea
where $\delta N_{\rm part}$ and $\delta T$ represent event-wise fluctuations, can be used to estimate the variance in the fluctuating global temperature at freeze-out caused by event-wise fluctuations in $N_{\rm part}$. In terms of inverse temperature $\beta = 1/T$, the relative variance of fluctuations in $\beta$ for a collection of similar events ({\em e.g.} same impact parameter) is given by
\bea
\langle (\delta\beta/\bar{\beta})^2 \rangle & = & \frac{\sigma_\beta^2}{\bar{\beta}^2} = \frac{1}{q_\beta}
= \left[ (12/\alpha)^2 T^8 N_{\rm part}^{1/3} \right]^{-1},
\label{Eq77}
\eea
where brackets ``$\langle ~  \rangle$'' denote an average over events, $\bar{\beta} = \langle \beta \rangle$, and the Poisson limit, $\langle (\delta N_{\rm part})^2 \rangle = N_{\rm part}$, was assumed. Using the BW fitted temperatures in  Eq.~(\ref{Eq77}) and the above value of $\alpha$, results in relative variances which are more than three orders of magnitude smaller than $1/q_\beta$ in Table~\ref{TableI}. The co-variations in relative variance, $\Delta(1/q)_{\rm cov}$ from the 2D BW model fits, are two-orders of magnitude larger than the above limit in Eq.~(\ref{Eq77}).

Both the $\beta$-fluctuations required to describe the single-particle $p_t$ distributions and the $(\beta_1,\beta_2)$ covariances required to describe the correlations are much larger than what can be accounted for with statistical fluctuations in  $N_{\rm part}$. The present results imply that much stronger, dynamical fluctuations are required in hydrodynamic approaches and that $\beta$-fluctuations within each collision event are much larger than event-wise fluctuations in mean-$\beta$. Furthermore, the dynamical fluctuation effects must persist, to some extent, until kinetic freeze-out and cannot be completely dissipated, implying that final-state temperatures at kinetic freeze-out cannot be uniform. These results impose significant constraints on the initial-state, on the effective partonic interactions in transport models, and on the parameters controlling hydrodynamic expansion.

%%%%%%%%%%%%%%%%%%%%%%%%%%%%%%%%%%
\begin{table*}[htb]
\caption{Blast-wave correlation model fit parameters to the 200~GeV Au+Au $(y_{t1},y_{t2})$ AS-CI correlation pseudodata. Statistical fitting errors are listed in parentheses. Relative expansion or contraction and relative co-variations in the thermal and transverse expansion parameters are also listed as explained in the text.}
\label{TableIII}
\begin{tabular}{c|cccc|ccc}
\hline \hline
Cent.(\%) & $\Delta(1/q)_{\rm Vol}$ & $\Delta(1/q)_{\rm cov}$ & $\Delta_{\eta_t}$ & $\frac{\chi^2}{\rm DoF}$ & 
     $\frac{\delta\sigma_\Sigma + \delta\sigma_\Delta}{2\bar{\beta}}$ &
     $\frac{\delta\sigma_\Sigma - \delta\sigma_\Delta}{2\bar{\beta}}$ & 
     $\frac{\Delta_{\eta_t}}{2\sigma_{\eta_t}}$ \\
\hline
 0-5   &  -0.00162(38)  &  0.000500(4) &  0.000400(135)  &  12.39  &  -0.00364  &  0.00112  &  0.0039 \\
 5-10  &  -0.00055(20)  &  0.000600(23) & 0.0$\pm$0.0009 & 14.46  & -0.00122   &  0.00133  &  0      \\
 10-20 &  -0.00106(25)  &  0.000650(8)  & 0.00175(54)   &  26.64  &  -0.00229  &  0.00141  &  0.0265 \\
 20-30 &  -0.00160(43)  &  0.000850(21) & 0.0$\pm$0.0026 & 26.29  &  -0.00335  &  0.00178  &  0      \\
 30-40 &  -0.00100(39)  &  0.00110(4)   & 0.0$\pm$0.0022 & 19.63  &  -0.00204  &  0.00225  &  0      \\
 40-60 &  +0.0020(6)    &  0.00140(5)   & 0.0$\pm$0.0008 & 18.09  &  +0.00391  &  0.00274  &  0      \\
 60-80 &  +0.0100(4)    &  0.00210(5)   & 0.0030(61)     & 11.19  &  +0.018    &  0.00377  &  0.075  \\
\hline \hline
\end{tabular}
\end{table*}
%%%%%%%%%%%%%%%%%%%%%%%%%%%%%%%%%%%%

%%%%%%%%%%%%%%%%%%%%%%%%%%%%%%
\begin{figure*}[t]

\includegraphics[keepaspectratio,width=6.5in]{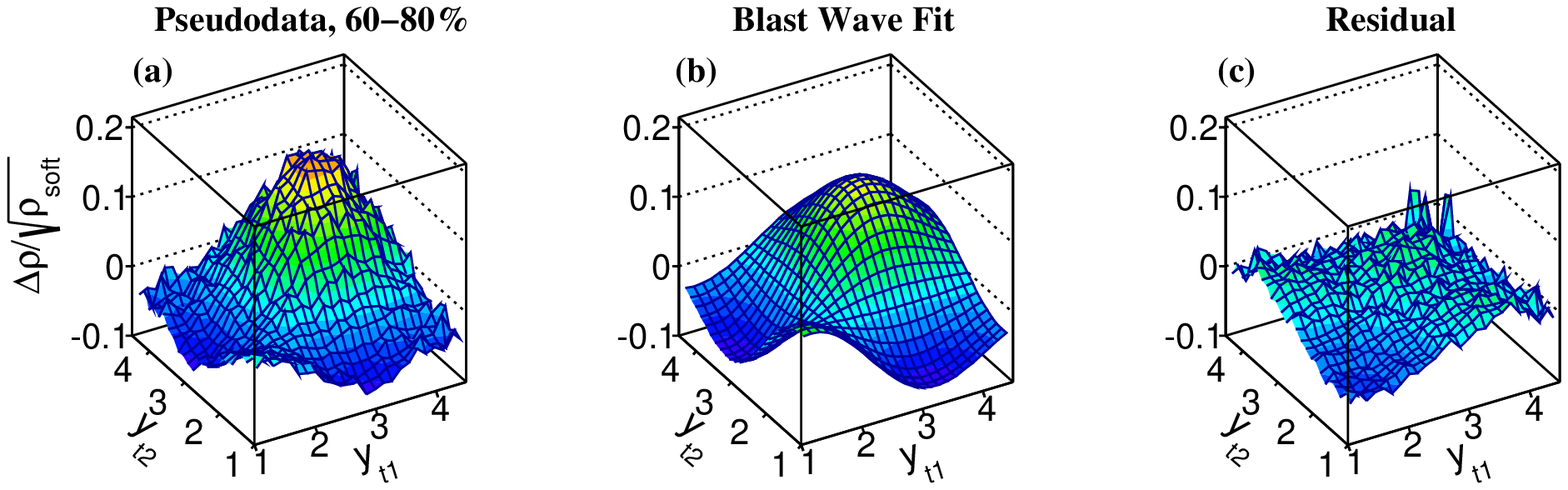} \\
%\put(-80,100){\bf (c)}
%\put(-230,100){\bf (b)}
%\put(-380,100){\bf (a)} \\
\includegraphics[keepaspectratio,width=6.5in]{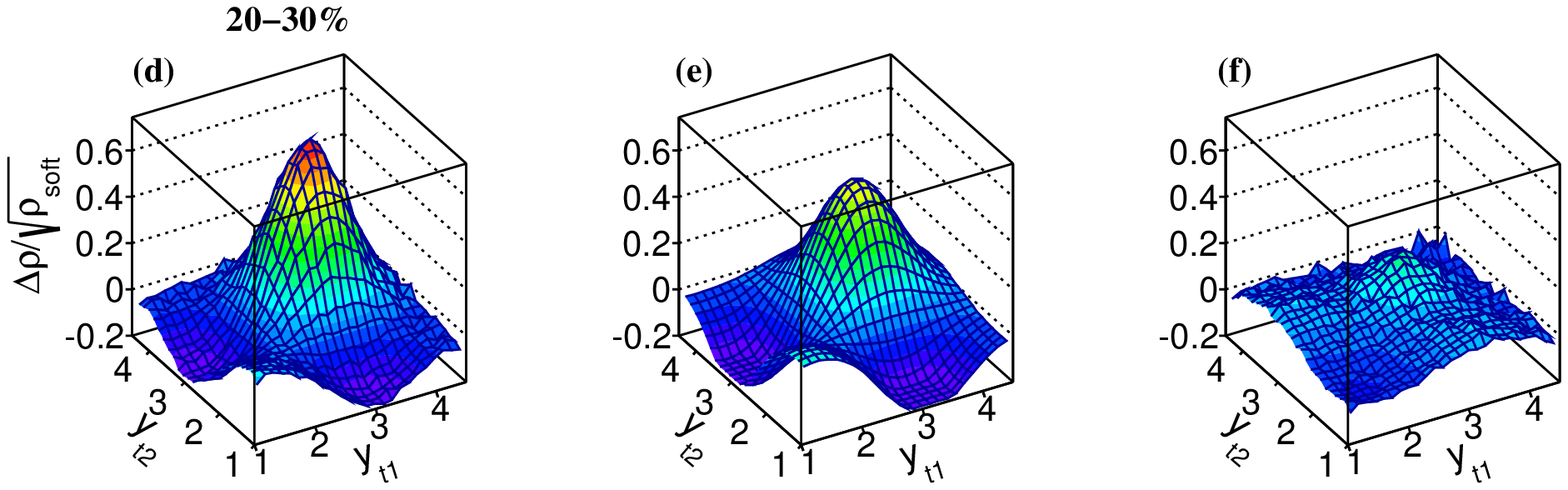} \\
%\put(-80,100){\bf (f)}
%\put(-230,100){\bf (e)}
%\put(-380,100){\bf (d)} \\
\includegraphics[keepaspectratio,width=6.5in]{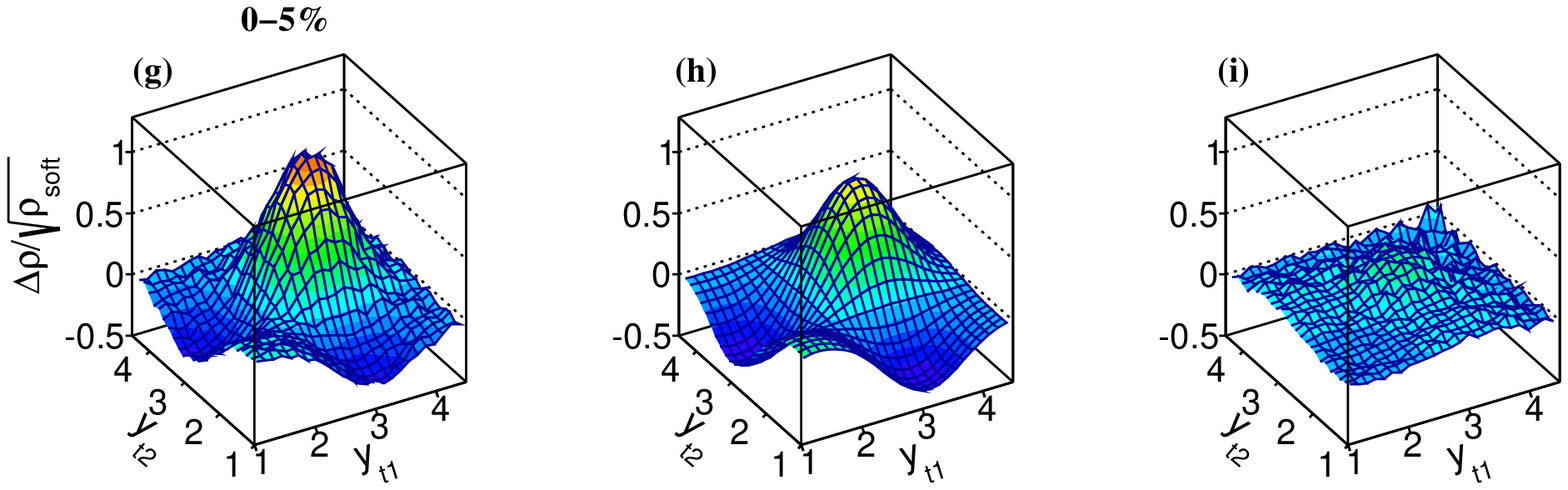}
%\put(-80,100){\bf (i)}
%\put(-230,100){\bf (h)}
%\put(-380,100){\bf (g)} 
\caption{\label{Fig3}
Fluctuating blast-wave model fits to the 200 GeV Au+Au away-side, charge-independent two-particle correlation pseudodata described in Appendix B for selected centralities 60-80\%, 20-30\% and 0-5\% in rows of panels from upper to lower, respectively. The left-hand column shows the pseudodata, the fitted model results are shown in the middle column, and the residuals (pseudodata - model) are presented in the right-hand column.}
\end{figure*}
%%%%%%%%%%%%%%%%%%%%%%%%%%%%%%%%%%

\subsection{Two-component fragmentation model description of correlations}
\label{SecVB}

The AS-CI correlation pseudodata were fit with the TCF model described in Sec.~\ref{SecIV} with parameters $\Delta(1/q)_{cs,{\rm Vol,cov}} = (1/2)[\Delta(1/q)_{cs\Sigma} \pm \Delta(1/q)_{cs\Delta}]$ [see Eqs.~(\ref{Eq36}) and (\ref{Eq59})], semi-hard multiplicity fluctuation variance $\sigma_h^2 = \sigma_s^2$ in Eq.~(\ref{Eq57}), semi-hard scattering correlation amplitude $\zeta$ in Eq.~(\ref{Eq61}), and $\hat{b}(y_{\rm max})$ probability distribution parameters $y_{\rm cut}^{\star}$ and $n_{\rm QCD}^{\star}$ in Eq.~(\ref{Eq63}). Other parameters of the TCF model were determined by fitting the single, charged-particle $p_t$ spectra data as discussed in Sec.~\ref{SecIV}, or were taken from Ref.~\cite{TomFrag}.

Ambiguities occurred in the $\chi^2$-minimization procedure in which discrete solutions were found for the color-string fragmentation parameter $\Delta(1/q)_{cs,{\rm cov}}$ corresponding to a normal saddle-shape correlation (positive value) as in Ref.~\cite{Ayamtmt}, or an inverted saddle-shape (negative value). An inherent assumption of the TCF model is that hadron fragments produced by the same color-string will sample a $p_t$ distribution with an overall slope ($\beta_{cs}$) whose value randomly fluctuates about a mean, resulting in a normal saddle shape correlation with $\Delta(1/q)_{cs,{\rm cov}} > 0$. Furthermore, it was found that acceptable descriptions of the correlations required relatively small absolute magnitudes for both $\Delta(1/q)_{cs,{\rm Vol}}$ and $\Delta(1/q)_{cs,{\rm cov}}$, similar to, or smaller than the corresponding 2D BW parameters in Table~\ref{TableIII}. To adhere to these restrictions on $\Delta(1/q)_{cs,{\rm Vol,cov}}$ and stabilize the $\chi^2$-minimization we fixed the $\Delta(1/q)_{cs,{\rm Vol,cov}}$ parameters to the fitted values given in Ref.~\cite{LizThesis} (see Appendix B). With the soft-component thus constrained, the correlation pseudodata were readily described by varying the remaining semi-hard scattering parameters $\sigma_s^2$, $\zeta$, $y_{\rm cut}^{\star}$ and $n_{\rm QCD}^{\star}$.

The model fits and residuals are compared with the correlation pseudodata in Fig.~\ref{Fig4}, and the fit parameters are listed in Table~\ref{TableIV}. Smooth, monotonic trends in the residuals and good, overall descriptions of the pseudodata were achieved. The fitted peak amplitudes at $(y_{t1},y_{t2}) \approx (3,3)$ are about 10\% below the pseudodata. Color-string fragmentation parameters $\Delta(1/q)_{cs,{\rm Vol}}$~\cite{LizThesis} are negative, indicating a slight, overall contraction in the widths of the distribution of $p_t$-slope parameter $\beta_{cs}$. This reduction is sufficiently small such that the marginals of the two-particle distributions remain within 1\% of the charged-particle distributions over the full $y_t$ range [1.0,4.5] for all centralities from 0 to 80\%. Parameters $\Delta(1/q)_{cs,{\rm cov}}$ from Ref.~\cite{LizThesis} monotonically decrease from peripheral to most-central collisions as was also found for the BW model fits (Table~\ref{TableIII}) where similar numerical values were found. Semi-hard scattering parameters $\sigma_s^2$ and $\zeta$ monotonically increase and decrease, respectively, from peripheral to most-central collisions, while parameters $y_{\rm cut}^{\star}$ and $n_{\rm QCD}^{\star}$ remain approximately constant with centrality. From the definition of the jet energy correlation function $\hat{b}(y_{\rm max})$ in Eq.~(\ref{Eq63}) we expect $n_{\rm QCD}^{\star} \approx 2(n_{\rm QCD} - 1)$ in the weak correlation limit. The fitted values of $n_{\rm QCD}^{\star}$ vary from about 11 to 12 which are smaller than this estimated range that varies from 12 to 16 using the values for $n_{\rm QCD}$ in Table~\ref{TableII}. This indicates that the distributions of correlated, semi-hard scattering maximum fragment rapidities, represented by distribution $\hat{b}(y_{\rm max})$, are weighted toward larger $y_{\rm max}$ values (more energetic jets) than the corresponding single-particle distribution $\hat{g}(y_{\rm max})$. Or, in other words, positively correlated jet fragment pairs are more likely to be associated with higher-energy jets, rather than lower. Overall, these results demonstrate that the TCF model is capable of providing qualitative descriptions of correlation data on transverse momentum resulting in smooth, monotonic centrality dependence in each of the fitting parameters.

In Eq.~(\ref{Eq73}) contributions to the correlated pair distribution $\Delta\bar{\rho}_{\rm TCF}(y_{t1},y_{t2})$ were separated into color-string fragmentation, semi-hard parton fragmentation, and semi-hard multiplicity fraction fluctuations. Neglecting the relatively small centrality dependence in the shape of the single-particle distribution $\hat{\rho}_{\rm [g]}(y_t)$, the centrality dependence of the last contribution in Eq.~(\ref{Eq73}) is approximately proportional to hard-scattering multiplicity variance $\sigma_h^2 = \sigma_s^2$. The centrality dependence of $\sigma_s^2$ from Table~\ref{TableIV} is approximately described by a power-law where
\bea
\sigma_s^2 & \approx & 0.17 N_{\rm bin}^{1.19}
\label{Eq78}
\eea
for the centrality range 0-60\%. This distribution somewhat exceeds binary scaling. The second contribution in Eq.~(\ref{Eq73}) can be expanded in powers of $\zeta$, which to leading-order is given by the combination of terms
\bea
\frac{\bar{N}-1}{\bar{N}} (\bar{N}_h^2 + \sigma_s^2) \zeta  & & \nonumber \\
 & & \hspace{-1.3in} \times \left[ \left( \hat{\rho}_{\rm 2D[b]}(y_{t1},y_{t2}) 
- \hat{\rho}_{\rm [b]}(y_{t1}) \hat{\rho}_{\rm [b]}(y_{t2}) \right) \right. \nonumber \\
 & & \hspace{-1.3in} \left. + \left( \hat{\rho}_{\rm [g]}(y_{t1}) - \hat{\rho}_{\rm [b]}(y_{t1}) \right)
         \left( \hat{\rho}_{\rm [g]}(y_{t2}) - \hat{\rho}_{\rm [b]}(y_{t2}) \right) \right],
\nonumber \\
\label{Eq79}
\eea
where $\hat{\rho}_{\rm [b]}$ is defined in analogy to Eq.~(\ref{Eq49}) and using Eq.~(\ref{Eq65}) is given by
\bea
\hat{\rho}_{\rm [b]}(y_{t}) & \equiv & \int_0^{\infty} dy_{\rm max} \hat{b}(y_{\rm max}) \hat{\rho}_{h}(y_{\rm max},y_{t})
\nonumber \\
 & = & \frac{1}{\zeta} \left[ \hat{\rho}_{\rm [g]}(y_{t}) - (1-\zeta) \hat{\rho}_{\rm [h]}(y_{t})
\right].
\label{Eq80}
\eea
If the small centrality dependences of $\hat{\rho}_{\rm [g]}(y_{t})$ and $\hat{\rho}_{\rm [b]}(y_{t})$ are neglected, the number of correlated pairs in this contribution is approximately proportional to $(\bar{N}_h^2 + \sigma_s^2) \zeta \approx \bar{N}_h^2 \zeta$. From Table~\ref{TableIV} we find that for the 0-40\% more-central collisions, where $\zeta$ becomes smaller, the dependence of $(\bar{N}_h^2 + \sigma_s^2) \zeta$ can be approximated by
\bea
(\bar{N}_h^2 + \sigma_s^2) \zeta & \approx & 0.43 N_{\rm bin}^{1.44}.
\label{Eq81}
\eea
Thus we find that empirical descriptions of the AS-CI correlation pseudodata, in terms of the TCF model, are consistent with a scenario in which the number of correlated particle-pairs from semi-hard scattering and fragmentation processes increases smoothly with centrality and at a rate somewhat in excess of N+N binary scaling. 

The contributions of the three terms in Eq.~(\ref{Eq73}) for the 60-80\%, 20-30\% and 0-5\% centrality bins are shown in Fig.~\ref{Fig5} in comparison with the correlation pseudodata. For the pure color-string fragmentation contribution, parameters $\sigma_s^2$ and $\zeta$ were set to zero. For the pure semi-hard multiplicity fluctuation result, parameters $\Delta(1/q)_{cs,{\rm Vol}}$, $\Delta(1/q)_{cs,{\rm cov}}$ and $\zeta$ were set to zero. For the pure semi-hard fragmentation result, $\Delta(1/q)_{cs,{\rm Vol}}$, $\Delta(1/q)_{cs,{\rm cov}}$ and $\sigma_s^2$ were set to zero. The results accurately represent the contributions of the first two terms in Eq.~(\ref{Eq73}) to the extent that $\sigma_s^2 \ll \bar{N}_s^2$ and $\sigma_s^2 \ll \bar{N}_h^2$ which are true at the 1\% amount or better (see Tables~\ref{TableII} and \ref{TableIV}), except for the 60-80\% results. The color-string fragmentation contributes from about 20\% of the predicted correlation peak amplitude at $(y_{t1},y_{t2}) \approx (3,3)$ in most-peripheral collisions to about 9\% in most-central collisions. The semi-hard scattering contributions [last two terms in Eq.~(\ref{Eq73})] together account for the remaining 80\% to 91\% of the predicted correlation peak in 60-80\% and 0-5\% centrality bins, respectively. The semi-hard parton fragmentation contribution $(\zeta > 0)$ dominates the correlation peak at (3,3) in more-central collisions.

%%%%%%%%%%%%%%%%%%%%%%%%%%%%%%%%%%
\begin{table*}[htb]
\caption{Two-component fragmentation correlation model fit parameters to the 200~GeV Au+Au $(y_{t1},y_{t2})$ AS-CI correlation pseudodata. Statistical fitting errors are in parentheses.}
\label{TableIV}
\begin{tabular}{cccccccc}
\hline \hline
Cent.(\%) & $\Delta(1/q)_{cs,{\rm Vol}}$ & $\Delta(1/q)_{cs,{\rm cov}}$ & $\sigma_s^2$ & $\zeta$ & $y_{\rm cut}^{\star}$ & $n_{\rm QCD}^{\star}$ & $\frac{\chi^2}{\rm DoF}$ \\
\hline
 0-5   & -0.000285(47)   &  0.000415(60)  & 620(87) & 0.042(2) & 4.06(4)  & 11.4(1.3) & 11.3  \\
 5-10  & -0.000333(62)   &  0.000475(72)  & 540(64) & 0.051(3) & 4.08(5)  & 11.8(1.8) & 13.5  \\
 10-20 & -0.000382(62)   &  0.000535(70)  & 310(39) & 0.061(2) & 4.12(4)  & 11.5(1.4) & 22.7  \\
 20-30 & -0.000449(115)   &  0.000631(68)  & 190(17) & 0.072(3) & 4.16(6)  & 11.6(2.2) & 22.7  \\
 30-40 & -0.000635(144)  &  0.000876(92)  & 115(11)  & 0.102(4)   & 4.22(9)  & 11.8(3.8) & 17.0  \\
 40-60 & -0.001106(492)  &  0.001452(110)  &  31(7)  & 0.158(6)   & 4.06(11)  & 11.6(4.8) & 13.8  \\
 60-80 & -0.001964(1440) &  0.002556(157) &  51(9)  & 0.44(3)    & 4.18(18) & 11.6(7.4) & 9.4  \\
\hline \hline
\end{tabular}
\end{table*}
%%%%%%%%%%%%%%%%%%%%%%%%%%%%%%%%%%%%

%%%%%%%%%%%%%%%%%%%%%%%%%%%%%%
\begin{figure*}[t]
\includegraphics[keepaspectratio,width=6.5in]{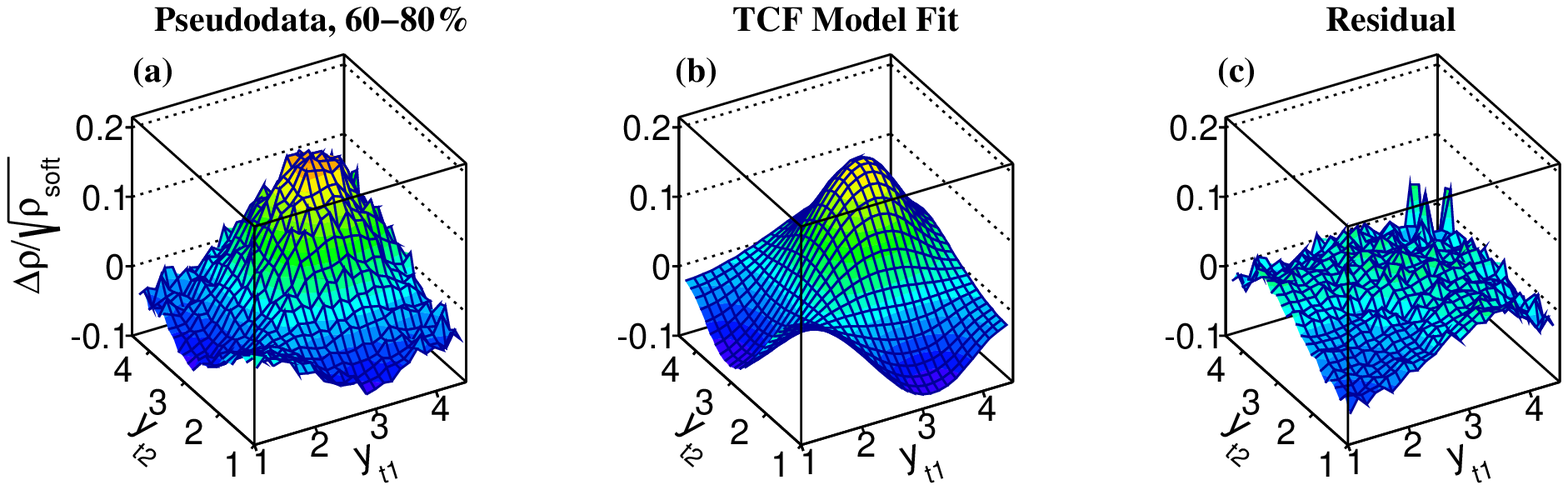} \\
%\put(-80,100){\bf (c)}
%\put(-230,100){\bf (b)}
%\put(-380,100){\bf (a)} \\
\includegraphics[keepaspectratio,width=6.5in]{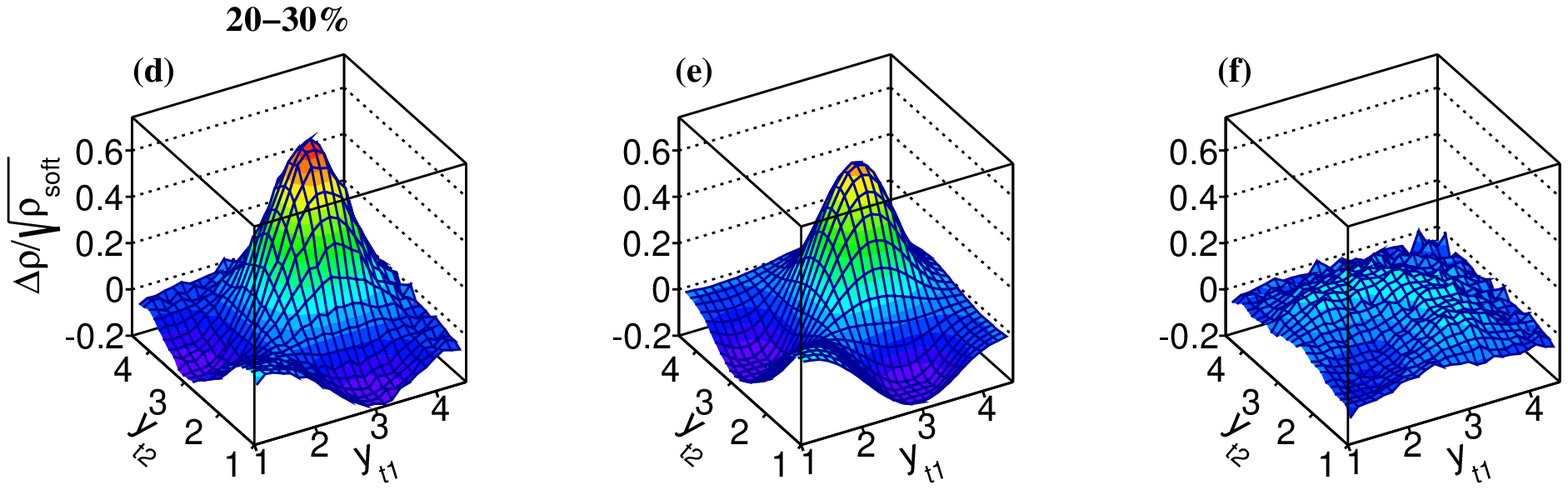} \\
%\put(-80,100){\bf (f)}
%\put(-230,100){\bf (e)}
%\put(-380,100){\bf (d)} \\
\includegraphics[keepaspectratio,width=6.5in]{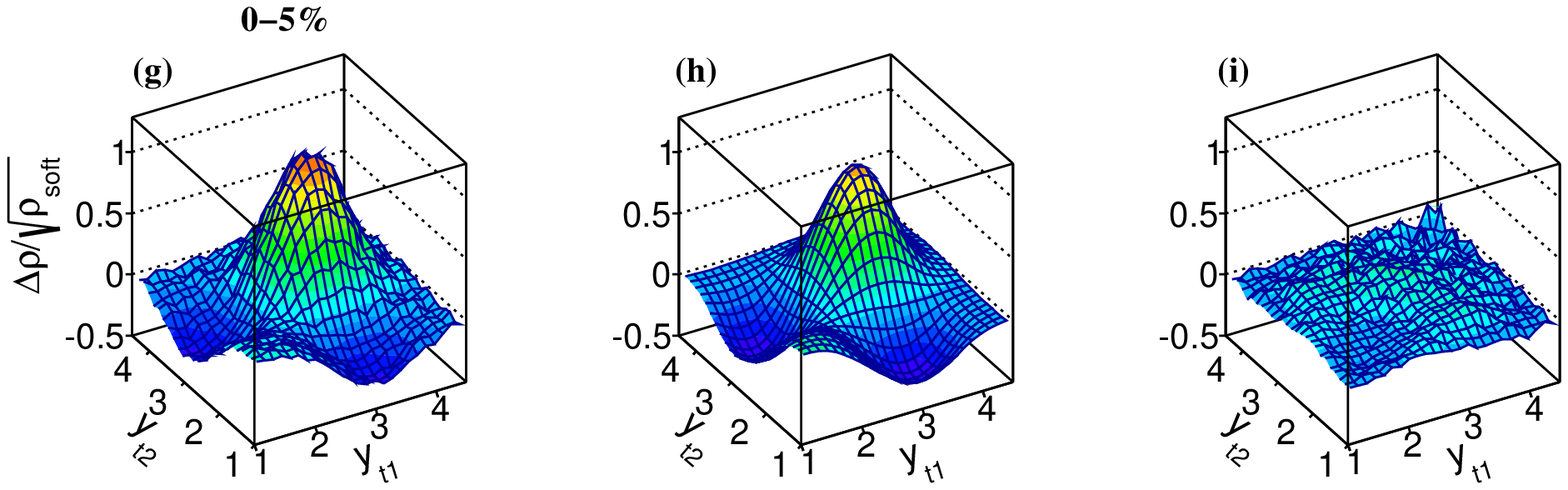}
%\put(-80,100){\bf (i)}
%\put(-230,100){\bf (h)}
%\put(-380,100){\bf (g)}
\caption{\label{Fig4}
Same as Fig.~\ref{Fig3} except for the TCF model.}
\end{figure*}
%%%%%%%%%%%%%%%%%%%%%%%%%%%%%%%%%%

%%%%%%%%%%%%%%%%%%%%%%%%%%%%%%
\begin{figure*}[t]

\includegraphics[keepaspectratio,width=6.7in]{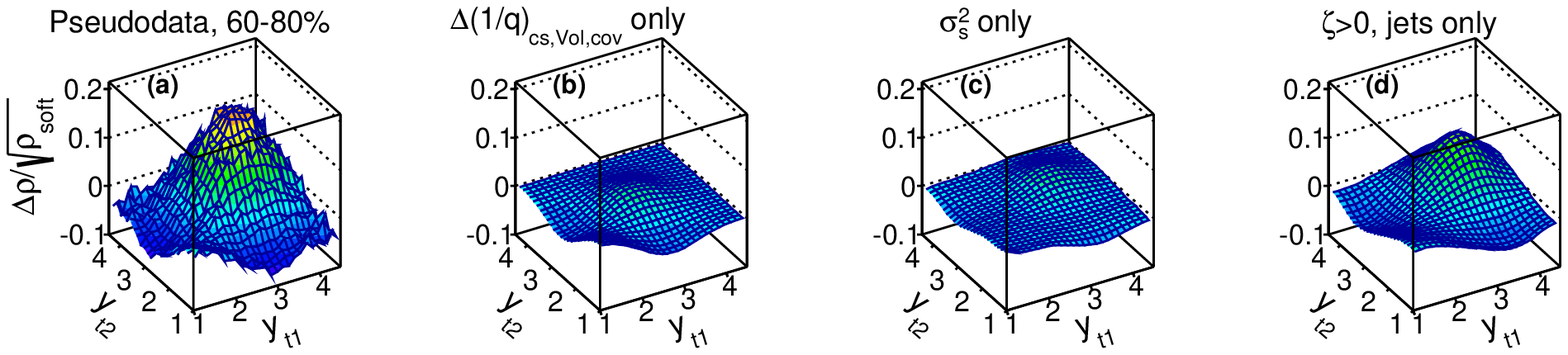}
%\put(-80,100){\bf (c)}
%\put(-230,100){\bf (b)}
%\put(-380,100){\bf (a)} \\
\includegraphics[keepaspectratio,width=6.7in]{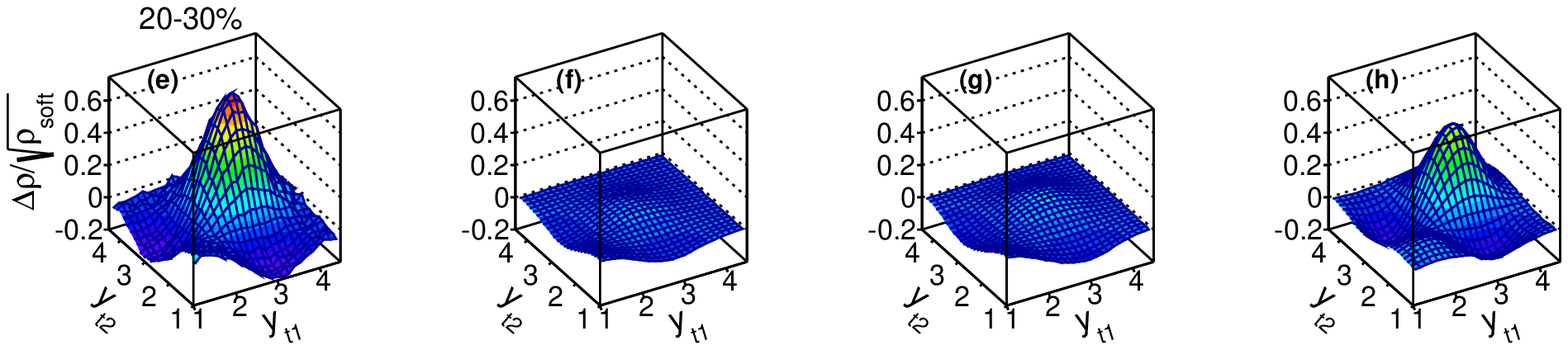}
%\put(-80,100){\bf (f)}
%\put(-230,100){\bf (e)}
%\put(-380,100){\bf (d)} \\
\includegraphics[keepaspectratio,width=6.7in]{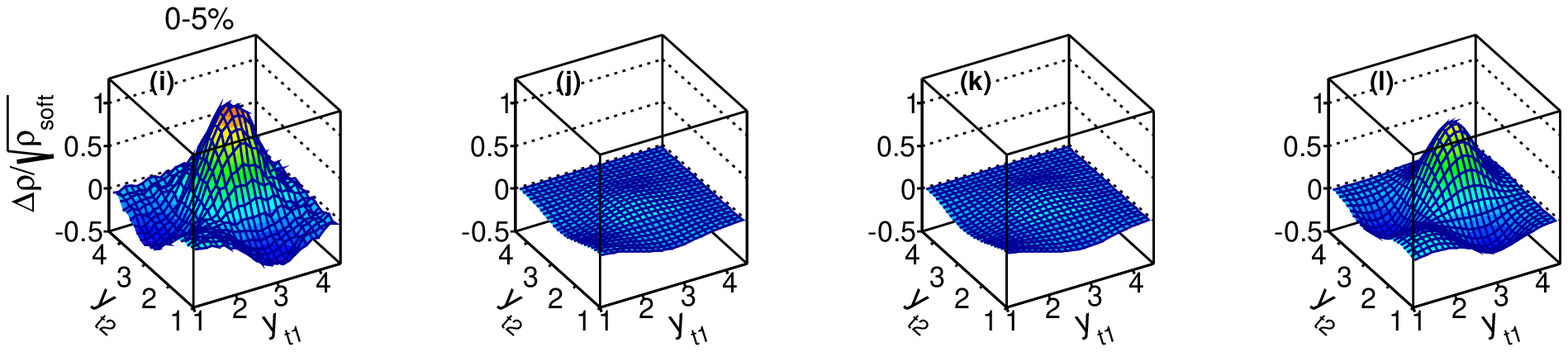}
%\put(-80,100){\bf (i)}
%\put(-230,100){\bf (h)}
%\put(-380,100){\bf (g)}
\caption{\label{Fig5}
Separate contributions to the fluctuating TCF model fits to the 200 GeV Au+Au away-side, charge-independent two-particle correlation pseudodata for selected centralities 60-80\%, 20-30\% and 0-5\% in rows of panels from upper to lower, respectively. The left-hand column of panels shows the pseudodata. Fluctuation contributions from color-strings, semi-hard multiplicity production, and semi-hard fragmentation are shown in the second, third and fourth columns of panels, respectively, as explained in the text.}
\end{figure*}
%%%%%%%%%%%%%%%%%%%%%%%%%%%%%%%%%%

\section{Summary and Conclusions}
\label{SecVI}

The study of relativistic heavy-ion collisions has greatly benefited from the plethora of two-particle correlation measurements and analysis over many years~\cite{Tomreview}. The vast majority of these correlation studies has focused on angular correlations. On the other hand, complementary correlation measurements on 2D transverse momentum are relatively scarce in the literature. In our opinion the scientific impact of the correlations on transverse momentum which do exist has been diminished by the lack of available theoretical predictions and the absence of phenomenologically based interpretations.

To address this deficiency we developed two phenomenological models based on fundamentally different frameworks for describing the dynamical evolution of the heavy-ion collision system. The first is based on hydrodynamic expansion as parametrized in the blast-wave model in which pair-wise correlated fluctuations in the temperature and transverse flow at kinematic freeze-out are included in order to generate two-particle correlations in the final state. The second model is based on soft-QCD, longitudinal color-string fragmentation and semi-hard QCD, transverse scattering and fragmentation in which fluctuations occur in the energies of the color-strings and in the four-momentum transfer in the QCD scatterings, as well as in the relative numbers of particles produced via soft and semi-hard processes.

We demonstrated that both models are capable of quantitatively describing the measured charged-particle $p_t$ spectra produced in $\sqrt{s_{\rm NN}}$ = 200~GeV Au+Au minimum-bias collisions. Using analytic representations of preliminary two-particle correlations on 2D transverse rapidity from the STAR Collaboration~\cite{LizHQ,LizThesis}, we further demonstrated that both models are capable of qualitatively describing the correlations, resulting in smooth, monotonic centrality dependent trends in most of the model parameters. The phenomenological model parameters and their resulting centrality trends can be interpreted in terms of the dynamical processes inherent in each model.

The results of this ``proof of principle'' study already provide some new physical insight and impose constraints on the two dynamical frameworks considered here. In the hydrodynamic, BW approach we found that statistical fluctuations in the number of participant nucleons from event-to-event, as the sole source of final-state fluctuations in the $p_t$ distribution, are much too small to account for the observed correlation structures. Much larger, dynamical fluctuations are required whose effects must persist until kinetic freeze-out, thus restricting the degree of dissipation in the collision medium. The BW results also imply that the magnitudes of intra-event temperature fluctuations far exceed the inter-event fluctuations in the mean temperature. This result may, for example, limit the allowed spatial scale for local, thermodynamic equilibrium in such models.

In the two-component fragmentation approach we found that the semi-hard scattering and fragmentation induced correlations required to describe the data appear to exceed binary scaling which suggests additional, multi-parton dynamics are required in the initial-state or during fragmentation within the dense medium. We also found that in more-central collisions the peak correlation structure at $(y_{t1},y_{t2}) \approx (3,3)$ is dominated by semi-hard parton fragmentation (minijets).

The connection between physical properties of the heavy-ion collision medium, {\em e.g.} temperature and flow velocity, inferred from analysis of single-particle $p_t$ spectrum data, and event-wise fluctuations was emphasized. Using the BW model we showed that fluctuations in the temperature and transverse flow affect the inferred, mean temperature and flow velocity by as much as a factor of two. Physical parameters inferred from fits to spectrum data using models without fluctuations are questionable.

The BW and TCF phenomenological models developed here can be used in future analyses of two-particle correlation measurements on transverse momentum or transverse rapidity to facilitate physical interpretation of the correlation structures and to better constrain theoretical models. Both phenomenologies can be used to estimate the magnitude and type of fluctuations required, within their respective frameworks, to describe correlation data. The magnitudes and centrality trends of those required fluctuations can be compared to the capabilities of theoretical models for producing such fluctuations. In this way, phenomenological analysis of two-particle correlations on transverse momentum may enable more informed estimates of the validity of different theoretical approaches for understanding relativistic heavy-ion collisions.

\vspace{0.2in}
{\bf Acknowledgements}
\vspace{0.1in}

The authors would like to thank Professor Thomas Trainor of the Univ. of Washington for many informative discussions relevant to this work and Professor Rainer Fries of Texas A\&M University for discussions related to the blast-wave model. This research was supported in part by the Office of Science of the U. S. Department of Energy under Grants No. DE-FG02-94ER40845 and No. DE-SC0013391.

\vspace{0.2in}
\begin{center}
{\bf Appendix A}
\end{center}
\vspace{0.1in}
The soft-reference prefactor for away-side pairs and all charged particles is given by
\bea
{\cal P}^{\rm AS-CI}_{\rm Fac,soft} & \equiv & \frac{1}{\sqrt{2}}
\frac{
      \frac{d^2N_{\rm ch}}{dy_{t1} d\eta_1} \frac{d^2N_{\rm ch}}{dy_{t2} d\eta_2}
     }
     {
\left[ \frac{d^2N_{\rm ch,soft}}{dy_{t1} d\eta_1} \frac{d^2N_{\rm ch,soft}}{dy_{t2} d\eta_2} \right]^{1/2}
     }
\label{EqA1}
\eea
where the distributions are calculated at the mid-points of each $y_t$-bin and factor $1/\sqrt{2}$ accounts for using only away-side pairs.
In this equation the charged particle distribution was parametrized with a Levy distribution where
\bea
\frac{d^2N_{\rm ch}}{dy_t d\eta} & = & 2\pi p_t \frac{dp_t}{dy_t} \left[
\frac{d^2N_{\rm ch}}{2\pi p_t dp_t d\eta} \right] \nonumber \\
 & = & \frac{2\pi p_t m_t A_{\rm ch}}{\left[1+(m_t - m_0)/(T_{\rm ch}q_{\rm ch}) \right]^{q_{\rm ch}}}.
\label{EqA2}
\eea
Fit parameters $A_{\rm ch}$, $T_{\rm ch}$ and $q_{\rm ch}$ for the 200~GeV Au+Au spectra data reported by the STAR Collaboration~\cite{STARspectra} were determined in the $y_t$ range from 1.34 to 4.36, corresponding to $p_t \in [0.25,5.5]$~GeV/$c$, and are listed in Table~\ref{TableV}. The $N_{\rm part}$ scaling, Kharzeev and Nardi soft-QCD process spectrum was also parametrized with the Levy distribution and is given by,
\bea
\frac{d^2N_{\rm ch,soft}}{dy_t d\eta} & = & \frac{2\pi p_t m_t A_{\rm soft}(N_{\rm part}/2)}{\left[1+(m_t - m_0)/(T_{\rm soft}q_{\rm soft}) \right]^{q_{\rm soft}}}.
\label{EqA3}
\eea
The number of participants for 200~GeV Au+Au minimum-bias collisions was estimated in Ref.~\cite{AxialCI} and interpolated to the present centrality bins (see Table~\ref{TableV}). A method for estimating the  $N_{\rm part}$ scaling, soft-QCD process spectrum was presented in Ref.~\cite{Trainorspectra}. In the present analysis the soft-QCD distribution was estimated by extrapolating the STAR~\cite{STARspectra} and PHENIX~\cite{PHENIXspectra} Collaborations' $p_t$ spectra data in each $p_t$ bin to the $\nu \rightarrow 1$, N+N collision limit and fitting the resulting distribution with the Levy model in Eq.~(\ref{EqA3}). The resulting fits gave $A_{\rm soft}$ = 5.81~($c$/GeV$^2$), $T_{\rm soft}$ = 0.169~GeV, and $q_{\rm soft}$ = 13.8.

%%%%%%%%%%%%%%%%%%%%%%%%%%%%%%%%%%%%
\begin{table}[htb]
\caption{Levy model fit parameters to the STAR Collaboration 200~GeV Au+Au minimum-bias charged particle $p_t$ spectra data in Ref.~\cite{STARspectra} in the range $y_t \in [1.34,4.36]$. Also listed are the number of participant nucleons, number of binary N+N collisions, and centrality measure $\nu$~\cite{AxialCI}. The estimated soft-process $p_t$ spectrum Levy model parameters in Eq.~(\ref{EqA3}) are: $A_{\rm soft}$ = 5.81~($c$/GeV$^2$), $T_{\rm soft}$ = 0.169~GeV, and $q_{\rm soft}$ = 13.8.}
\label{TableV}
\begin{tabular}{ccccccc}
\hline \hline
Centrality & $\nu$ & $N_{\rm part}$ & $N_{\rm bin}$ & $A_{\rm ch}$ & $T_{\rm ch}$ & $q_{\rm ch}$ \\
  (\%)     &       &                &               & ($c$/GeV$^2$)  & (GeV)    & \\
\hline
0-5   &   5.95   &  350.3  &  1042  &  1154.8  &  0.2176  &  17.41  \\
5-10  &   5.50   &  299.5  &   824  &   935.8  &  0.2167  &  17.15  \\
10-20 &   4.98   &  233.7  &   582  &   724.2  &  0.2129  &  16.00  \\
20-30 &   4.34   &  166.4  &   361  &   503.4  &  0.2090  &  15.09  \\
30-40 &   3.75   &  116.1  &   218  &   350.3  &  0.2036  &  14.28  \\
40-60 &   2.87   &  59.8   &   85.7 &   205.76 &  0.1882  &  12.49  \\
60-80 &   1.97   &  19.5   &   19.2 &   77.96  &  0.1695  &  11.06  \\
\hline \hline
\end{tabular}
\end{table}
%%%%%%%%%%%%%%%%%%%%%%%%%%%%%%%%%%%%

\vspace{0.2in}
\begin{center}
{\bf Appendix B}
\end{center}
\vspace{0.1in}

Analytic representations of preliminary charged particle correlations on $(y_{t1},y_{t2})$ are described here and in Refs.~\cite{LizHQ,LizThesis}. Preliminary, charged-particle correlations on $(y_{t1},y_{t2})$ in the range $y_t \in [1.0,4.5]$ for minimum-bias Au+Au collisions at $\sqrt{s_{NN}}$ = 200~GeV from the STAR Collaboration were reported by Oldag~\cite{LizHQ,LizThesis}. The same-event and mixed-event pair densities were both normalized to the total number of pairs as this analysis predates the methods developed in Ref.~\cite{MCBias}. The data were fitted with a 2D-Levy distribution [Eq.~(\ref{Eq59})] plus a constant offset and a 2D Gaussian. The correlations described with this model include all away-side, charged-pair combinations. The AS angular selection eliminates the enhanced correlation structure along the $y_{t1} = y_{t2}$ main-diagonal caused by quantum correlations between identical bosons~\cite{HBT} as discussed in~\cite{Ayamtmt}. The analytical fitting function is given by
\bea
\frac{\Delta\rho}{\sqrt{\rho_{\rm soft}}}|_{\rm AS-CI} & = & {\cal P}^{\rm AS-CI}_{\rm Fac,soft}
\left( \frac{\hat{\rho}_{\rm 2D-Levy}^{\rm data} - \hat{\rho}_{\rm 2D-mix}}{\hat{\rho}_{\rm 2D-mix}} \right) \nonumber \\
 &  & \hspace{-0.5in} + A_0 + A_1 e^{-y_{t\Delta}^2/2\sigma_{\Delta}^2} e^{-(y_{t\Sigma}-2y_{t0})^2/2\sigma_{\Sigma}^2}
\label{EqB1}
\eea
where the 2D-Levy distribution is the same as in Eq.~(\ref{Eq59}) with parameters $\beta_0$, $q_{\Sigma}$ and $q_{\Delta}$ in Ref.~\cite{LizThesis} replacing parameters $\bar{\beta}_{cs}$, $q_{\beta_{cs\Sigma}}$ and $q_{\beta_{cs\Delta}}$ in Eq.~(\ref{Eq59}). The corresponding variance difference quantities are given by $\Delta(1/q)_{\Sigma,\Delta} = 1/q_{\Sigma,\Delta} - 1/q_{\rm fluct}$. Also in Eq.~(\ref{EqB1}) we introduced sum and difference variables $y_{t\Sigma,\Delta} = y_{t1} \pm y_{t2}$. The marginal of $\hat{\rho}_{\rm 2D-Levy}^{\rm data}$ is given by
\bea
\hat{\rho}_{\rm marg}(y_{t1}) & = & \int dy_{t2} \hat{\rho}_{\rm 2D-Levy}^{\rm data}(y_{t1},y_{t2})
\label{EqB2}
\eea
and the mixed-event reference $\hat{\rho}_{\rm 2D-mix}(y_{t1},y_{t2})$ is the product of marginals for particles 1 and 2.

The 2D-Levy distribution alone did not produce satisfactory descriptions of the data and was supplemented with a constant offset ($A_0$) plus a 2D Gaussian. Fit parameters $\Delta(1/q)_{\Sigma,\Delta}$, $q_{\rm fluct}$, $A_0$, $A_1$, $y_{t0}$, $\sigma_\Delta$ and $\sigma_\Sigma$ were interpolated from the trends plotted in  Fig.~5.14 of Ref.~\cite{LizThesis}, at the mid-points of the centrality bins studied here. The 2D Gaussian widths along the difference direction $y_{t\Delta}$ exceeded the corresponding widths along $y_{t\Sigma}$. Physically, for the AS correlations, this could be caused by initial-state transverse momentum, $K_T$, in the parton-parton collision frame which would impart more $p_t$ to the fragments of one jet than the other, resulting in a broadening along $y_{t\Delta}$ when averaged over many dijets. Such additional, initial-state dynamics could be included in both the BW and TCF models but, for simplicity, was not accounted for in this initial ``proof-of-principle'' model study. The width $\sigma_\Delta$ in the pseudodata was therefore set equal to $\sigma_\Sigma$.

The correlation pseudodata were assigned statistical errors corresponding to the number of pairs per bin expected for the 9.5 million, 200~GeV minimum-bias Au+Au collisions in the data volume reported in \cite{LizThesis}, for the observed charged-particle $p_t,\eta$ distributions in centrality bins 0-5\%, 5-10\%, 10-20\%, 20-30\%, 30-40\%, 40-60\% and 60-80\%, for single-particle acceptance $|\eta| \leq 1$, $p_t \geq 0.15$~GeV/$c$, full $2\pi$ azimuth, and assuming symmetric correlations with respect to $\pm|y_{t1} - y_{t2}|$. The latter symmetrization is valid when particles 1 and 2 are taken from the same collection of particles, for example all charged particles.  This step was implemented by counting each unique particle pair in both bins with coordinates $(y_{t1},y_{t2})$ and $(y_{t2},y_{t1})$. For diagonal bins $(y_{t},y_{t})$, only the $y_{t1} \geq y_{t2}$ half was used for calculating the statistical errors. Typical statistical errors (for $y_t \leq 3$) in more-central collisions vary from approximately 1\% to 3\% relative to the correlation amplitude at the peak near $(y_{t1},y_{t2})=(3,3)$. The errors increase to the range 3\% to 5\% in more-peripheral collisions. The pseudodata were generated in each $(y_{t1},y_{t2})$ bin by sampling a Gaussian distribution whose mean equals the calculated value in Eq.~(\ref{EqB1}) and whose width parameter ($\sigma$) was equal to the estimated statistical error. The correlation pseudodata were binned on a uniform $25\times25$ 2D grid for $y_t \in [1.0,4.5]$ corresponding to $p_t \in [0.16,6.3]$~GeV/$c$. Pseudodata were generated for $(y_{t1},y_{t2})$ bins with $y_{t1} \geq y_{t2}$, and then copied to the $(y_{t2},y_{t1})$ bin.

%%%%%%%%%%%%%%%%%%%%%%%%%%%%%%%%%%%%%%%%%%%%%%%%%%%%%%%%%%%%%%%%%%%%%%%%%%%

\end{document}